\documentclass[11pt]{article}
\usepackage{rotating}
\usepackage{graphicx}
\usepackage{float}
\usepackage{lscape}
\usepackage[T1]{fontenc}
\usepackage[usenames,dvipsnames,svgnames,table]{xcolor}
\usepackage[colorlinks=true,
            linkcolor=red,
            urlcolor=blue,
            citecolor=blue]{hyperref}
            \usepackage[export]{adjustbox}
\usepackage{multirow}            
\usepackage{fancyhdr}
\usepackage{dlfltxbcodetips}
\usepackage{graphicx}
\usepackage{graphics}
\usepackage{color}
\usepackage{wrapfig}
\usepackage{lscape}
\usepackage{pifont}
\usepackage[T1]{fontenc}
\usepackage[utf8]{inputenc}
\usepackage{mathtools}   % 
\usepackage{rotating}
\usepackage{natbib}
\usepackage{setspace}
\usepackage{amsfonts}
\usepackage{amssymb}
\usepackage{amsmath}
\usepackage{amsthm}
\usepackage{mathrsfs}
\title{Semiparametric Estimation of First-Price Auction Models \footnote{ We thank St\'ephane Bonhomme, Han Hong (the co-editor), Isabelle Perrigne and three anonymous referees for their insightful comments and suggestions that improved the content and the exposition of the paper. The usual caveats apply.}}
\author{Gaurab Aryal \\ University of Chicago\\ \texttt{aryalg@uchicago.edu} 
\and 
Maria Florencia Gabrielli \footnote{ Corresponding Author}\\
CONICET \&\\
Universidad Nacional de Cuyo\\
\texttt{florgabrielli@gmail.com}
\and 
Quang Vuong \\
 New York University\\  \texttt{qvuong@nyu.edu}}
\newcommand{\Real}{\mathbb{R}}

\usepackage{bbm}
\newcommand{\Unit}{\mathbbm{1}}
\newcommand{\ds}{\displaystyle{\frac{1}{L}\sum_{\ell=1}^L\frac{1}{I_\ell}\sum_{p=1}^{I_\ell}}}
\newcommand{\dsnew}{\displaystyle{\frac{1}{L}\sum_{\{\ell:I_\ell=I\}}^L\frac{1}{I}\sum_{p=1}^I}}
\newcommand{\dsh}{1/L\sum_{\ell=1}^L1/I_\ell\sum_{p=1}^{I_\ell}}
\newcommand{\dsj}{\displaystyle{\frac{1}{L}\frac{L}{n_I}\sum_{\{j:I_j=I\}}^L\sum_{q=1}^{I}}}
\newcommand{\dsi}{\displaystyle{\frac{1}{L}\sum_{\ell=1}^L\frac{1}{I_\ell(I_\ell-1)}\sum_{p=1}^{I_\ell}}}
\newcommand{\dsinew}{\displaystyle{\frac{1}{L}\sum_{\{\ell:I_\ell=I\}}^L\frac{1}{I(I-I)}\sum_{p=1}^I}}
\newcommand{\dsils}{\displaystyle{\frac{1}{L}\frac{L}{n_I}\sum_{\{\ell:I_\ell=I\}}^L\frac{1}{I(I-1)}\sum_{p=1}^{I}}}
\newcommand{\dstwo}{\displaystyle{\frac{2}{L}\sum_{\{\ell:I_\ell=I\}}^L\frac{1}{I}\sum_{p=1}^{I}}}

\newcommand{\ls}{\displaystyle{\frac{1}{L}\sum_{\ell=1}^L}}
\newcommand{\dsls}{\displaystyle{\frac{1}{L^2}\frac{L}{n_I}\sum_{\{\ell:I_\ell=I\}}^L\frac{1}{I(I-1)}\sum_{p=1}^{I}}}
\newcommand{\dslsnew}{\displaystyle{\frac{1}{L^2}\frac{L}{n_I}\sum_{\{\ell:I_\ell=I\}}^L\frac{1}{I(I-1)}\sum_{p=1}^{I}}}

\newcommand{\dskGj}{{\displaystyle{\frac{1}{Lh_{G}}\frac{L}{n_I}\sum_{\{j:I_j=I\}}^L\sum_{q=1}^{I}}e_1^T \left(\frac{X_{I,R+1}^T W_x^G X_{I,R+1}}{L_I}\right)^{-1} X_{R+1,j} K_G\left(\frac{X_j-X_\ell}{h_{G}}\right)\Unit(B_{qj}\leq B_{p\ell})}}

\newcommand{\dskgj}{{\displaystyle{\frac{1}{Lh_{g}^2}\frac{L}{n_I}\sum_{\{j:I_j=I\}}^L\sum_{q=1}^{I}} e_1^T \left(\frac{X_{I,R}^T W_x^g X_{I,R}}{n_I}\right)^{-1} X_{R,j} K_{1g}\left(\frac{X_j-X_\ell}{h_{g}}\right)K_{2g}\left(\frac{B_{qj}-B_{p\ell}}{h_{g}}\right)}}

\newcommand{\tslminus}{{\displaystyle{\frac{1}{L (L-1)}\frac{L}{n_I}\sum_{\{\ell:I_\ell=I\}}^L
\sum_{\{j: I_j=I,j\neq \ell\}}^L\frac{1}{I(I-1)}\sum_{p=1}^{I}\sum_{q=1}^{I}}}}

\newcommand{\tslminusone}{{\displaystyle{\frac{1}{L (L-1)}\sum_{\{\ell:I_\ell=I\}}^{L-1}
\sum_{\{j: I_j=I,j=\ell+1\}}^L\frac{1}{I}\sum_{p=1}^{I}\sum_{q=1}^{I}}}}

\newcommand{\tslminusonetwo}{{\displaystyle{\frac{2}{L (L-1)}\sum_{\{\ell:I_\ell=I\}}^{L-1}
\sum_{\{j: I_j=I,j=\ell+1\}}^L\frac{1}{I}\sum_{p=1}^{I}\sum_{q=1}^{I}}}}

\newcommand{\ts}{{\displaystyle{\frac{1}{L^2}\frac{L}{n_I}\sum_{\{\ell:I_\ell=I\}}^L
\sum_{\{j: I_j=I\}}^L\frac{1}{I(I-1)}\sum_{p=1}^{I}\sum_{q=1}^{I}}}}

\newcommand{\tsd}{{\displaystyle{\frac{1}{L^2}\frac{L}{n_I}\sum_{\{\ell:I_\ell=I\}}^L
\sum_{\{j: I_j=I,j\neq \ell\}}^L\frac{1}{I(I-1)}\sum_{p=1}^{I}\sum_{q=1}^{I}}}}

\newcommand{\tsnd}{{\displaystyle{\sum_{\{\ell:I_\ell=I\}}^{L-1}
\sum_{\{j: I_j=I,j=\ell+1\}}^L\frac{1}{I}\sum_{p=1}^{I}\sum_{q=1}^{I}}}}

\newcommand{\sumj}{{\displaystyle{\sum_{I}\frac{1}{(I-1)}}}}

\newcommand{\htr}{{\displaystyle{\frac{\partial S_L^{T}}{\partial \theta}}}}

\newcommand{\Htr}{{\displaystyle{\frac{\partial \hat S_L^T}{\partial \theta}}}}
\newcommand{\htt}{{\displaystyle{\frac{\partial S_L}{\partial \theta^{T}}}}}
\newcommand{\Htt}{{\displaystyle{\frac{\partial \hat S_L}{\partial \theta^{T}}}}}

\newcommand {\imfl}{\frac{{L}}{n_I(I-1)}}

\newcommand {\ikGqj}{{\omega_{I,R+1,j}^G    K_{G,h_G}(X_j-X_\ell)\Unit(B_{qj}\leq B_{p\ell})}}

\newcommand {\imoneratioKGu}{\frac{m_1(\xi(uh_{G}+Y_{qj},I),u_2h_{G}+X_j,I;\theta_0)}{g_0(u_1h_{G}+B_{qj}|u_2h_{G}+X_j,I)}}
\newcommand {\izmone}{\frac{m_1(V_{p\ell},X_\ell,I;\theta_0)}{g_0(B_{p\ell}|X_\ell,I)}}
\newcommand {\izmonexi}{\frac{m_1(\xi(B_{p\ell},X_\ell,I),X_\ell,I;\theta_0)}{g_0(B_{p\ell}|X_\ell,I)}}
\newcommand {\izjmone}{\frac{m_1(V_{qj},X_j,I;\theta_0)}{g_0(B_{qj}|X_j,I)}}
\newcommand {\izjmoneu}{m_1(\xi(u h_{g}+Y_{qj},I),u_2h_{g}+X_j,I;\theta_0)}

\newcommand {\ratioigj}{\frac{m_1(V_{qj},X_j,I;\theta_0)}{g_0(B_{qj}|X_j,I)}}
\newcommand {\ratioGg}{\frac{G_0(B_{p\ell}|X_\ell,I)}{g_0(B_{p\ell}|X_\ell,I)}}
\newcommand {\ratioGgj}{\frac{G_0(B_{qj}|X_j,I)}{g_0(B_{qj}|X_j,I)}}
\newcommand {\ratioGgju}{\frac{G_0(u_1h_{g}+B_{qj}|u_2h_{g}+X_j,I)}{g_0(u_1h_{g}+B_{qj}|u_2h_{g}+X_j,I)}}

\newcommand {\kgqj}{{\omega_{I,R,j}^g K_{1g,h_g}(X_j-X_\ell)K_{2g,h_g}(B_{qj}-B_{p\ell})}}

\newcommand {\kG}{{\omega_{I,R+1,j}^G K_G\left(\frac{X_j-X_\ell}{h_{G}}\right) \Unit(B_{qj}\leq B_{p\ell})}}
\newcommand {\kGj}{{\omega_{I,R+1,j}^G K_G\left(\frac{X_\ell-X_j}{h_{G}}\right) \Unit(B_{p\ell}\leq B_{qj})}}
\newcommand {\kgz}{{\omega_{I,R,j}^g K_{1g}\left(\frac{X_j-X_\ell}{h_{g}}\right) K_{2g}\left(\frac{B_{qj}-B_{p\ell}}{h_{g}}\right)}}
\newcommand {\kgzj}{{\omega_{I,R,j}^g K_{1g}\left(\frac{X_\ell-X_j}{h_{g}}\right) K_{2g}\left(\frac{B_{p\ell}-B_{qj}}{h_{g}}\right)}}

\newcommand {\hGd}{{{\frac{1}{h_{G}}}}}

\newcommand {\hGtwod}{{{\frac{1}{2h_{G}^{2}}}}}

\newcommand {\hgdone}{{{\frac{1}{h_{g}^2}}}}
\newcommand {\hgdonehalf}{{{\frac{1}{2h_{g}^2}}}}
\newcommand {\hgtwod}{{{\frac{1}{2h_{g}^{4}}}}}
\newcommand {\mo}{{m(V_{p\ell},Z_\ell;\theta_0)}}
\newcommand {\mhato}{{m(\hat V_{p\ell},Z_\ell;\theta_0)}}
\newcommand {\moneo}{{m_1(V_{p\ell},Z_\ell;\theta_0)}}

\newcommand {\moneob}{{m_1(V^*_{p\ell},Z_\ell;\theta_0)}}
\newcommand {\moneu}{{m_1(\xi(uh_{G}+Y_{qj},I),u_2h_{G}+X_j,I;\theta_0)}}

\newcommand {\monet}{{m_1(V_{p\ell},Z_\ell;\theta)}}

\newcommand {\Gtilde}{{\hat{G}(B_{p\ell}|Z_\ell)}}
\newcommand {\Gtildei}{{\hat{G}(B_{p\ell}|X_\ell,I)}}
\newcommand {\gtilde}{{\hat{g}(B_{p\ell}|Z_\ell)}}
\newcommand {\gtildei}{{\hat{g}(B_{p\ell}|X_\ell,I)}}
\newcommand {\G}{{G_0(B_{p\ell}|Z_\ell)}}
\newcommand {\Gi}{{G_0(B_{p\ell}|X_\ell,I)}}
\newcommand {\g}{{g_0(B_{p\ell}|Z_\ell)}}
\newcommand {\gi}{{g_0(B_{p\ell}|X_\ell,I)}}
\newcommand {\gj}{{g_0(B_{qj}|X_j,I)}}

\newcommand {\Gj}{{G_0(B_{qj}|X_j,I)}}

\newcommand {\gz}{{g_0(Y_{p\ell}|I)}}
\newcommand {\gzj}{{g_0(Y_{qj}|I)}}

\newcommand {\gjoint}{{g_0(B_{p\ell},X_\ell,I)}}
\newcommand {\gjointj}{{g_0(B_{qj},X_j,I)}}
\newcommand {\guhG}{{g_0(uh_{G}+Y_{qj}|I)}}

\newcommand {\gju}{{g_0(uh_{g}+Y_{qj}|I)}}
\newcommand {\Gzju}{{G_0(u_1h_{g}+B_{qj}|u_2h_{g}+X_j,I)}}
\newcommand {\pl}{{p_L((Y_{p\ell},I),(Y_{qj},I))}}
\newcommand {\plj}{{p_L((B_{pj},X_j,I),(B_{qj},X_j,I))}}
\newcommand {\dypl}{{dY_{p\ell}}}
\newcommand {\dypj}{{dY_{qj}}}
\newcommand{\Ch}{{\mathcal F}}

\newtheorem{proposition}{Proposition}

\date{\today}

\begin{document}

\begin{titlepage}

\maketitle

\normalsize

\begin{abstract}
\thispagestyle{empty}

We propose a semiparametric
method to estimate the density of private values in first-price auctions. 
Specifically, we
model private values through a set of conditional
moment restrictions and use a two-step procedure.
In the first step we recover a sample of pseudo
private values using Local Polynomial Estimator. In the second step
we use a GMM procedure to estimate the parameter(s) of interest. We show that the proposed semiparametric estimator is consistent, has an asymptotic normal distribution, and attains the parametric (``root-n'') rate of convergence.

\bigskip


\end{abstract}
\end{titlepage}

\newpage
\setcounter{page} {1}
\normalsize

\doublespacing
\section{Introduction}

\indent
From a theoretical point of view, auctions are modeled as games of
incomplete information in which asymmetric information among players
(seller/buyer and bidders) is one of the key features, \cite{Krishna2002, McAfeeMcMillan1987, Wilson1992}.
 From an applied perspective, as auction is a widely used
mechanism to allocate goods and services, many data sets are available for empirical research. 
By assuming
that observed bids are the equilibrium outcomes of an underlying
auction model under consideration, the structural approach to
provides a framework analyze auction data in which the theoretical
model and its empirical counterpart are closely related. The main
objective of this approach is then to recover the structural
elements of the auction model. This line of research has been
considerably developed in the last fifteen years. The difficulties
in estimating auction models are many. First, auction models lead to
nonlinear econometric models through the equilibrium
strategies. Second,  auction models may not lead to a closed-form solution making the derivation of an
econometric model even more difficult. Third, the estimation of auction models often
requires the numerical computation of the equilibrium strategies.
Some important work in this are documented by \cite{PerrigneVuong1999, PaarschHong2006, AtheyHaile2007, PerrigneVuong2008}, among others.

We distinguish two methods for estimating structural
auction models: direct method and indirect method. Direct methods
were developed first, and they rely on parametric
econometric models. Starting from a specification of the underlying
distribution of private values, the objective of direct methods is
to estimate the parameter vector characterizing such a distribution.
Within this class of methods, there are two major estimation
procedures. The first methodology, introduced by \cite{Paarsch1992,DonaldPaarsch1993}, 
 is a fully parametric setup that uses
Maximum Likelihood based estimation procedures requiring the
computation of the equilibrium strategy. 
Since it is computationally demanding, \cite[see][]{DonaldPaarsch1993}, only very simple distributions are considered in
practice. 
Moreover, because the support of the bid
distribution depends on the estimated parameter(s), it has a nonstandard limiting distribution, \cite[see][]{HiranoPorter2003}. In view of
this, \cite{DonaldPaarsch1993} develop a piecewise pseudo
maximum-likelihood estimator requiring the computation of the equilibrium strategy
that can be obtained using specific parametric distribution(s).
\cite{LaffontOssardVuong1995} introduced the second methodology,
which is computationally more convenient. 
Relying on the revenue
equivalence theorem,  the authors propose a simulation-based method
that avoids computation of the equilibrium strategy, and therefore
allows for more general parametric specifications for the 
value distribution.

More recently \cite{GPV} (hereafter, GPV (2000)) developed an alternative, fully nonparametric indirect  procedure. This methodology relies on a simple but crucial observation that, using the first-order condition of the bidder's optimization problem, the value can be expressed as a function of the (corresponding) bid, and the distribution and density of observed bids.
This function, which is the inverse of the equilibrium strategy, identifies the model nonparametrically.
Therefore, in contrast to the direct method, this method starts from the distribution of observed bids in order to estimate the distribution of unobserved private values without computing the Bayesian Nash equilibrium strategy or its inverse.
This naturally calls for a two-step procedure. In the first step, a sample of pseudo private values is obtained while using (say) kernel estimators for the distribution and density  of observed bids. 
In the second step, this sample of pseudo values is used to nonparametrically estimate its density.\footnote{ GPV (2000) also establish uniform consistency and, using the minimax theory as developed by \cite{IbragimovHasminskii1981}, determine the optimal rate of convergence of this estimator.}

Though a fully nonparametric (Kernel) estimator is flexible and robust to misspeficiation, it has few drawbacks.
It has slow rate of convergence, which makes it hard to accommodate a multidimensional auction covariates (
curse of dimensionality) and it is ill-behaved at the boundaries of the support. 
To address these problems, we we propose a semiparametric procedure where the first step is 
fully nonparametric, in that we use Local Polynomial Estimation (LPE) of \cite{FanGijbels1996}, instead of Kernel, to obtain the bid density and distribution, and in the second step we model private values through a set of conditional moment restrictions and estimate the (finite) parameters using generalized method of moments (GMM). We then derive the asymptotic properties of the estimator. 
The advantage of using LPE is that it is well-behaved at the boundary, and by using conditional moment restrictions we can accommodate a large number of covariates, making our method useful for applied work. See for example \cite{BonetPesendorfer2003, Rezende2008, LiZheng2009, AtheyLevinSeira2011, KrasnokutskayaSeim2011, AtheyCoeyLevin2013, Groeger2014} who have used similar, either fully-parametric or semi parametric,  
indirect moment based procedure to accommodate a large number of covariates.  
None of them, however, provide any asymptotic properties for their
estimator. 
We contribute to this literature by showing that our procedure is consistent, asymptotically normal and achieves parametric rate of convergence. 

For notational tractability and relatively cleaner exposition we focus primarily on symmetric first-price
sealed-bid auction models with independent private value and a non-binding reserve price. 
Once the asymptotic properties of this simple case has been characterized, extending the estimation procedure to accommodate more general auction environment is tedious but conceptually straightforward -- only the asymptotic variance will change, not the rate of convergence. 
More generally, our method extends to models estimated using a nonparametric indirect procedure including auctions with asymmetric bidders. 
%In Section \ref{section:extensions} we consider some important extensions such as the auctions with binding reserve price, symmetric and asymmetric affiliated private value.

Let $V_{p\ell}$, $p=1,\ldots, I_\ell$, $\ell=1,\dots, L$
denote the private value of the $p$th bidder for the $\ell$th
auctioned object. Let $Z_{\ell}\equiv(X_\ell,I_\ell)\in \Real^{d+1}$
denote the vector of exogenous variables, it includes auction covariates $X_\ell$ and the number of
bidders $I_\ell$. 
To model the private values, we posit that there is some known and sufficiently smooth function $M(\cdot,\cdot;\theta):\Real^{d+2}\rightarrow \Real^{q}$ and parameter vector $\theta
\in \Real^p$ such that, $q\geq p$ and at some true parameter $\theta_{0}$ the values satisfy the following set of conditional moment
restrictions
\begin{eqnarray}
{\rm E}[M(V,Z;\theta_0)|Z]=0,\label{eq:1}
\end{eqnarray}
where the expectation is with respect to the value distribution $F(\cdot|Z;\theta_0,\gamma_0)$ with $\gamma_0$ as the (possibly infinite dimensional) nuisance parameter. 
These moment conditions are, however, infeasible because $V$ are unobserved. 
But, in equilibrium, the bid $B=s(V,Z;\theta_0,\gamma_0),$ where $s(\cdot)$ is the bidding strategy that depends on the parameter vector $\theta_0$ both
directly through $B$, since $B\sim G(\cdot|Z;\theta_0,\gamma_0)$ (say), and
indirectly through $V$, since $V\sim F(\cdot|Z;\theta_0,\gamma_0)$. 
This means (\ref{eq:1}) can be naturally expressed as 
$
{\rm E}\big\{M[s^{-1}(B,Z;\theta_0,\gamma_0),Z;\theta_0]\big|Z\big\}=0,
$
which requires the computation of the equilibrium
strategy as well as of its inverse. This could be computationally demanding for two different reasons. First, such
computation has to be carried out for any trial value of the
parameters $(\theta,\gamma)$. %\footnote{ When $\gamma$ is infinite
%dimensional the model falls within the class of models
%analyzed by \cite{AiChen2003}.} 
Second, in a more general class of auction 
models, such as when values are affiliated or when bidders are asymmetric, the computation of the equilibrium strategy
$s(\cdot,\cdot;\theta_0,\gamma_0)$ and of its inverse is much more
involved and costly. 
Therefore, we propose to replace $V$ in Equation (\ref{eq:1}) by its nonparametric (local polynomial) estimator $\hat V=\hat \xi(B,Z)$ (and not inverse strategy) to make the moment condition feasible and operational. Thus the feasible conditional moment restriction becomes
\begin{eqnarray*}
{\rm E}\big\{M[\hat\xi(B,Z),Z;\theta_0]\big|Z\big\}\approx0.
\end{eqnarray*}

We propose a two-step semiparametric procedure: first, we use LPE to obtain the nonparametric estimator of the value $\hat V=\hat \xi(B,Z)$; second we  
use GMM procedure to obtain an estimate for $\theta_0$. 
Unlike the most widely used Parzen-Rosenblatt Kernel based estimator, LPE is not ill-behaved close to
the support boundaries \cite[see][]{FanGijbels1996} and hence we do not have to trim any bids. 
This provides a remarkable advantage to our procedure, since otherwise we would have to trim bids, which are endogenous variables, which would then 
imply an automatic trimming on private values, thereby affecting the moments. In a standard econometric framework only exogenous variables are trimmed,  \cite{LavergneVuong1996,Robinson1988}.
We show that our estimator is consistent, asymptotically normal and
converges uniformly at the parametric $\sqrt{L}$ rate. 

As it is well
known that nonparametric estimators converge at a slower rate than
$\sqrt{L}$ and their rates are negatively related to the
dimension of the vector of exogenous variables, the so-called curse of dimensionality. 
This makes these
estimators less desirable in applications, especially when a limited
number of observations is available and/or when the number of
exogenous variables is relatively large.\footnote{ Examples of semiparametric estimators attaining $\sqrt{L}$ rate can be found in \cite{NeweyMcFadden1994,Powell1994}. An example of a semiparametric estimator
converging at a slower than the parametric rate but not subject to the curse of dimensionality -- its rate is
independent of $d$ -- is given by \cite{CGPV2011}.} 
Our estimator does not have this drawback because its
convergence rate is independent of the dimension of
the exogenous variables.
A second major advantage of our estimation
procedure is that, even though we focus on symmetric, inpdendent private value auctions without reserve price, our method provides a framework for a (moment based) semiparametric procedure that can be used to estimate more general auction such as auctions with binding announced or random reserve price, \cite{LiPerrigne2003}, symmetric and asymmetric affiliated values, \cite{LiPerrigneVuong2002,CampoPerrigneVuong2003}, as long as the moment conditions are sufficiently smooth (defined later). 
This rules out moment conditions that are based on quantiles.\footnote{ For examples of use of quantiles in empirical auctions see \cite{HaileHongShum2006, MarmerShneyerovXu2013, Gimenes2014}.}
%Indirect methods in general do not require computation neither of the
% equilibrium strategy nor of its inverse.
%Therefore these methods are specially convenient when there is no
%closed form solution to the differential equation(s) characterizing
%the equilibrium strategy like auctions with asymmetric bidders. 

In an short extension, we show how the semiparametric procedure can be applied to these auctions, including auctions with unobserved heterogeneity \cite{Krasnokutskaya2011}. 
As it will be clear, allowing these features will affect the asymptotic variance but not the rate of convergence, except when auctions have unobserved heterogeneity. 
This is because to accommodate unobserved heterogeneity we need a three-step semiparametric procedure -- the new step is to estimate the density of the unobserved heterogeneity using empirical characteristics function. 
So it is not clear whether we can even achieve the $\sqrt{L}$ consistency, but a proper analysis of asymptotic properties of such semiparametric estimator beyond the scope of our paper and is left for future research.  

The rest of the paper is organized as follows. In Section \ref{section:model}, we
introduce the theoretical model, from which  the
structural econometric model and our semiparametric estimator is derived. Section \ref{section:asymptotic} establishes the asymptotic properties of our estimator, and \ref{section:MCMC} presents some Monte Carlo experiments to illustrate the properties of our procedure. Section \ref{section:extensions} proposes some extension, and we conclude in Section \ref{section:conclusion}. The 
Appendix collects the proofs of our results.

\section{The Model\label{section:model}}

\subsection{The Symmetric IPV Model}

\indent
We present the benchmark theoretical model underlying our structural
econometric model, namely the symmetric IPV model with a non-binding
reserve price. Although this is somehow restrictive for
applications, it allows us to develop our
econometric procedure in a more transparent way. 
A single and indivisible object is auctioned to $I_\ell$
risk neutral bidders who are assumed to be ex ante identical. 
The
total number of bidders may vary across auctions. 
Private values are denoted by $V$ and we assume that
each valuation $V_{p\ell}$, $\ell=1,\ldots,L$, $p=1,\ldots,I_\ell$,
is distributed according to $F(\cdot|Z_\ell;\theta_0,\gamma_0)$,
where $\theta_0 \in \Real^p$ is the parameter of interest and
$\gamma_0$ is a nuisance parameter that could be infinite or finite dimensional or
even an empty set. The support of $F(\cdot|\cdot)$ is
$[\underline{V}_\ell, \overline{V}_\ell]$, with
$0\leq\underline{V}_\ell=\underline{V}(Z_\ell)< \overline{V}_\ell=
\overline{V}(Z_\ell)<\infty$. 
Among others, \cite{RileySamuelson1981} have shown that for every $\ell$,
$I_\ell\geq2$ the equilibrium bid $B_{p\ell}$ in the $\ell$th
auction is given by 
\begin{eqnarray}
B_{p\ell}=s_0(V_{p\ell},Z_\ell)=V_{p\ell}-\frac{1}{F(V_{p\ell}|Z_\ell;\theta_0,\gamma_0)^{I_\ell-1}}\int_{\underline{V}_\ell}^{V_{p\ell}}F(v|Z_\ell;\theta_0,\gamma_0)^{I_\ell-1} dv,
\label{BNE}
\end{eqnarray}
where $s(\cdot,\cdot)$ is the unique symmetric Bayes Nash Equilibrium strategy that  is monotonic and differentiable. 
Let $G(\cdot|Z_\ell; \theta_0, \gamma_0)\equiv G_0(\cdot|Z_\ell)$ and $g(\cdot|Z_\ell; \theta_0, \gamma_0)\equiv g_0(\cdot|Z_\ell)$ be the distribution and density of observed bids in the $\ell^{th}$ auction, respectively. 
 From GPV (2000), values $V$ can be identified as
\begin{eqnarray}
V_{p\ell}=\xi_0(B_{p\ell},Z_\ell)=B_{p\ell}+\frac{1}{I_\ell-1}\frac{G_0(B_{p\ell}|Z_\ell)}{g_0(B_{p\ell}|Z_\ell)},\quad p=1,\ldots,I_\ell; \ell=1,\ldots, L.\label{eq:gpv}
\end{eqnarray}

\subsection{The Two Step Estimator}

\indent
Similar to GPV (2000), (\ref{eq:gpv}) forms the basis for our
econometric model. The difference with GPV (2000) is to model
private values as a set of moment conditions. Therefore knowledge of
$G_0(\cdot|\cdot)$ and $g_0(\cdot|\cdot)$ would lead us to a
GMM framework. However, these functions are unknown in practice but
can be easily estimated from observed bids. This suggests the
following two-step procedure.

In the first step 
we recover a sample of pseudo private values by using nonparametric
LPE. The second step departs from the nonparametric
second step of GPV (2000) since we use (parametric) GMM procedure to
obtain an estimator for $\theta_0$ instead. Before presenting our two-step
estimator, it is worth mentioning that some of our assumptions are
similar or even identical to those in GPV (2000). This is not
surprising since our methodology follows closely their methodology. In particular we follow GPV (2000) and indicate when some
modifications are necessary. Our first two assumptions deal with the data generating process and the smoothness of the latent
joint distribution of $(V_{p\ell},Z_\ell)$.

\medskip\noindent
{\bf Assumption A1:} {\em
\begin{itemize}
\item[(i)]
$Z_\ell=(X_\ell,I_\ell)\in \Real^{d+1}$, $\ell=1,2,\ldots,L$ are
independently and identically distributed as $F_m(\cdot,\cdot)$ with
density $f_m(\cdot,\cdot)$.
\item[(ii)]
For each $\ell$, $V_{p\ell}$, $p=1,\ldots,I_\ell$ are independently and
identically distributed conditionally on $Z_\ell$ as
$F(\cdot|\cdot;\theta_0,\gamma_0)$ with density
$f(\cdot|\cdot;\theta_0,\gamma_0)$, where $\theta_0 \in \Real^p$
and $\gamma_0$ can be finite or infinite dimensional or empty.
\end{itemize}        }

\medskip\noindent
Let $\cal I$ be the set of possible values for $I_\ell$. 
We use ${\mathcal S}(*)$ to denote the support of $*$, and use ${\mathcal S}_I(*)$ to denote the support when there are $I$ bidders.\footnote{ We use the notation $I_\ell$ (with the subscript $\ell$) to denote that there are $I_{\ell}$-many bidders in the $\ell^{th}$ auction, and $I$ (without the subscript $\ell$) to denote an auction with $I$-many bidders. For example, suppose there are $L=3$ auctions, with 2, 3 and 2 bidders in auction $\ell=1,2$ and $3$, respectively. Here $\ell \in \{1,2,3\}, I_1=2, I_2=3$ and $I_3=2$ and simply $I=2$ refers to auctions with 2 bidders, which is either auction 1 or 3.}

\medskip\noindent
{\bf Assumption A2:} {\em
${\cal I}$ is a bounded subset of $\{2, 3 ,\ldots \}$, and
\begin{itemize}
  \item[(i)]
    For each $I \in {\cal I}$, $ {\mathcal S}_i (F) =
      \{ (v,x): x \in [\underline{x},\overline{x}],
        v \in [\underline{v} (x), \overline{v} (x) ] \} $,
    with $\underline{x}<\overline{x}$,
 \item[(ii)]
    For $(v,x,I) \in {\mathcal S}(F)$, $f(v|x,I;\theta_0,\gamma_0) \geq c_f > 0$, and
    for $(x,I) \in {\mathcal S}(F_m)$, $f_m(x,I) \geq c_f > 0$,
 \item[(iii)]
    For each $I \in {\cal I}$, $F(\cdot | \cdot,I;\theta_0,\gamma_0 )$ and $f_m(\cdot,I)$ admit
    up to $R+1$ continuous bounded partial
    derivatives on ${\mathcal S}_I(F)$ and ${\mathcal S}_I(F_m)$, with $R > d+1$.
\end{itemize}}
\noindent These assumptions can be found in GPV (2000) as well, though  A2-(iii) is stronger in our case. That is, we require $R$ to be sufficiently large with respect to the dimension of $X$, i.e. $R > d+1$, which is commonly used in the semiparametric literature, see \cite{PowellStockStoker1989} among others. 
The next two assumptions are on kernels and bandwidths used in the first stage.

\medskip\noindent
{\bf Assumption A3:}{\em

\begin{itemize}
\item [(i)]
The kernels $K_G(\cdot)$, $K_{1g}(\cdot)$ and $K_{2g}(\cdot)$ are
symmetric with bounded hypercube supports and twice continuously
bounded derivatives.

\item [(ii)]
$\int K_G(x) dx =1$,
$\int K_{1g}(x) dx =1$, $\int K_{2g}(b) db=1$.

\item [(iii)]
$K_G(\cdot)$, $K_{1g}(\cdot)$ and $K_{2g}(\cdot)$ are of the order $(R-1)$.%\footnote{ So the moments of order strictly smaller than $R-1$ vanish. This is similar to the assumption A3 in GPV (2000), and is a standard assumption in the nonparametric literature.
%}
\end{itemize}}

\medskip\noindent
{\bf Assumption A4:} {\em The bandwidths $h_{G}$, $h_{1g}$ and $h_{2g}$ satisfy:

\begin{itemize}

\item [(i)]
$h_{G}\rightarrow 0$ and $\displaystyle{\frac{L h_{G}^d}{\log L} }\rightarrow \infty$, as $L\rightarrow\infty$,

\item [(ii)]

$h_{1g}\rightarrow 0$, $ h_{2g}\rightarrow 0$ and $\displaystyle{\frac{L h_{1g}^dh_{2g}}{\log L} }\rightarrow \infty$, as $L\rightarrow\infty$.

\end{itemize}}

\medskip

For simplicity of presentation and tractability of the notations, in the remainder of the paper we will consider only univariate $X$, i.e., $d=1$, except in the Monte Carlo section when we consider $d=2$.  
Since we prove that the rate of convergence is independent of $d$ (Proposition \ref{prop:2}) all the asymptotic results will work for $d>0$ except for the form of asymptotic variance, because 
when we move from $d=1$ to $d>1$, we only have to adapt the dimension of the regressor, the degree of polynomial and the asymptotic variance. See \cite[][section 3]{RuppertWand1994} for an example of how to specify a polynomial with $d=2$.

In order to describe our two--step estimator, we observe first that, our objective is
to estimate the ratio
$\psi(\cdot|\cdot)=G_0(\cdot|\cdot)/g_0(\cdot|\cdot)$ by
$\hat{\psi}=\hat{G}(\cdot|\cdot)/\hat{g}(\cdot|\cdot)$ (see equation (\ref{eq:gpv})) using  
LPE for each function.
From Proposition 1 in GPV (2000) we know that $G_0(\cdot|\cdot)$ is $R+1$ times continuously differentiable on its entire support and
therefore $g_0(\cdot|\cdot)$ is $R$ times continuously differentiable on its entire support as well.\footnote{Observe that by Proposition 1 in GPV (2000) we also know that the conditional density $g_0(\cdot|\cdot)$ is $R+1$
times continuously differentiable on a closed subset of the interior of the support. Thus the degree of smoothness close
to the boundaries and at the boundaries of the support is not $R+1$.} Given the smoothness of each function we propose to
use a LPE($R$), i.e. a LPE of degree $R$, for $G_0(\cdot|\cdot)$ and a LPE($R-1$) for $g_0(\cdot|\cdot)$.
For consistency of the first step  
it is possible to choose the optimal bandwidths \`{a} la \cite{Stone1982}. 
However, unlike GPV (2000) we do not need to specify a ``boundary bandwidth'' since the local polynomial method does not require knowledge of the location of the endpoints of the support. Therefore, it is not necessary to estimate the boundary of the support of the bid distribution. 

Let $P_\rho(X;\beta)$ denote a polynomial of degree $\rho$ in $X$ with parameter $\beta$. Then for each each $I$,
\begin{eqnarray*}
\hat{G}(b|x)&=&\arg\min_{\beta_G} \sum_{\{\ell:I_\ell=I\}}^L\sum_{p=1}^I\Big\{Y_{p\ell}^G-P_R(X_\ell-x;\beta_G)\Big\}^2 \frac{1}{h_{G}}K_G\left(\frac{X_\ell-x}{h_G}\right)\\
\hat{g}(b|x)&=&\arg\min_{\beta_g} \sum_{\{\ell:I_\ell=I\}}^L\sum_{p=1}^I\Big\{Y_{p\ell}^g-P_{R-1}(X_\ell-x;\beta_g)\Big\}^2 \frac{1}{h_{1g}}K_{1g}\left(\frac{X_\ell-x}{h_{1g}}\right),
\label{LPE_G}
\end{eqnarray*}
where $Y_{p\ell}^G=\Unit(B_{pl}\leq b)$ and$Y_{p\ell}^g=\frac{1}{h_{2g}}K_{2g}\left(\frac{B_{p\ell}-b}{h_{2g}}\right).$
More precisely we have,
\begin{eqnarray}
\hat G(b|x,I)&=&\frac{1}{h_{G}}\sum_{\{\ell:I_\ell=I\}}^L\sum_{p=1}^{I}e_1^T (X_{I,R+1}^T W_{x}^GX_{I,R+1})^{-1} X_{R+1,\ell}K_G\left(\frac{X_\ell-x}{h_{G}}\right) \Unit(B_{p\ell}\leq b)\nonumber\\
&=&\!\!\frac{1}{Lh_{G}}\frac{L}{n_I}\!\!\sum_{\{\ell:I_\ell=I\}}^L\sum_{p=1}^{I}e_1^T \left(\frac{X_{I,R+1}^T W_{x}^GX_{I,R+1}}{n_I}\right)^{-1}\! \!\!X_{R+1,\ell} K_G\left(\frac{X_\ell-x}{h_{G}}\right) \Unit(B_{p\ell}\leq b);\label{eq:Ghat}\\
\hat g(b|x,I)&=&\frac{1}{h_{1g}h_{2g}}\sum_{\{\ell:I_\ell=I\}}^L\sum_{p=1}^{I}e_1^T (X_{I,R}^T W_{x}^gX_{I,R})^{-1} X_{R,\ell}K_{1g}\left(\frac{X_\ell-x}{h_{1g}}\right) K_{2g}\left(\frac{B_{p\ell}-b}{h_{2g}}\right)\nonumber\\
&=&\frac{1}{L h_{1g}h_{2g}}\frac{L}{n_I}\!\!\!\sum_{\{\ell:I_\ell=I\}}^L\!\sum_{p=1}^{I}e_1^T \left(\frac{X_{I,R}^T W_{x}^gX_{I,R}}{n_I}\right)^{-1}\!\!\!\!\! X_{R,\ell}K_{1g}\left(\frac{X_\ell-x}{h_{1g}}\right) K_{2g}\left(\frac{B_{p\ell}-b}{h_{2g}}\right),\label{eq:ghat}
\end{eqnarray}
where for $\iota\in\{R,R+1\}$,
$e_1$ is the unit vector in $\Real^{\iota}$ containing a 1 in its first entry,
$n_I=I L_I$,
$L_I=\#\{\ell:I_\ell=I\}$,
$X_{\iota,\ell}=[1 \quad (X_\ell-x)\ldots (X_\ell-x)^{\iota-1}]^T$  is a $\iota\times1$ vector,
$$X_{I,\iota}=\left(
\begin{array}{cccc}
1&(X_1-x) &\ldots &(X_1-x)^{\iota-1}\\
\vdots &\vdots &\vdots &\vdots \\
1&(X_{n_I}-x) &\ldots &(X_{n_I}-x)^{\iota-1}
\end{array}\right)$$
is the $n_I\times \iota$ matrix of regressors with the first $I$ rows identical and similarly for the other rows,
\begin{eqnarray*}
W^G_{x}={\rm{diag}}\left\{\frac{1}{h_{G}}K_G\left(\frac{X_\ell-x}{h_{G}}\right)\right\};\quad
W^g_{x}={\rm{diag}}\left\{\frac{1}{h_{1g}}K_{1g}\left(\frac{X_\ell-x}{h_{1g}}\right)\right\},
\end{eqnarray*}
where $K_G(\cdot)$, $K_{1g}(\cdot)$ and $K_{2g}(\cdot)$ are some kernels with
bounded support and $h_{G}$, $h_{1g}$,$h_{2g}$ are some
bandwidths (see Assumptions A3 and A4). Given (\ref{eq:Ghat}) and (\ref{eq:ghat}), the (pseudo) private value is given by
\begin{eqnarray}
\hat V_{p\ell}=B_{p\ell}+\frac{1}{I_\ell-1}
\hat{\psi}(B_{p\ell}|Z_\ell).\label{eq:vhat}
\end{eqnarray}
Unlike in GPV (2000), $\hat{\psi}$ is not subject to the so-called boundary effect, a typical problem encountered in kernel estimation, and hence we do not need to trim out observations that are ``too close'' to the boundary of the support of the joint distribution of $(B_{p\ell},Z_\ell)$.
The second step of our estimation procedure is as follows. We propose to use the sample of pseudo private values in the following conditional moment restrictions, namely
\begin{eqnarray*}
{\rm E}\big[M(\hat V,Z;\theta_0)\big|Z\big]\approx0,
\end{eqnarray*}

\noindent for some \emph{known} function
$M(\cdot,\cdot;\theta):\Real^{3}\rightarrow\Real^{q}$ and $\theta
\in \Real^p$ with $q\geq p$.
For example, we could use 
\begin{eqnarray}
{\rm E}[\ln ( V_{p\ell}) \mid Z_{\ell}]&=& \theta_{0,1}'Z_{\ell}.\\
Var[\ln (V_{p\ell}) \mid Z_{\ell}]&=& [\exp(\theta_{0,2}'Z_{\ell})]^{2}.\label{eq:moment-example}
\end{eqnarray}
as the moment conditions, like in \cite{ KrasnokutskayaSeim2011, BajariHoughtonTadelis2014}.
This set of conditional moment restrictions translates into the following set of unconditional moment restrictions,
\begin{eqnarray}
{\rm E}\big[m(\hat V,Z;\theta_0)\big]\approx0,\label{eq:moment}
\end{eqnarray}

\noindent where $m(\cdot,\cdot;\theta):\Real^{3}\rightarrow
\Real^q$ is known.
In view of (\ref{eq:moment}), we propose to estimate $\theta_0$ by $\hat\theta$, where 
\begin{eqnarray}
\hat{\theta}=\arg\min_{\theta\in\Theta} \hat S_L^T(\theta)\Omega \hat
S_L(\theta),\label{eq:thetahat}
\end{eqnarray}

\noindent where $\hat S_L(\theta)=\dsh m(\hat V_{p\ell},Z_\ell;\theta)$ and $\Omega$ is a
positive definite matrix of order $q$.
Ideally, one would like to specify the following set of conditional moment restrictions
${\rm E}[M(V,Z;\theta_0)|Z]=0,
$
which would lead to the unconditional moment
restrictions
$
{\rm E}[m(V,Z;\theta_0)]=0.
$
Therefore, if $S_L(\theta)=\dsh m(V_{p\ell},Z_\ell;\theta)$ the infeasible estimator $\tilde \theta$, (say), is such that
\begin{eqnarray*}
\tilde{\theta}=\displaystyle{\arg\min_{\theta\in\Theta}
S^T_L(\theta)\Omega S_L(\theta)}.\end{eqnarray*}
%For example, we could use the following conditional moment conditions: 
%\begin{eqnarray}
%{\rm E}[\ln ( V_{p\ell}) \mid Z_{\ell}]&=& \theta_{0}'Z_{\ell}.\\
%Var[\ln (V_{p\ell}) \mid Z_{\ell}]&=& [\exp(\theta_{0}'Z_{\ell})]^{2},\label{eq:moment-example}
%\end{eqnarray}
%The sample analog of the unconditional moment conditions corresponding to (\ref{eq:moment-example}) can be  
%\begin{eqnarray*}
%m_{1}&=&\frac{1}{L}\sum_{\ell=1}^{L}\frac{1}{I_{\ell}}\sum_{p=1}^{I_{\ell}}Z_{\ell}' (\ln({V}_{p\ell})- \theta'Z_{\ell}) =0, \\
%m_{2}&=&\frac{1}{L}\sum_{\ell=1}^{L}\frac{1}{I_{\ell}}\sum_{p=1}^{I_{\ell}}Z_{\ell}' ( (\ln(V_{p\ell})- \theta'Z_{\ell})^{2}- [\exp(\theta'Z_{\ell})]^{2} )  =0.
%\end{eqnarray*}
%Replacing $\ln (V_{p\ell})$, in above, with $\ln (\hat{V}_{p\ell})$ from (\ref{eq:vhat}), gives feasible moment conditions. 

\noindent\textbf{Remark--} The asymptotic distributions of the
feasible estimator $\hat\theta$ and the infeasible estimator $\tilde\theta$ are closely related, but are not the
same, see Proposition \ref{prop:2}.

%The empirical counterparts of this moment conditions are can then be used in a second step can be easily stated. 
\section{Asymptotic Properties\label{section:asymptotic}}

\indent
In this section we show that our two-step semiparametric estimator
$\hat\theta$ of $\theta_0$ is consistent and asymptotically normal
distributed. Moreover, we establish that our estimator attains the
parametric uniform rate of convergence given an appropriate choice
of the bandwidths used in the first step to estimate
$G_0(\cdot|\cdot)$ and $g_0(\cdot|\cdot)$. As we will discuss below
the optimal bandwidths, given by \cite{Stone1982}, i.e. the one-step bandwidths, cannot
be chosen, instead our choice implies that in practice one needs to
undersmooth.
We also discuss the assumptions under which our results
hold.

\subsection{Consistency}

\indent
Our first result establishes that $\hat\theta$ is a
(strongly) consistent estimator for $\theta_0$. Moreover this is the case even if one uses
the optimal bandwidths for estimating $G_0(\cdot|\cdot)$ and
$g_0(\cdot|\cdot)$ in the first step, i.e. the bandwidths proposed
by \cite{Stone1982}. To see this, we notice that the ``optimal one-step'' bandwidths satisfy our assumption A4 above (with $d=1$) since they are of the form,
\begin{eqnarray*}
h_{G}=\lambda_{G}\left( \frac{\log L}{L}\right)^{1/(2R+3)};\quad h_{1g}=\lambda_{1g}\left( \frac{\log L}{L}\right)^{1/(2R+1)};\quad h_{2g}=\lambda_{2g}\left( \frac{\log L}{L}\right)^{1/(2R+1)},
\end{eqnarray*}
where $\lambda_{G}$, $\lambda_{1g}$ and $\lambda_{2g}$ are strictly positive constants.
As observed by GPV (2000), $h_{G}$, $h_{1g}$ and $h_{2g}$, as given above are
optimal bandwidth choices to estimate $G_0(\cdot|\cdot)$ and
$g_0(\cdot|\cdot)$ given Proposition \ref{prop:1} and A2-(iii) in that
paper.\footnote{As pointed out before, A2-(iii) in our case is
stronger than A2-(iii) in GPV (2000). Thus their Proposition 1 also
holds in our framework.} 
Thus, A4 implies that our consistency
result can be established when using LPE in the first
stage that converge at the best possible rate.

\medskip\noindent
{\bf Assumption A5:} {\em

\begin{itemize}

\item[(i)]
The parameter space $\Theta\subset\Real^p$ is compact and $\theta_0$
is in the interior of $\Theta$,

\item [(ii)]
Identifying assumption: ${\rm E}[m(V,Z;\theta)]=0$ if and only if
$\theta=\theta_0$,

\item[(iii)]
$\displaystyle{\sup_{\theta \in \Theta}\left\vert\ds \left\Vert
m(V_{p\ell},Z_\ell; \theta)\right\Vert- {\rm E}\left\Vert
m(V,Z;\theta)\right\Vert\right\vert}=o_{as}(1)$,

\item[(iv)]
$m(V,Z;\theta)$ is Lipschitz in $V$-- there exists a measurable
function $K_1(Z), {\rm
E}[K_1]<\infty$ such that 

\[\forall V,V'\in[\underline{V},\overline{V}],\forall\theta\in\Theta,\quad\!\!\left\Vert m(V,Z;\theta)-m(V',Z;\theta)\right\Vert\leq K_1(Z)
\left\vert V-V'\right\vert.\]
\end{itemize}}

Let  $m_{k}(\cdot,\cdot)$ be the partial derivative of $m(\cdot,\cdot)$ with respect to its $k^{th}$ argument. 

\medskip\noindent
{\bf Assumption A6:} {\em

\begin{itemize}

\item[(i)]

$m_3(V,Z;\theta)$ is Lipschitz in $V$: there exists a measurable
function $K_3(Z), {\rm
E}[K_3]<\infty$, such that $$\forall V, V' \in [\underline{V},\overline{V}], \forall \theta\in\Theta,\quad\!\!
\left\Vert m_3(V,Z;\theta)-m_3(V',Z;\theta)\right\Vert\leq K_3(Z)\vert V-V'\vert.$$
\item[(ii)]

$m_3({V},Z;\theta)$ is Lipschitz in $\theta$: there exists a
measurable function $K_4(Z), {\rm
E}[K_4]<\infty$ such that 
$$\forall\theta,\theta'\in \Theta,V \in [\underline{V},\overline{V}],\quad\!\! \left\Vert m_3({V},Z;\theta)-m_3({V},Z;
\theta')\right\Vert\leq K_4(Z)\left\Vert \theta-
\theta'\right\Vert.$$
%
%
%for every $\theta$, $\theta' \in \Theta$ and $V \in [\underline{V},\overline{V}]$. Moreover ${\rm
%E}[K_4(Z)]<\infty$,

\item[(iii)]

$\displaystyle{\sup_{\theta \in\Theta} \left\Vert \ds
m_3(V_{p\ell},Z_\ell;\theta)-{\rm
E}[m_3(V,Z;\theta)]\right\Vert}=o_{as}(1)$
and 
${\rm E}[m_3'(V,Z;\theta)]\Omega {\rm E}[m_3(V,Z;\theta)]$ is non
singular.

\item [(iv)]

$\displaystyle{\sup_{\theta \in\Theta}\left\Vert
m_3(V,Z;\theta)\right\Vert\leq K_5(V,Z)}$ with ${\rm
E}[K_5(V,Z)]<\infty$,

\item[(v)]

$m_1(V,Z;\theta)$ is Lipschitz in $V$: there exists a measurable
function $K_6(Z), {\rm
E}[K_6]<\infty$ such that 
$$\forall V,V'\in[\underline{V},\overline{V}],\theta \in \Theta,\quad\!\! \left\Vert m_1(V,Z;\theta)-m_1(V',Z;\theta)\right\Vert\leq K_6(Z)\vert V-V'\vert.$$
\item [(vi)]

$\displaystyle{\sup_{\theta \in \Theta}\left\Vert
m_1(V,Z;\theta)\right\Vert \leq K_7(V,Z)}$ with ${\rm
E}[K_7(V,Z)^2]<\infty$.

\item [(vii)]  ${\rm{E}}[m_1(V,Z;\theta_0)]<\infty$, where the expectation is with respect to the joint cdf of $(V,Z)$. 

\end{itemize}}

\noindent Assumptions 5 and 6 are implied by the regularity conditions used in GMM estimators, \cite[see][]{NeweyMcFadden1994}.
These regularity conditions impose appropriate differentiability restrictions on the moment functions, which rule out certain kinds of moment conditions. 
For instance these assumptions rule out moment conditions that define quantiles.  
Let $\rho_{\tau} (s)= \mid s\mid +(2\tau - 1)$ where $\tau\in(0,1)$. 
Then the conditional and unconditional moment conditions, \cite{Koenker2005},  in this case are respectively given by
\begin{eqnarray*}
{\rm E}[\rho(V,Z;\theta_{0}(\tau))]&=&{\rm E}[\mid V-Z'\theta_{0}(\tau) \mid+ (2\tau-1)]=0,\\
{\rm E}[Z\rho(V,Z;\theta_{0}(\tau))]&=&{\rm E}[Z(\mid V-Z'\theta_{0}(\tau) \mid+ (2\tau-1))].
\end{eqnarray*}
and the corresponding sample moment condition is
\begin{eqnarray*}
\ds Z_{\ell} (\mid V_{p\ell}- Z_{\ell}'\theta(\tau) \mid +(2\tau-1)).
\end{eqnarray*}
Since a function $|s|$ is not everywhere differentiable, our method does not apply because we use Taylor's series expansion.\footnote{ One can, however, use results from empirical processes to allow for non-smooth moment conditions. Typically those conditions impose sufficient differentiability of the distribution function $F(\cdot|\cdot;\cdot)$ and stochastic equicontinuity. Since one of the arguments of the moment conditions is estimated nonparametrically, verifying stochastic equicontinuity in our framework can be difficult. We want to thank one of the referees for this observation.}
Now, we show that the estimator is consistent. 
%Intuitively, our proof uses the fact that our moment functions are differentiable in its arguments to apply the usual Taylor's series expansion of the sample first-order condition for $\hat \theta$ around $\hat\theta=\theta_{0}$. 
%If the criterion function is defined through a quantile, then it is no longer true that this function is continuously differentiable at values of $\theta$ for which $v=\theta(\tau)'z$, moreover, as is well known the (discontinuous) subgradient itself has a derivative that is identically zero wherever it is defined. Thus, the usual Taylor expansion is not applicable to this problem.\footnote{ In order to allow for non-differentiable moments we could use empirical process theory to establish the stochastic equicontinuity condition. However, our framework has the difficulty that one of the arguments of the moment condition is a non-paramatric estimator. This makes establishing stochastic equicontinuity not a straightforward task.}
%We now state our consistency result.
\begin{proposition}\label{prop:1}Let $\hat\theta$  be defined as in (\ref{eq:thetahat}).
Then, under A1-A5 $\hat{\theta}\stackrel{a.s}{\longrightarrow}\theta_0$. 
%\[\hat\theta\stackrel{a.s}{\longrightarrow}\theta_0.\]
\end{proposition}
The proof is in the Appendix. This is the first step in order to be able to establish the asymptotic distribution of the estimator. Moreover, there is no need to under-smooth the distribution and density functions in the first step in order for $\hat\theta$ to be consistent.

\subsection{Asymptotic Normality}

\indent
Given that $\hat\theta$ is a (strongly) consistent estimator for $\theta_0$, in Proposition \ref{prop:2} we establish its asymptotic distribution and its uniform
convergence rate, under some additional regularity conditions.
Since optimal bandwidth choice requires under-smoothing in semiparametric procedures we have to modify our choice of bandwidths.\footnote{ Another typical property usually encountered
has to do with a sufficiently large degree of smoothness relative to
the dimension of the exogenous variables, as reflected by A2-(iii).} 
Thus, for $\hat\theta$
to achieve the parametric uniform rate of convergence we need to
specify bandwidths for our first step that rule out the optimal
choice and moreover that imply under-smoothed estimates for $\hat
G(\cdot|\cdot)$ and $\hat g(\cdot|\cdot)$, as is made clear by A4.AN below.

\medskip\noindent
{\bf Assumption A4.AN:} {\em The bandwidths $h_{G}$, $h_{1g}$ and $h_{2g}$ satisfy

\begin{itemize}

\item[(i)]

$\sqrt{L}h_G^{R+1}\rightarrow 0$ and $\displaystyle{\frac{\log L}{\sqrt{L} h_{G}} }\rightarrow 0$, as $L\rightarrow\infty$,

\item[(ii)]

$\sqrt{L}h_{1g}^{R}\rightarrow 0$, $ \sqrt{L}h_{2g}^{R}\rightarrow 0$ and $\displaystyle{\frac{\log L}{\sqrt{L} h_{1g}h_{2g}} }\rightarrow 0$, as $L\rightarrow\infty$,

\item[(iii)]

$h_{1g}=h_{2g}$.

\end{itemize}}
\noindent The assumption that $h_{1g}$ and $h_{2g}$ vanish at the same rate, is to simplify the notation in the proof. In fact it is enough to choose any pair of bandwidths strictly smaller than their optimal counterparts.\footnote{For the multivariate case ($d>1$), these conditions become: \begin{itemize}

\item[(i)]

$\sqrt{L}h_G^{R+d}\rightarrow 0$ and $\displaystyle{\frac{\log L}{\sqrt{L} h_{G}^d} }\rightarrow 0$, 
as $L\rightarrow\infty.$

\item[(ii)]

$\sqrt{L}h_{1g}^{R+d-1}\rightarrow 0$, $ \sqrt{L}h_{2g}^{R}\rightarrow 0$ and $\displaystyle{\frac{\log L}{\sqrt{L} h_{1g}^dh_{2g}} }\rightarrow 0$, as $L\rightarrow\infty$.

\end{itemize}}

%Next, we establish our main result.
\begin{proposition}\label{prop:2}
Let $\hat\theta$  be defined as in (\ref{eq:thetahat}). Then, under A1-A3, A4.AN and A5-A6, we have
\[\sqrt{L}(\hat\theta-\theta_0)\stackrel{d}{\longrightarrow}{\mathcal N}(0,\Sigma),\]
where for each $I\in \mathcal{I}, \Sigma={\rm{Var}}(\psi_1)$, with 
\begin{eqnarray*}
\psi_1&=&-{1}/{I}\sum_{p=1}^{I}\Bigg\{(C^T\Omega C)^{-1}C \Omega
m(V_{p1},X_1,I;\theta_0) +2 \Bigg[\sum_{I}\frac{1}{I(I-1)}N(Y_{p1},I)\\
&& f_m^{-1}(X_1,I) g_0(Y_{p1},I)-{\rm{E}}\Big[\sum_{I}\frac{1}{I(I-1)}N(Y_{p1},I)f_m^{-1}(X_1,I)
g_0(Y_{p1},I)\Big]\Bigg]\Bigg\},\\
C&=&{\rm E}[\partial m(V,X,I; \theta_0)/\partial\theta], Y_{p\ell}\equiv (B_{p\ell}, X_{\ell})\\
N(Y_{p1},I)&=& [m_1(V_{p1},X_1,I;\theta_0)/g_0(B_{p1}|X_1,I)^2] G_0(B_{p1}|X_1,I).
\end{eqnarray*}
\end{proposition}
\noindent Proposition \ref{prop:2} is important for several reasons. First it establishes that our semiparametric estimator has a standard limiting distribution. Asymptotic Normality is fundamental since most of the econometric tests rely on it. Second, although slow estimators are used in the first step of our estimation procedure to recover pseudo private values, the estimator of the parameter of interest converges at the best possible rate. Third, our semiparametric estimator is not subject to the curse of dimensionality. Finally, Proposition \ref{prop:2} can be used to conduct inference on $\theta_0$.
There are already some empirical papers in the literature that fit in our framework. For example, \cite{ KrasnokutskayaSeim2011, AtheyLevinSeira2011} estimate auction models using semiparametric or fully parametric procedure, and if we ignore the unobserved heterogeneity, both these papers satisfy all our assumptions.

 We conclude this section with few points about our procedure, especially that of the LPE, that deserve mention. 
First, LPE regression is more computationally complex than the standard least squares method,
because a model must be fit for each observed data point. 
With ``brute force'' methods, it
would take approximately $L\times I_{\ell}$ times longer to fit a local linear regression than it would take to fit a
``global'' linear regression; see \cite{SeifertBrockmannEngelGasser1994}. 
This is without factoring all the calculations that go in kernel evaluations and choosing bandwidths. 
Many methods for choosing the bandwidth $h$ rely
on cross-validation, \cite{FanGijbels1995,PrewittLohr2006}.
 This necessitates solving the LPE minimization repeatedly. 
 The complexity multiplies as $d$ increases, for then we need higher degree polynomial which are difficult to evaluate. 
 So care must be given to using a ``quicker'' method for solving the minimization. 
\cite{FanMarron1994, HallWand1996} propose to use ``updating'' and linear ``binning'' for this purpose.\footnote{ All of these methods are now easily implemented using statistical programming language like $\textbf{\textsf{R}}$.}

Second since in the equilibrium, and as mentioned in the introduction,  $G(b|Z)\equiv F(s^{-1}(b;\theta_{0},Z)|Z)$, the parameter of interest $\theta_{0}$ enters the moment conditions directly and indirectly through the first stage nonparametric estimate of $\psi(\cdot)$. 
This makes our estimation procedure different from the widely studied semiparametric method, for example \cite{Chamberlain1992}, where the parameter of interest does not enter the nuisance (nonparametric) first stage estimate, and as a result we lose some efficiency. 
Had the first-step also been parametric then for a candidate $\theta$ the nuisance function would be calculated and then in the minimization step, $\theta$ would enter the moment conditions twice. 
With nonparametric first-step, however, we did not have to fix $\theta$, but at a cost of higher variance or lower efficiency.\footnote{ We owe this obsevration to one of the referees.}
But determining the exact loss of efficiency would require us to determine semiparametric efficiency for the non-regular case which is considered to be a hard problem, \cite{IbragimovHasminskii1981,Newey1990,Chamberlain1992, BickelKlaassenRitovWellner1993}, and is beyond the scope of this paper.

\section{Monte Carlo Experiments\label{section:MCMC}}
Now, we want to see the performance of the estimator we proposed through two sets of Monte Carlo exercises. 
In the first set, we consider one dimensional auction characteristics, i.e., $d=1$ and in the second set we consider $d=2$. 
For both cases we fix number of bidders $I_{\ell}=5$ for all $\ell=1,\ldots, L$, where $L=200$ when $d=1$ but when $d=2$, we let $L$ to be either $200, 100$ or $50$. 

We use two yardsticks to evaluate the performance of our estimator. 
The first is the visual method where we present the estimated density using our procedure, and to facilitate comparison we also present the true density along with the estimated density that uses GPV (2000) method. 
The second method is to compare the optimal ex-ante expected revenue for the seller. 
To compute the revenue we first use the plug-in method to choose the optimal reserve price, $r=\frac{1-\hat{F}(r)}{\hat{f}(r)}$, \cite{Myerson1981} and then calculate the corresponding (maximized) expected revenue \cite{Krishna2002}
\begin{eqnarray}
\Pi(r)=I\left[r (1-\hat{F})(\hat{F}(r))^{I-1}+\int_{r}^{\overline{v}}(1-\hat{F}(t))t(I-1)(\hat{F}(t))^{(I-2)}\hat{f}(t)dt\right].\label{eq:revenue}
\end{eqnarray}
Like with the figures, we calculate the revenue corresponding to our semiparametric estimate, GPV (2000) estimate and the true density. 
Since the final goal of estimating the value density is to choose optimal auction, comparing revenues across different estimators is a good way to judge the performance of the estimators -- the closer the revenue to the truth the better the estimates. 
We present all of these results while fixing $X$ at its median value and find that our estimator performs well according to both measures. 

\subsection{One Dimensional Covariate}
 Let $X\sim \ln {\mathcal N} (0,1)$ truncated at 0.055 and 30 to satisfy A2-(i), and $V|X\sim F(\cdot|Z;\theta_0,\gamma_0)=\ln {\mathcal N}(1+X, 1)$ truncated at 0.055 and 30, so $\theta_0=(1,1)^T$ and $\gamma_0=\{\emptyset\}$. 
 While estimating, we assume that $R=3$. 
 In line with assumption  A3, we choose the triweight kernel $(35/32)(1-u^2)^3 \Unit(|u|\leq 1)$ for the
three kernels involved in our first step estimators.
We choose the bandwidths according to A4.AN. In particular we use $h_G=2.978\times1.06\hat\sigma_x (IL)^{-1/6.5}$,
$h_{1g}=2.978\times1.06 \hat\sigma_x (IL)^{-1/4.5}$, $h_{2g}=2.978\times1.06 \hat\sigma_b (IL)^{-1/4.5}$, where $\hat\sigma_b$ and $\hat\sigma_x$ are
the estimated standard deviations of observed bids and  object heterogeneity, respectively. The factor $2.978\times1.06$ follows from the so-called rule of thumb \cite[see][]{Hardle1991}. The use of $I$ arises because we have $I$ bidders per auction.

To replicate the GPV (2000) estimator we choose the bandwidths according to the optimal rates. Thus, the order of the bandwidths
is $L^{-1/9}$ for $h_G$ and the second step bandwidth $h_x$ and $L^{-1/10}$ for $h_{gb}$ and $h_{gx}$ and the second step bandwidths $h_{fv}$ and $h_{fx}$. Specifically we use $h_G=1.06\hat\sigma_x (IL)^{-1/9}$,
$h_{gx}=1.06 \hat\sigma_x (IL)^{-1/10}$, $h_{gb}=1.06 \hat\sigma_b (IL)^{-1/10}$ where $\hat\sigma_b$ and $\hat\sigma_x$ are as defined above. The second step bandwidths are $h_{fv}=1.06 \hat\sigma_{\hat v} (n_t)^{-1/10}$, $h_{fx}=1.06 \hat\sigma_{x} (n_t)^{-1/10}$ and $h_x=1.06 \hat\sigma_{x} (L)^{-1/9}$, where $n_t$ is the number of observations remaining after trimming. See Table \ref{hD1} for all the bandwidths.\footnote{For GPV(2000) we also need to compute the boundary bandwidths.}
\begin{table}[t!]
  \centering
  \scalebox{1}{
    \begin{tabular}{lccc}
    \hline
          & Notation & Constant & Rate  \\
    \hline
    \multicolumn{1}{l}{\multirow{3}[2]{*}{LPE}} & $h_G$ & $2.978\times 1.06\times \hat\sigma_x$ & $(IL)^{-1/6.5}$ \\
    \multicolumn{1}{l}{} & \multicolumn{1}{c}{$h_{1g}$} & $2.978\times 1.06\times  \hat\sigma_x$ & $(IL)^{1/4.5}$  \\
    \multicolumn{1}{l}{} & \multicolumn{1}{c}{$h_{2g}$} & $2.978\times 1.06\times  \hat\sigma_b$ & $(IL)^{-1/4.5}$ \\
   \hline
    \multicolumn{1}{l}{\multirow{3}[2]{*}{GPV 1st step}} & $h_G$ & $1.06 \times \hat\sigma_x$ & $(IL)^{-1/9}$  \\
    \multicolumn{1}{l}{} & $h_{gb}$ & $1.06\times  \hat\sigma_b$ & $(IL)^{-1/10}$ \\
    \multicolumn{1}{l}{} & $h_{gx}$ & $1.06\times \hat\sigma_x$ & $(IL)^{-1/10}$ \\
   \hline
    \multicolumn{1}{l}{\multirow{3}[2]{*}{GPV 2nd step}} & $h_{f_{\hat v}}$ & $1.06\times \sigma_{\hat v}$ & $n_t^{-1/10}$  \\
    \multicolumn{1}{l}{} & $h_{f_{x}}$ & $1.06\times  \hat\sigma_x$ & $n_t^{-1/10}$ \\
    \multicolumn{1}{l}{} & $h_x$ & $1.06\times  \hat\sigma_x$ & $L^{-1/9}$  \\
   \hline
    \multicolumn{1}{l}{Boundary} & $h_\delta$ & $\lambda_{\delta}>0 $ & $n^{-1/2}$ \\
    \hline
    \end{tabular}}%
    \caption{\footnotesize\textbf{Bandwidths when d=1.}}
  \label{hD1}%
\end{table}%
%We generate $IL$ private values $V_{p\ell}$ from $F_{0}(\cdot|\cdot,\cdot)$ and compute bids $B_{p\ell}$ using (\ref{BNE}). 
%Then, we estimate the distribution and density functions of observed bids using (\ref{eq:Ghat}) and (\ref{eq:ghat}).
%Next, we determine the pseudo private values $\hat V_{p\ell}$ from (\ref{eq:vhat}), and estimate $\hat{\theta}$ using 
%the sample moment condition as
We use 1000 replications for estimation where in each replication we: (i) generate randomly $IL$ private values using the truncated normal
distribution; (ii) compute the corresponding bids $B_{p\ell}$ using (\ref{BNE}); (iii) use these bids to estimate the distribution and density functions using (\ref{eq:Ghat}) and (\ref{eq:ghat}); (iii) determine the pseudo private values $\hat V_{p\ell}$ corresponding to $B_{p\ell}$; (iv) use this sample of pseudo private values to obtain $\hat{\theta}$ using 
the sample moment condition
\begin{eqnarray}
\frac{1}{IL}\sum_{\ell=1}^L \sum_{p=1}^I \nabla_{\theta}\ln f(\hat V_{p\ell}|I,X;\theta) =0.\label{eq:score}
\end{eqnarray}
We now quickly verify that this data generating process satisfies Assumptions A1 -- A6. 
It is immediate to verify that Assumption A1--A4 are satisfied because of the way we have designed the experiment. 
In our estimation we will restrict our attention at finding parameters from a compact set, and since log-likelihood is concave and smooth Assumption A5 (i)--(iii) are satisfied. Although we do not show the derivation, we can use the mean-value theorem to bound the slope of the moment conditions with respect to $V$. This slope is highest but bounded when $V=0.055$, which satisfies A5 (iv).   
Although tedious, we can still use the mean-value theorem to verify A6 (i), (ii) and (v). 
Again, the first part of A6 (iii) follows from the regularity conditions and the law of large numbers, the second part and the rest of A6 are satisfied by design.

\begin{figure}[t!]
\centering
%\caption{Densities of bidders' private values}
%\includegraphics[width=16.0cm]{mcpaper.eps}
\includegraphics[width=3.4in]{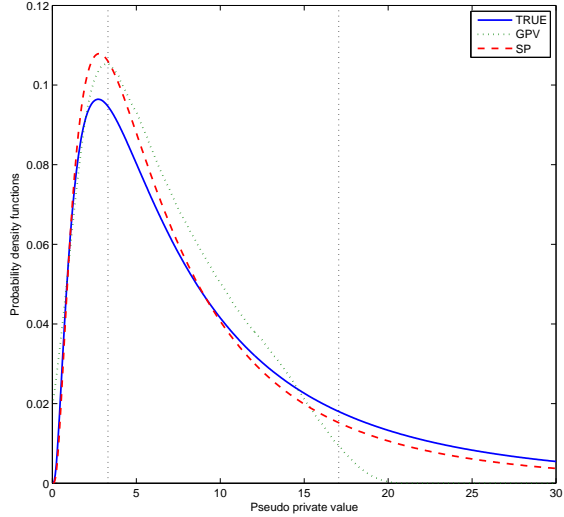}
\caption{{\bf Valuation Densities}. TRUE refers to the true density, while SP and GPV refer to the semiparametric and GPV(2000) estimators, respectively. The vertical lines  correspond to the trimming that is required only for the GPV estimator.}   \label{fig:1}
\end{figure}

We present our estimator (labeled SP) along with GPV (2000) estimated density and the true density, all evaluated at the median $X$ in Figure \ref{fig:1}. 
As is evident, our estimator is very close to the true density suggesting that it performs reasonably well. 
Next, we calculate the expected revenue given in (\ref{eq:revenue}) with $\overline{v}=30$. 
The true expected revenue is $\Pi_{True}=5.9$ and the revenue using our estimate gives $\Pi_{SP}= 5.8$ while using GPV gives $\Pi_{GPV}=4.6$, which means our estimate is much closer to the true value. 

\subsection{ Two Dimensional Covariate}
Let $X=(X_{1}, X_{2})^{T}\sim \log {\mathcal N}\left(\left(\begin{array}{c}1\\1\end{array}\right), \left(\begin{array}{cc}1 & 0.8\\0.8 &1\end{array}\right)\right)$, and  $V|X\sim \log{\mathcal N}(\mu_{v}(X), \sigma_{v}^{2}(X))$, both truncated at 0.055 and 30. 
As before we fix $I=5$ bidders in all auctions, but consider three different specifications for $(\mu_{v}(X), \sigma_{v}^{2}(X))$: (i) $(1+X_{1}/X_{2}, 1)$; (ii) $(1+X_{1}+X_{2}, 1)$; and (iii) $(1+X_{1}/X_{2}, \exp(0.01(X_{1}+X_{2})))$.
To compare the performance of our estimator we use simulated data from $L=200, 100, 50$ auctions. 

%Since $d=2$, to satisfy A2.(iii)  we require that the distribution of private values has 5 continuous bounded partial derivatives, so that $R=4$. 
In line with the assumption  A3, we choose the products of tri-weight kernel $(35/32)(1-u^2)^3 \Unit(|u|\leq 1)$ in our first-step.
Like before, we choose two bandwidths, one for each covariate, according to A4.AN, while ensuring under smoothing when compared to GPV estimator. %In particular we use $h_{G1}=2.978\times1.06\hat\sigma_{x_1} (IL)^{-1/8.5}$, $h_{G2}=2.978\times1.06\hat\sigma_{x_2} (IL)^{-1/8.5}$
%$h_{1g1}=2.978\times1.06 \hat\sigma_{x_1} (IL)^{-1/9.5}$, $h_{1g2}=2.978\times1.06 \hat\sigma_{x_2} (IL)^{-1/9.5}$, $h_{2g}=2.978\times1.06 \hat\sigma_b (IL)^{-1/9.5}$, where $\hat\sigma_b$ and $\hat\sigma_{x_j}$ are the estimated standard deviations of observed bids and  object heterogeneity, respectively. 
%It is important to recall, that we must under smooth when compared to the optimal bandwidth that we use to simulate the GPV estimator. 
To replicate the GPV (2000) estimator we choose the bandwidths according to the optimal rates. Thus, the order of the bandwidths
is $L^{-1/12}$ for $h_G$ and the second step bandwidth $h_{x_j}$ and $L^{-1/13}$ for $h_{gb}$ and $h_{gx_j}$ and the second step bandwidths $h_{fv}$ and $h_{fx_j}$, for $j=1,2$, see Table \ref{hD2}.
%Specifically we use $h_{G_k}=1.06\hat\sigma_{x_j} (IL)^{-1/12}$,
%$h_{gx_j}=1.06 \hat\sigma_{x_j} (IL)^{-1/13}$, $h_{gb}=1.06 \hat\sigma_b (IL)^{-1/13}$ where $\hat\sigma_b$ and $\hat\sigma_{x_j}$ are the corresponding standard deviations. The second step bandwidths are $h_{fv}=1.06 \hat\sigma_{\hat v} (n_t)^{-1/13}$, $h_{fx_j}=1.06 \hat\sigma_{x} (n_t)^{-1/13}$ and $h_{x_j}=1.06 \hat\sigma_{x} (L)^{-1/12}$, where $n_t$ is the number of observations remaining after trimming.

\begin{table}[t]
  \centering
  \scalebox{0.95}{
    \begin{tabular}{lccc}
    \hline
          & Symbol & Constant & Rate  \\
    \hline
    \multicolumn{1}{l}{\multirow{3}[2]{*}{LPE}} & $h_{Gj}$ & $2.978\times 1.06\times \hat\sigma_{xj}$ & $(IL)^{-1/8.5}$ \\
    \multicolumn{1}{l}{} & \multicolumn{1}{c}{$h_{1gj}$} & $2.978\times 1.06\times  \hat\sigma_{xj}$ & $(IL)^{1/9.5}$  \\
    \multicolumn{1}{l}{} & \multicolumn{1}{c}{$h_{2g}$} & $2.978\times 1.06\times  \hat\sigma_b$ & $(IL)^{-1/9.5}$ \\
   \hline
    \multicolumn{1}{l}{\multirow{3}[2]{*}{GPV 1st step}} & $h_{Gj}$ & $1.06 \times \hat\sigma_xj$ & $(IL)^{-1/12}$  \\
    \multicolumn{1}{l}{} & $h_{gb}$ & $1.06\times  \hat\sigma_b$ & $(IL)^{-1/13}$ \\
    \multicolumn{1}{l}{} & $h_{gxj}$ & $1.06\times \hat\sigma_{xj}$ & $(IL)^{-1/13}$ \\
   \hline
    \multicolumn{1}{l}{\multirow{3}[2]{*}{GPV 2nd step}} & $h_{f_{\hat v}}$ & $1.06\times \sigma_{\hat v}$ & $n_t^{-1/13}$  \\
    \multicolumn{1}{l}{} & $h_{f_{xj}}$ & $1.06\times  \hat\sigma_{xj}$ & $n_t^{-1/13}$ \\
    \multicolumn{1}{l}{} & $h_{xj}$ & $1.06\times  \hat\sigma_{xj}$ & $L^{-1/12}$  \\
   \hline
    \multicolumn{1}{l}{Boundary} & $h_\delta$ & $\lambda_{\delta}>0 $ & $n^{-1/3}$ \\
    \hline
    \end{tabular}}%
    \caption{\footnotesize\textbf{Bandwidths when d=2}.}
  \label{hD2}%
\end{table}%
We follow exactly the same steps as with $d=1$ to estimate the parameters, except that now we have three different sample sizes, $n=L\times I\in\{250,500,1000\}$ and three different specifications for mean and variance, and we use $\Theta= [0.055, 30]\times[0.001,2]$.
In total, there are 9 different cases, and hence 9 different densities. 
Figure \ref{fig:2} below shows the true density of private values against our estimator (the dashed line) and the GPV estimator (the dotted line). As can be seen in Figure \ref{fig:2} our estimator performs really well, even when there are only 50 auctions, while GPV (2000) is infeasible because after trimming we had very few observations left.
The estimated revenues for each density is presented in Table \ref{table:revenue}.
And as before, our estimator still performs relatively well. 
\begin{table}[h!]
\begin{center}
\scalebox{0.95}{
\begin{tabular}{|l|l|l|l|l|l|l|l|l|l|}
\hline
$(\mu(X),\sigma^{2}(X))$            & \multicolumn{3}{l|}{$\qquad$1} & \multicolumn{3}{l|}{$\qquad$2} & \multicolumn{3}{l|}{$\qquad$3} \\ \hline
 Profit$\backslash$ L & 200    & 100   & 50    & 200    & 100   & 50    & 200    & 100   & 50    \\ \hline
$\Pi_{True}$      & 6.2    & 6.1   & 6.0   & 5.9    & 5.7   & 5.7   & 6.9    & 7.0   & 7.0   \\ \hline
$\Pi_{SP}$        & 5.5    & 5.2   & 4.8   & 5.4    & 5.5   & 5.5   & 6.6    & 6.5   & 6.5   \\ \hline
$\Pi_{GPV}$       & 4.2    & 3.6   & --    & 4.8    & 2.9   & --    & 4.6    & 4.0   & --      \\ \hline
\end{tabular}}
\caption{\footnotesize \textbf { Optimal Revenue}: Each column corresponds to the three sets of mean and variance, and each cell contains the optimal revenue as defined in (\ref{eq:revenue}), one for each density in Figure \ref{fig:2}.} \label{table:revenue}
\end{center}
\end{table}

\begin{landscape}
\begin{figure}[ht!]
\centering
\includegraphics[width=8.5in]{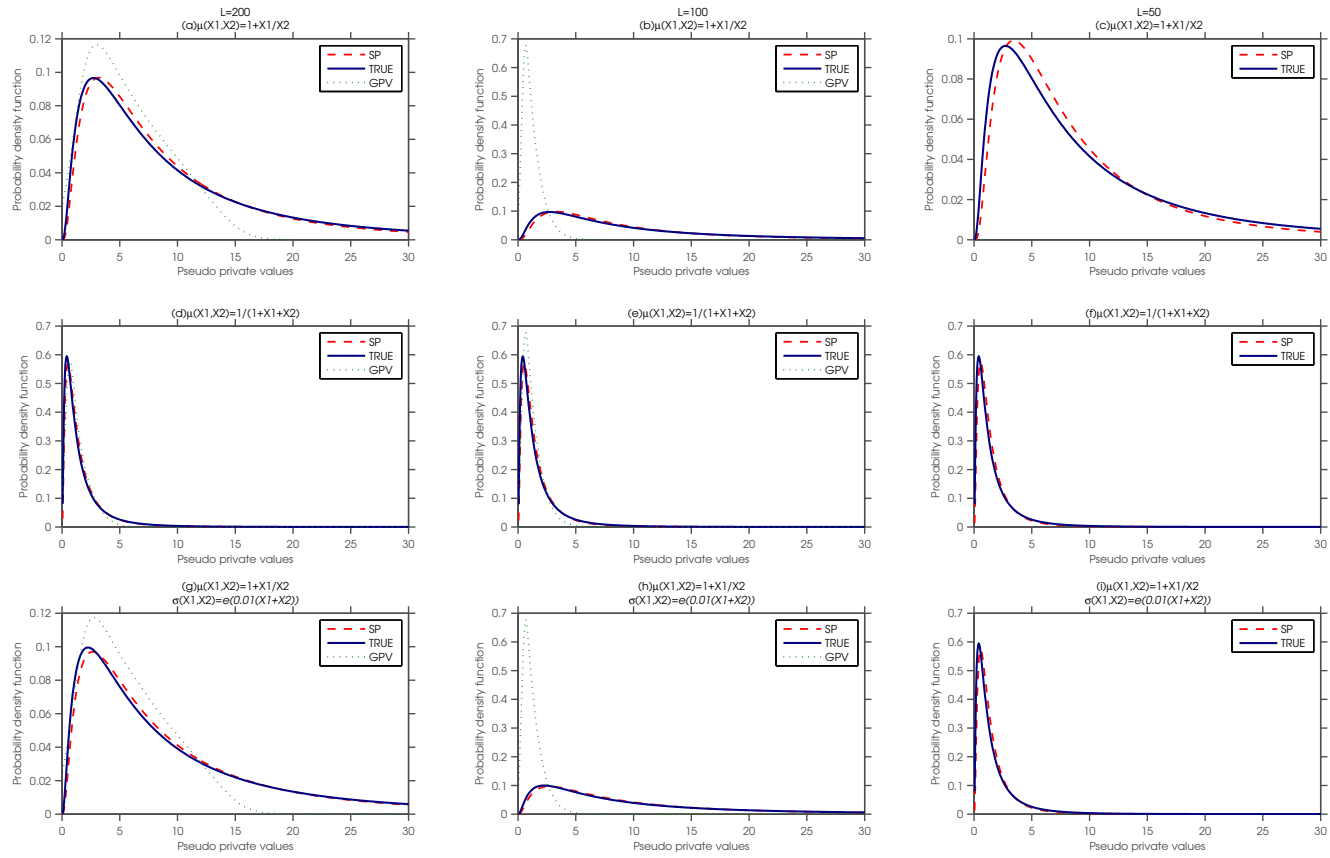}
\caption{\footnotesize\textbf{Valuation Densities}. TRUE refers to the true density, while SP and GPV refer to the semiparmaetric and GPV(2000) estimators, respectively. Each row refers to a different DGP while each column refers to different number of auctions. }  \label{fig:2}
\end{figure}
\end{landscape}

\section{Extensions\label{section:extensions}}

\indent
In this Section we indicate how to extend our procedure to a more general class of auction models. 
%To keep the notation as simple as possible we consider models without observed object heterogeneity. This is not restrictive since relaxing this assumption implies that the distribution and density functions have to be replaced by their conditional counterparts.

%\newpage
%\bigskip

\subsection{Binding Reserve Price}
%{\bf{Binding Reserve Price}}

\medskip

The first natural extension of the model considered in Section 2 is the symmetric IPV first-price auction model with a binding reserve price, announced or random.

%\medskip
\subsubsection{Announced Reserve Price}
%{\it{Announced Reserve Price}}

%\medskip

An announced binding reserve price $(r_{0}>\underline{V})$ constitutes
a screening device for participating in the auction. As pointed out
by GPV (2000) the Bayesian Nash equilibrium strategy is still given
by (2) in this set up, but the number $I$ of potential bidders
becomes unobserved and typically different from the observed number,
$I^*$, of actual bidders who have submitted a bid $(\geq r_{0})$.
Hence the model has a new structural element, namely $I$, in
addition to the latent distribution of bidders' private values. As
shown in GPV (2000), the differential equation defining the
equilibrium strategy can be rewritten as
\begin{eqnarray*}
V_{p}=\xi_0(B_{p},G^*_0,F(r_{0}),I)=B_{p}+\frac{1}{I-1}\left(\frac{G^*_0(B_{p})}{g^*_0(B_{p})}+\frac{F(r_{0})}{1-F(r_{0})}\frac{1}{g^*_0(B_{p})}\right),
\end{eqnarray*}

\noindent for $p=1,\ldots,I^*$ and where $G^*_0(\cdot)$ is the
truncated distribution of an observed bid conditional upon the fact
that the corresponding private value is grater than or equal to
$r_{0}$. Provided one can estimate $I$ and $F(r_{0})$ this equation is
the basis for a two step procedure analogous to that of Section
\ref{section:model}.
In particular, for $\ell=1\ldots,L$ and $p=1\ldots,I$ if $B_{p\ell}\geq r_{0\ell}$ the above equation becomes

\begin{eqnarray*}
V_{p\ell}=B_{p\ell}+\frac{1}{I_{\ell}-1}\left(\frac{G^*_0(B_{p\ell}\vert I_{\ell},Z_{\ell},\theta_{0})}{g^*_0(B_{p\ell}\vert I_{\ell},Z_{\ell},\theta_{0})}+\frac{F(r_{0}\vert Z_{\ell},\theta_{0}))}{1-F(r_{0}\vert I_{\ell},Z_{\ell},\theta_{0}))}\frac{1}{g^*_0(B_{p\ell}\vert I_{\ell},Z_{\ell},\theta_{0}))}\right),
\end{eqnarray*}
where $Z_{\ell}=(r_{0\ell},X_{\ell})$ and $\theta_{0}$ is the unknown true parameter vector. Let $h(\cdot\vert X_{\ell},\gamma_{0})$ be the probability mass function of the number of potential bidders, $I_{\ell}$, which is known up to a finite parameter $\gamma_{0}$. 
At the $\ell$-th auction, a potential bidder $p$ will bid if an only if his private valuation is above the reserve price. Thus, at this auction, the number of actual bidders is $I^{*}_{\ell}=1/I_{\ell}\sum_{p=1}^{I_{\ell}} \Unit(V_{p\ell} \geq r_{0\ell})$, which is a binomial random variable with parameters $(I_{\ell}, 1-F(r_{0\ell}\vert X_{\ell},\theta_{0})).$ In view of these, we propose to use the following moment conditions for bids and the observed number of bidders
\begin{eqnarray*}
{\rm E}[V_{p\ell}\vert V_{p\ell}\geq \xi_{0}(r_{0\ell}),X_{\ell};\gamma]&=&\sum_{I_{\ell}\geq2} m(I_{\ell},r_{0\ell},X_{\ell};\theta_{0}) h(I_{\ell} \vert X_{\ell};\gamma),\\
{\rm E}[I_{\ell}^{*}\vert I_{\ell}, Z_{\ell}]&=&\sum_{I_{\ell}\geq2} I_{\ell} [1-F(r_{0\ell}\vert X_{\ell};\theta)] h(I_{\ell} \vert X_{\ell};\gamma) , \\
{\rm E}[I_{\ell}^{*2}\vert I_{\ell},Z_{\ell}]&=&\sum_{I_{\ell}\geq2}  [1-F(r_{0\ell}\vert X_{\ell};\theta)]  [F(r_{0\ell}\vert X_{\ell};\theta) +I_{\ell} (1-F(r_{0\ell}\vert X_{\ell};\theta))],
\end{eqnarray*}
where the moment function $m(\cdot,\cdot,\cdot;\cdot)$ can be similar to the ones in (\ref{eq:moment-example}).\footnote{ We thank an anonymous referee for suggesting us to give the moment conditions for this model.} 
 
%These moment conditions constitute the structural econometric model corresponding to the theoretical model in which there is a binding reserve price and a varying number of potential bidders across auctions. %\footnote{ We thank an anonymous referee for suggesting us to give the moment conditions for this model.} 

%{\it{Random Reserve Price}}
\subsubsection{Random Reserve Price}
%\medskip

In some cases, as in timber and wine auctions, the seller may decide not to announce the reserve price at the time the auction takes place. Hence, the reserve price is said to be secret or random. Since bidders do not know it when submitting their bids, this fact brings into the model a new kind of uncertainty that has to be taken into account. To present the basic equation underlying our two-step procedure in this model we need first to introduce additional notation. 
To keep the notation as simple as possible we consider models without observed object heterogeneity. This is not restrictive since relaxing this assumption implies that the distribution and density functions have to be replaced by their conditional counterparts.

Let $V_0$ be the private value of the risk-neutral seller for the auctioned object. Moreover, we assume that $V_0$ is distributed according to $H(\cdot)$ defined on the same support as $F(\cdot)$ and that $H(\cdot)$ is common knowledge. \cite{ElyakimeLaffontLoiselVuong1994} have shown that in a first-price sealed bid auction $r_0=V_0$. In addition the bidders' equilibrium strategy is the solution of a differential equation which in general cannot be solved explicitly. See \cite{LiPerrigne2003}. However, this differential equation can be rewritten as follows
\begin{eqnarray*}
V_p=\xi_0(B_p,H,G_0,I)=B_{p}+\frac{1}{(I-1)\left(\frac{g_0(B_p)}{G_0(B_p)}+\frac{h(B_p)}{H(B_p)}\right)}\:,
\end{eqnarray*}
\noindent for $p=1,\ldots,I$. As mentioned by \cite{PerrigneVuong1999} since the reserve price is kept secret, all potential bidders
submit a bid. Hence $I$ is typically observed. The above equation
can be used as the basis of a two-step procedure similar to the one
described in Section 2. Namely, in a first step observed bids and
reserve prices can be used to estimate nonparametrically the
distribution $G_0(\cdot)$, its density $g_0(\cdot)$ as well as the
distribution $H(\cdot)$ and its density $h(\cdot)$. Next, pseudo
private values can be recovered using the equation above in order to
define a set of moment conditions for estimating the parameter of
interest $\theta_0$ in a second step.

We conclude this part  by noting that because of the revenue equivalence principle, \cite{Myerson1981,RileySamuelson1981}, our method is also useful to study other standard auctions such as third-price auctions, \cite{KagelLevin1993}, all-pay auctions.\footnote{ We thank one of the referees for the suggestion. These results are available upon request.}

\subsection{The Symmetric Affiliated Private Value (APV) Model}
%{\bf{The Symmetric Affiliated Private Value (APV) Model}}
%\medskip

To assume independence across private value can be restrictive since one can expect some degree of affiliation or positive correlation among private values. Thus, another natural extension of our framework is to consider the more general class of model encompassed by symmetric APV models. Affiliation means that if one bidder draws a high valuation for the auctioned object, then others bidders are likely to draw higher valuations too. \cite{LaffontVuong1996} study the problem of identification and theoretical restrictions in a general framework, namely in Affiliated Value (AV) models. 
In particular they show that any symmetric AV model is observationally equivalent to some symmetric APV model because the utility function is not identified from observed bids only.
Therefore when only data on observed bids are available, the result in \cite{LaffontVuong1996} implies that APV models can be considered, provided that we have identification.

We briefly indicate here how to adapt our estimation procedure to this kind of models, when all bids are observed and the reserve price is non-binding. Let $Y_p=\max_{p'\neq p}V_j$. The differential equation defining the equilibrium strategy in the APV model can be written as follows
\begin{eqnarray*}
V_p=\xi_0(B_p,G_0)\equiv B_p+\frac{G_{0,B_1|B_1}(B_p|B_p)}{g_{0,B_1|B_1}(B_p|B_p)},
\end{eqnarray*}
 subject to the boundary condition $s(\underline{V})=\underline{V}$, where $G_{0,B_1|B_1}(X_1|X_1)=F_{Y_1|V_1}(s^{-1}(X_1))$, $B_1=s_{0}(Y_1)$. The index ``1'' refers to any bidder since bidder are assumed to be ex-ante symmetric.This equation is again the basis for the identification result and estimation procedure. See \cite{PerrigneVuong1999} for details.
The theoretical restrictions as shown by \cite{LiPerrigneVuong2002} indicate that the joint distribution of bids $G_0(\cdot)$ can be rationalized by a symmetric APV model if and only if (i) $G_0(\cdot)$ is symmetric and affiliated and (ii) the function $\xi_0(\cdot,G_0)$ is strictly increasing on its support. Moreover, if these two conditions are satisfied, then the joint distribution $F(\cdot)$ of private values is identified.
In view of their results, \cite{LiPerrigneVuong2002} propose a two step fully nonparametric procedure in the same sprite as GPV(2000). Nevertheless, for the affiliated model the procedure has to be performed for each size $I$ (i.e. for each given number of bidders). 
Regarding estimation, the equation above suggests a two-step procedure analogous to the one described in Section \ref{section:model} for each size $I$ of bidders. In the first step the ratio $G_{{B_1|B_1}}(\cdot|\cdot)/g_{{B_1|B_1}}(\cdot|\cdot)$ can be estimated nonparametrically and then pseudo private values can be recovered. 
 In the second step a GMM procedure can be implemented to estimate the parameters of the underlying distribution of private values for a given $I$.
 % As suggested by \cite{LiPerrigneVuong_2002} if the underling structure doe not depend on $I$ one can pool the pseudo private values obtained from different sizes $I$. 
It is known that with affiliation, the rate of convergence is slower than with independence; this follows from Proposition 2 in \cite{LiPerrigneVuong2002}. In particular this Proposition gives explicit forms for the bandwidths that can be used in our framework since these choices satisfy our assumption A4.AN.\footnote{They consider homogenous auctions; see footnote 10 in \cite{LiPerrigneVuong2002}.}

%\medskip
\subsection{Asymmetric Models}
%{\bf{Asymmetric Models}}
%\medskip

Assuming that bidders are ex ante identical may constitute a limitation, and in some cases one needs to relax this assumption. Asymmetric auction models, however, lead to systems of differential equations without a closed form solution. Hence, the direct approach becomes extremely difficult to implement. Nevertheless, using our indirect procedure, parameters can be structurally estimated without solving for the equilibrium strategy or its inverse. 
%\medskip
\subsubsection{The Asymmetric IPV Model}
Following the exposition in \cite{PerrigneVuong2008} we assume that asymmetry is ex ante known to all bidders. Let $F_1(\cdot),\ldots,F_I(\cdot)$ be the private value distributions of the $I$ bidders whose identities are observed and let $G_{01}(\cdot),\ldots,G_{0I}(\cdot)$ be the corresponding bid distributions; see \cite{FlambardPerrigne2006}. We can then express the system of differential equations as
\begin{eqnarray*}
V_p=B_p+\frac{1}{\sum_{p'\neq p}\frac{g_{0p'}(B_p)}{G_{0p'}(B_p)}},\quad p,p'=1,\ldots, I,
\end{eqnarray*}
which lead, naturally to a two-step procedure, similar as before. 

%\medskip
\subsubsection{The Asymmetric APV Model}
%{\it{The Asymmetric APV Model}}
%\medskip

For simplicity we consider only two types of bidders. That is, the model assumes that the $I$-dimensional vector $(V_{11},\ldots,V_{1I_1},V_{21},\ldots,V_{2I_2})$ is distributed jointly as $F(\cdot)$ which is exchangeable in its first $I_1$ and last $I_2$ arguments. We can interpret this structure as follows. There is symmetry within each subgroup, and since $F(\cdot)$ is affiliated, there is general positive dependence among private values. \cite{PerrigneVuong1999} show that type specific equilibrium bidding strategies are characterized as the solution of the following system of differential equation,
\begin{eqnarray*}
V_{1p}&=&\xi_{1}(B_{1p},G_0)\equiv B_{1p}+\frac{G_{0B_1^*,B_0|B_1}(B_{1p},B_{1p}|B_{1p})}{\partial G_{0B_{1p}^*,B_0|B_1}(B_{1p},B_{1p}|B_{1p})/\partial (B^*_1,B_2)}, \hspace{6pt} p=1,2,\ldots,I_1\\
V_{2p}&=&\xi_{0}(B_{2p},G_0)\equiv B_{2p}+\frac{G_{0B_1,B_2^*|B_2}(B_{2p},B_{2p}|B_{2p})}{\partial G_{0B_1,B_2^*|B_2}(B_{2p},B_{2p}|B_{2p})/\partial (B_1,B_2^*)}, \hspace{6pt} p=1,2,\ldots,I_2,
\end{eqnarray*}
where $B_t^*=\max _{p\neq1,p\in I_t}B_{tp}$, $B_t=\max _{p\in I_t}B_{tp}$, for $t=1,2$. \cite{CampoPerrigneVuong2003} show that this identifies $F(\cdot,\ldots,\cdot)$, and use a nonparametric two-step procedure to estimate the model. 
Similar to above, the two-step semiparametric procedure would involve using the above system of equations to recover the pseudo private values after obtaining nonparametric estimates for $G_{0B_1^*,B_2|B_1}(\cdot,\cdot|\cdot)$ and $G_{0B_1,B_2^*|B_2}(\cdot,\cdot|\cdot)$, and then estimating the parameters of $\theta$ through a set of moment conditions.
As \cite{CampoPerrigneVuong2003} have shown, the choice of bandwidths for asymmetric APV is similar to bandwidths for symmetric case, like  \cite{LiPerrigneVuong2002}. Which means we can follow the same steps as in symmetric APV to choose our bandwidths.

\subsection{Unobserved Heterogeneity}
In some auctions it is possible that even after conditioning on auction covariates $Z_{\ell}$, the bids are still correlated.
Other than affiliation, such correlation could be a result of an auction characteristic $U_{\ell}\in\mathbb{R}_{++}$ that is missing in the data, but is observed by the bidders.  Such auctions are known as auctions with unobserved heterogeneity. 
In this subsection, we propose one possible way to adapt our semiparametric procedure to auctions with multiplicative unobserved heterogeneity as studied by \cite{Krasnokutskaya2011}.
As it will be clear later, in this case the rate of convergence and the asymptotic variance derived in our Proposition \ref{prop:2} will not be applicable. 
But determining the exact asymptotic properties is beyond the scope of this paper.  
We begin by introducing new and relevant notations and assumptions. 

\medskip\noindent
{\bf Assumption A8:} {\em 

\begin{itemize}
\item[(i)] Let $V_{p\ell}=\tilde{V}_{p\ell}\times U_{\ell}$ be bidder $p$'s value in $\ell^{th}$ auction such that $U_{\ell}\perp \tilde{V}_{p\ell}$  and $U_{\ell}\perp Z_{\ell}$.
\item[(ii)] Given $Z$ the random variables $V_{p\ell},\tilde{V}_{p\ell}$ are independently and identically distributed as $F(\cdot|Z)$ and  $\tilde{F}(\cdot|Z)$, respectively. 
\item[(iiii)]  $U_{\ell}$ is independently and identically distributed as $F_{U}(\cdot)$ across auctions with $E(\ln (U))=0$.
\end{itemize}
}
\medskip\noindent
In summary, $U_{\ell}$ is independent across auctions, and in every auction with covariates $Z_{\ell}$ each bidder draws his/her ``true value'' $\tilde{V}_{p\ell}$ from $\tilde{F}(\cdot|Z_{\ell})$ and bids according to $V_{p\ell}$. 
Let $s(\cdot|Z_{\ell},U_{\ell})$ denote the bidding strategy when the observed and unobserved covariates are, $Z_{\ell}$ and $U_{\ell}$, respectively, and let $\tilde{s}(\cdot|Z_{\ell})=s(\cdot|Z_{\ell},U_{\ell}=1)$ be the bidding strategy when the $U_{\ell}=1$, i.e., without unobserved heterogeneity. 
 \cite{Krasnokutskaya2011} shows that under Assumption A8: $s(V_{p\ell}|Z_{\ell},U_{\ell})=\tilde{s}(\tilde{V}_{p\ell}|Z_{\ell})\times U_{\ell}$ so that the bids satisfy $B_{p\ell}=\tilde{B}_{p\ell}\times U_{\ell}$, where $\tilde{B}_{p\ell}$ is the bid by bidder $p$ in auction $\ell$ when $U_{\ell}=1$. 
If we use $G_{0}(\cdot|Z,U)$ and $\tilde{G}_{0}(\cdot|Z)$ to denote the conditional distribution of $B$ given $(Z,U)$ and the conditional distribution of $\tilde{B}$ given $Z$, respectively, then $B_{p\ell}=\tilde{B}_{p\ell}\times U_{\ell}$ implies 
\begin{eqnarray*}
\tilde{G}_{0}(b|Z)&=&\Pr(\tilde{B}\leq b|Z)=\Pr(B\leq b\times U|Z,U)= {G}_{0}(b\times U|Z,U);\\
\tilde{g}_{0}(b|Z)&=&{g}_{0}(b\times U|Z,U)\times U.
\end{eqnarray*}
To simplify notation we suppress $Z$, and everything is to be understood as conditional on $Z$, unless stated otherwise. 
The first step is to identify $F_{U}(\cdot)$. For auction $\ell=1,\ldots, L$ select any two bids and call them $(B_{1\ell},B_{2\ell})$, and let $\Ch_{(\ln B_{1},\ln B_{2})}(\cdot,\cdot)$ be the joint characteristic function of $(\ln B_{1}, \ln B_{2})$. 
Then under a normalization $E[\ln \tilde{B}_{1}]=0$, \cite{Krasnokutskaya2011} shows that we can use \cite{Kotlarski1966} to identify the characteristic function of $ \ln U$ as 
$$
\Ch_{\ln U}(t)=\exp\left(\int_{0}^{t}\frac{\partial \Ch_{(\ln B_{1},\ln B_{2})}(0,c)/\partial \ln B_{1}}{\Ch_{(\ln B_{1},\ln B_{2})}(0,c)}dc\right),
$$
which identifies $F_{\ln U}(\cdot)$ as the Fourier inverse of $\Ch_{\ln U}(t)$.
 So for the remainder we treat $F_{U}(\cdot)$ as known.
The second step consists on estimating $\tilde{G}_{0}(\cdot)$ and $\tilde{g}_{0}(\cdot)$ of $\tilde{B}$ from: 
$$
\ln B_{p\ell}=\ln \tilde{B}_{p\ell}+ \ln U_{\ell}, \quad p=1,\ldots, I_{\ell}, \ell=1,\ldots, L.
$$ 
Since $\ln U_{\ell}$ is unobserved, the estimators defined in (\ref{eq:Ghat}) and (\ref{eq:ghat}) are infeasible. 
We can, however, replace the unobserved $(\ln \tilde{B}_{j}-b)^{\rho}K_{h}(\ln \tilde{B}_{j}- b)$ in (\ref{eq:Ghat}) and (\ref{eq:ghat}) with $(\ln {B}_{j}-b)^{\rho}\tilde{K}_{h}(\ln {B}_{j}- b)$,  \cite{FanTruong1993, DelaigleFanCarroll2009}, where $\tilde{K}_{h}(b)=h^{-1}\tilde{K}(b/h)$, and satisfies 
$$
{\rm E}\left\{(\ln {B}_{j}-b)^{\rho}\tilde{K}_{h}(\ln {B}_{j}- b)|\ln \tilde{B}_{j}\right\}=(\ln \tilde{B}_{j}-b)^{\rho}K_{h}(\ln \tilde{B}- b),\quad \rho=0,1.
$$  
\cite{DelaigleFanCarroll2009} propose using the Fourier transformation of the above equation to determine $\tilde{K}_{h}(\cdot)$, and show that the presence of $U_{\ell}$  only affect the
variance of the estimator.\footnote{ The final result also depends on the smoothness of the density of the unobserved heterogeneity, see \cite{DelaigleHallMeister2008,DelaigleFanCarroll2009}.}
% whether the density  prpose a  
%replace  Under the assumption that $\Ch_{\ln U}(t)\neq 0, \forall t\in\mathbb{R}$, the inverse Fourier transform, \cite{Lukcas1960}, gives 
%\begin{eqnarray*}
%\tilde{G}_{\ln \tilde{B}}(\tilde{b})&=&\frac{1}{2}-\frac{1}{\pi}\int_{0}^{\infty}\frac{1}{t}Im\left[e^{-{\bf i} t \tilde{b}}\Ch_{\ln \tilde{B}}(t)\right]dt=\frac{1}{2}-\frac{1}{\pi}\int_{0}^{\infty}\frac{1}{t}Im\left[e^{-{\bf i} t \tilde{b}}\frac{\Ch_{\ln {B}}(t)}{\Ch_{\ln U}(t)}\right]dt;\\
%\tilde{g}_{\ln\tilde{B}}(\tilde{b})&=&\frac{1}{2\pi}\int e^{-{\bf i}t\tilde{b}}\Ch_{\ln \tilde{B}}(t)dt=\frac{1}{2\pi}\int e^{-{\bf i}t\tilde{b}}\frac{\Ch_{\ln {B}}(t)}{\Ch_{\ln U}(t)}dt. 
%\end{eqnarray*}
%Instead of using the sample average $\frac{1}{L}\sum_{\ell=1}^{L}\frac{1}{I_{\ell}}\sum_{p=1}^{I_{\ell}}\exp({\mathbf i}t B_{p\ell})$ as the estimator for ${\Ch}_{\ln B}(t)$ we propose to use 
%$$
%\hat{\Ch}_{\ln B}(t)=\int e^{{\bf i}t\ln B} \hat{g}_{\ln B}(b)db,
%$$ 
%where $\hat{g}_{\ln B}(\cdot)$ is the LPE(R-1) of the density of $\ln B$ (see Section 2.2), because the former is unstable since the characteristic function has large fluctuations at its tail \cite[see][]{DelaigleHallMeister2008}. 
This gives us the estimates of the distribution and density of $\tilde{B}$. 
Even though we cannot recover $\ln \tilde{B}_{p\ell}$, from $\ln B_{p\ell}$, we simulate the former  from ${\tilde{G}}_{0}(\cdot|Z_{\ell})$ and determine
$$
\tilde{V}_{p\ell}^{\iota}=\tilde{B}_{p\ell}^{\iota}+\frac{1}{I_{\ell}-1}\frac{\tilde{G}_{0}(\tilde{B}_{p\ell}^{\iota}|Z_{\ell})}{\tilde{g}_{0}(\tilde{B}_{p\ell}^{\iota}|Z_{\ell})},\quad p=1,\ldots, I_{\ell}, \ell=1,\ldots, L, \iota=1,\ldots, {\mathbf S},
$$
where ${\bf S}$ is large. 
Then in the third-step we estimate $\hat{\theta}$ using the appropriate empirical moment conditions: 
$$
\frac{1}{{\bf S}}\sum_{\iota=1}^{{\bf S}}  \frac{1}{L}  \sum_{\ell=1}^{L}  \frac{1}{I_{\ell}} \sum_{p=1}^{I_{\ell}} m(\tilde{V}_{p\ell}^{\iota}, Z_{\ell};\theta)\approx0.
$$
We conjecture that the estimator is consistent and asymptotically normal.
Since the rate of convergence of the estimator in the first step depends heavily on the smoothness of $f_{U}(\cdot)$ -- the smoother the density, the slower the convergence \cite{Fan1991} -- and because the error in steps 1 and 2 affect the asymptotic variance of $\hat{\theta}$, Proposition \ref{prop:2} does not apply here. 
Full characterization of the asymptotic properties of $\hat{\theta}$ needs careful consideration and is left for future research.  

\section{Conclusions\label{section:conclusion}}

\indent
In this paper we develop an indirect procedure to estimate first-price sealed-bid auction models, contributing in this way to the structural analysis of auction data that has been developed in the last fifteen years. Following GPV (2000) our procedure is in two steps. The difference with GPV (2000) is that our second step is implemented using a GMM procedure so that our resulting model is semiparametric. We show that our semiparametric estimator converges uniformly at the parametric $\sqrt{L}$ rate while the nonparametric estimator in GPV (2000) was shown to converge at the best possible rate according to the minimax theory which is slower than the parametric rate. Moreover, our procedure is not subject to the so-called curse of dimensionality or in other words the convergence rate is independent of the dimension of the exogenous variables. We establish consistency and asymptotic normality of our estimator.

Given the nature of our procedure it is not necessary to solve explicitly for the equilibrium strategy or its inverse. This is a valuable advantage with respect to direct methods specially when estimating models that lead to intractable first-order conditions, such as asymmetric auction models. More generally, our method extends to models which have been estimated using a nonparametric indirect procedure. In this respect, we briefly outline how this can be done in models with a binding reserve price (announced or random), affiliated private value models and asymmetric models.

Finally, we conducted a set of Monte Carlo simulations. The main purpose for this was to asses the performance of our estimator in finite samples relative to the nonparametric estimator proposed by GPV (2000). Our semiparametric estimator does a good job in matching the true density. When comparing with the nonparametric GPV (2000) estimator, we can see that the estimator developed in this paper is not subject to boundary effects. Moreover, using our estimator generates optimal revenue that is closer to the revenue, if we had used the true density, than using GPV (2000) estimator. 

Since \cite{Krasnokutskaya2011} unobserved auction heterogeneity has become important in empirical auction -- ignoring it can lead to serious misspecification error. Moreover, it is known that the nonparmetric estimation is precarious, more so that the auctions without unobserved heterogeneity. Although we touched on this subject in this paper, we believe that determining asymptotic and efficiency properties of a semiparametric estimator is important.
We hope our paper provides the necessary impetus and motivation for someone to explore this problem.  

\newpage
\small
\setcounter{equation}{0}
\renewcommand{\theequation}{A.\arabic{equation}}
\begin{center}
{\bf Appendix \\
Proofs of Asymptotic Properties}
\end{center}
\bigskip
This Appendix gives the proofs of our asymptotic results
(Propositions \ref{prop:1} and \ref{prop:2}).
First, we present two important results.

\medskip

\noindent {\bf{Results:}} Under A4 we have,

\begin{itemize}
\item[(i)]
$\displaystyle{\sup_{(b,x,I)}}\left\vert \hat g(b|x,I)-g_0(b|x,I)\right\vert=O_{as}\left(h_{1g}^R+h_{2g}^R+ \sqrt{\frac{\log L}{L h_{1g}h_{2g}}}\quad\right)$

\item[(ii)]
$\displaystyle{\sup_{(b,x,I)}}\left\vert \hat G(b|x,I)-G_0(b|x,I)\right\vert=O_{as}\left(h_{G}^{R+1}+ \sqrt{\frac{\log L}{L h_{G}}}\quad\right)$

\medskip
\noindent For a proof of the above results we refer the reader to \cite{KorostelevTsybakov1993}.
We observe that the above results imply that $\displaystyle{\sup_{p\ell}}\vert\hat V_{p\ell}-V_{p\ell}\vert=o_{as}(1).$

\end{itemize}
{\bf{Proposition \ref{prop:1}}}
\begin{proof} 
It suffices to show that
$\displaystyle{\sup_{\theta \in \Theta}\parallel S_L(\theta)-\hat
S_L(\theta)\parallel}=o_{as}(1)$. From the triangle inequality, A5-(iv) it follows that
\begin{eqnarray}
\sup_{\theta \in \Theta}\Vert S_L(\theta)-\hat S_L(\theta)\Vert&=&\sup_{\theta \in \Theta}\Bigg\Vert\ds m(V_{p\ell},Z_\ell;\theta)-\ds m(\hat V_{p\ell},Z_\ell;\theta)\Bigg\Vert\nonumber\\
&=&\sup_{\theta \in \Theta}\Bigg\Vert\ds [m(V_{p\ell},Z_\ell;\theta)-m(\hat V_{p\ell},Z_\ell;\theta)]\Bigg\Vert\nonumber\\
&\leq & \ds \sup_{\theta \in \Theta}\Big\Vert m(V_{p\ell},Z_\ell;\theta)-m(\hat V_{p\ell},Z_\ell;\theta)\Big\Vert \nonumber\\
&\leq & \ds K_1(Z_\ell)\vert\hat{V}_{p\ell}-V_{p\ell}\vert \notag\\
&=&\{{\rm E}[K_1(Z)]+o_{as}(1)\} \sup_{p\ell}\vert\hat V_{p\ell}-V_{p\ell}\vert \nonumber\\
&=&o_{as}(1)
\end{eqnarray}
Where we use the fact that $\hat V_{p\ell}$ is a consistent estimator of $V_{p\ell}$, i.e. we make use of the 2 results stated at the beginning of this Appendix.
Therefore, the desired result follows. 
\end{proof}
\newpage
{\bf {Proposition \ref{prop:2}}}
\begin{proof}
Recall that $\hat{\theta}$ is the feasible estimator while $\tilde{\theta}$ is the infeasible estimator. 
We want to show: 
\begin{eqnarray*}
\sqrt{L}(\hat\theta-\theta_0)&=&\sqrt{L}(\tilde{\theta}-\theta_0)-\frac{L(L-1)}{L^2}\frac{2}{\sqrt{L}}
\sum_{\{\ell:I_\ell=I\}}^L\frac{1}{I}\sum_{p=1}^I\Bigg\{\sum_{I}\frac{1}{I(I-1)} N(Y_{p\ell},I) f_m^{-1}(X_\ell,I) \\
&&g_0(Y_{p\ell},I)+{\rm{E}}\Bigg[\sum_{I}\frac{1}{I(I-1)} N(Y_{p\ell},I) f_m^{-1}(X_\ell,I) g_0(Y_{p\ell},I)\Bigg]\Bigg\}+o_p(1)\\
&=&-\frac{1}{\sqrt{L}}\sum_{\{\ell:I_\ell=I\}}^L\frac{1}{I}\sum_{p=1}^I\Bigg\{(C^T\Omega C)^{-1}C \Omega m(V_{p\ell},X_\ell,I;\theta_0)
+2\frac{L(L-1)}{L^2}\\
&&\sum_{I}\frac{1}{I(I-1)} N(Y_{p\ell},I) f_m^{-1}(X_\ell,I) g_0(Y_{p\ell},I)\\
&&-{\rm{E}}\Bigg[\sum_{I}\frac{1}{I(I-1)} N(Y_{p\ell},I) f_m^{-1}(X_\ell,I) g_0(Y_{p\ell},I)\Bigg]\Bigg\}+o_p(1),
\end{eqnarray*}
where $C={\rm{E}}\left[{\partial m(V,X,I;\theta_0)}/{\partial \theta}
\right], Y_{p\ell}\equiv (B_{p\ell}, X_{\ell}), \Omega$ is the p.d weighting matrix and
$$
N(Y_{p\ell},I)=[m_1(V_{p\ell},X_\ell,I;\theta_0)/ g_0(B_{p\ell}|X_\ell,I)^2]
G_0(B_{p\ell}|X_\ell,I).$$
The terms inside $\{\}$ in the second equality is the influence function.\footnote{The idea behind that is the following observation. Suppose after a Taylor expansion, we have: 
\begin{eqnarray*}
\left\{\begin{array}{c}\sqrt{n}(\hat{\theta}-\theta_{0})=\frac{1}{\sqrt{n}}\sum_{i=1}^{n}\psi(z_{i})+o_{p}(1)\\E(\psi(z))=0, var(\psi(z))<\infty\end{array}\right.
\end{eqnarray*}
where $\psi(\cdot)$ is the influence function, then the asymptotic variance is $V=var(\psi(z))$. An alternative method would have been to follow \cite{Newey1994} and use the path derivative approach.} 
Once we have shown this asymptotic linear representation the result follows from the Central Limit Theorem. 

From the FOCs that characterize $\tilde{\theta}$ and $\hat\theta$
respectively, we have
\begin{eqnarray}
\frac{1}{2}\frac{\partial Q_L}{\partial
\theta}(\tilde{\theta})&=&\htr(\tilde{\theta})\Omega S_L(\tilde{\theta})=0\label{eq:a2}\\
\frac{1}{2}\frac{\partial \hat Q_L}{\partial
\theta}(\hat\theta)&=&\Htr(\hat\theta)\Omega \hat S_L(\hat\theta)=0.\label{eq:a3}
\end{eqnarray}
We can use a Taylor expansion around $\theta_0$ to obtain
\begin{eqnarray}
S_L(\tilde{\theta})&=&S_L(\theta_0)+\htt(\overline{\theta})(\tilde{\theta}-\theta_0)\label{eq:a4}\\
\hat S_L(\hat\theta)&=&\hat
S_L(\theta_0)+\Htt(\overline{\theta}^*)(\hat\theta-\theta_0),\label{eq:a5}
\end{eqnarray}
where $\overline{\theta}$ and $ \overline{\theta}^*$ are
vectors between $\tilde{\theta}$ and $\theta_0$, and $\hat\theta$
and $\theta_0$, respectively.
Thus using (\ref{eq:a4}) in (\ref{eq:a2}) we get
\begin{eqnarray*}
\htr(\tilde{\theta})\Omega\left[S_L(\theta_0)+\htt(\overline{\theta})(\tilde{\theta}-\theta_0)\right]
=\htr(\tilde{\theta})\Omega S_L(\theta_0)+\htr(\tilde{\theta})\Omega\htt(\overline{\theta})(\tilde{\theta}-\theta_0)=0.
\end{eqnarray*}

Therefore, we have
%\begin{eqnarray}
\[\sqrt{L}(\tilde{\theta}-\theta_0)=-\left[\htr(\tilde{\theta})\Omega\htt(\overline{\theta})\right]^{-1}\htr(\tilde{\theta})\Omega\sqrt{L}S_L(\theta_0)
=-\tilde{A}^{-1}\tilde{B}\sqrt{L}S_L(\theta_0).\label{eq:infeasible}\tag{**}\]
%\end{eqnarray}

Similarly using (\ref{eq:a5}) in (\ref{eq:a3}) yields
\begin{eqnarray*}
\sqrt{L}(\hat\theta-\theta_0)&=&-\left[\Htr(\hat\theta)\Omega\Htt(\overline{\theta}^*)\right]^{-1}
\Htr(\hat\theta)\Omega \sqrt{L} \hat S_L(\theta_0)
=-\hat A^{-1} \hat B\sqrt{L} \hat S_L(\theta_0).
\end{eqnarray*}
Next, we show: (i) $\tilde{B}-\hat B=o_{as}(1)$ ( Step \ref{step1});
 (ii) $\tilde{A}-\hat A=o_{as}(1)$, which together with A6-(iii) imply $\tilde{A}^{-1}-\hat A^{-1}=o_{as}(1)$(Step \ref{step2});
  and finally (iii) $\sqrt{L}[S_L(\theta_0)-\hat S_L(\theta_0)]=O_p(1)$ (Step \ref{step3}).
%The proof consists of three steps:
%\begin{description}
%\item[ i.] Step 1:  $\tilde{B}-\hat B=o_{as}(1)$, pages 30-31.
%\item[ ii.] Step 2: $\tilde{A}-\hat A=o_{as}(1)$, pages 31-33
%\item[ iii.] Step 3: $\sqrt{L}[S_L(\theta_0)-\hat S_L(\theta_0)]=O_p(1)$. pages 33- end.
%\end{description}
\setcounter{equation}{5}
\setcounter{section}{0}
\renewcommand{\theequation}{A.\arabic{equation}}
\section{Step 1\label{step1}} We prove $\tilde{B}-\hat B=o_{as}(1)$. The term
$\tilde{B}-\hat B$ can be written as
\begin{eqnarray*}
\tilde{B}-\hat
B&=&\htr(\tilde{\theta})\Omega-\Htr(\hat\theta)\Omega=\left(\htr(\tilde{\theta})-\Htr(\hat\theta)\right)\Omega\\
&=&\left(\ds \left(m_3^T(V_{p\ell},Z_\ell,\tilde{\theta})-m_3^T( \hat
V_{p\ell},Z_\ell;\hat\theta)\right)\right)\Omega.
\end{eqnarray*}

It suffices to show that the norm of the term between brackets is
$o_{as}(1)$ since $\Omega$ is a positive definite matrix. Namely
\begin{eqnarray}
\lefteqn{\left\Vert \ds m_3^T(V_{p\ell},Z_\ell,\tilde{\theta})-m_3^T(\hat V_{p\ell},Z_\ell;\hat\theta))\right\Vert}\nonumber\\
&=&\Bigg\Vert\ds \Big[(m_3^T(V_{p\ell},Z_\ell,\tilde{\theta})-m_3^T(\hat V_{p\ell},Z_\ell;\tilde\theta))+m_3^T(\hat V_{p\ell},Z_\ell;\tilde{\theta})-m_3^T(\hat V_{p\ell},Z_\ell;\hat\theta))\Big]\Bigg\Vert\nonumber\\
&\leq& \left \Vert \ds\left[(m_3^T(V_{p\ell},Z_\ell;\tilde{\theta})-m_3^T(\hat
V_{p\ell},Z_\ell;\tilde\theta))\right]\right\Vert+\left\Vert\ds m_3^T(\hat V_{p\ell},Z_\ell;\tilde{\theta})-m_3^T(\hat V_{p\ell},Z_\ell;\hat\theta))\right\Vert\nonumber\\
&=&C+D,\label{eq:a6}
\end{eqnarray}

\noindent where the last line follows from the triangle inequality.
The term $C$ in (\ref{eq:a6}) is
\begin{eqnarray*}
C&=&\left\Vert \ds\left[(m_3^T(V_{p\ell},Z_\ell;\tilde{\theta})-m_3^T(\hat
V_{p\ell},Z_\ell;\tilde\theta))\right]\right\Vert\leq\ds\left\Vert\left[m_3^T(V_{p\ell},Z_\ell;\tilde{\theta})-m_3^T(\hat
V_{p\ell},Z_\ell;\tilde{\theta})\right]\right\Vert\\
&\leq& \ds K_3(Z_\ell) \vert V_{p\ell}-\hat V_{p\ell}\vert\leq\{{\rm{E}}[K_3(Z)]+o_{as}(1)\}\sup_{p\ell}\vert \hat V_{p\ell}-V_{p\ell}\vert=o_{as}(1),
\end{eqnarray*}
\noindent where we use assumption A6-(i) and the fact that $\hat V_{p\ell}$ is uniformly consistent-- these results are stated at the beginning of this Appendix.
We consider now the term $D$ in (\ref{eq:a6}):
\begin{eqnarray*}
D&=&\left\Vert \ds \left[m_3^T(\hat V_{p\ell},Z_\ell;\tilde{\theta})-m_3^T(\hat V_{p\ell},Z_\ell;\hat{\theta})\right]\right\Vert\leq \ds \left\Vert m_3^T(\hat V_{p\ell},Z_\ell;\tilde{\theta})-m_3^T(\hat V_{p\ell},Z_\ell;\hat{\theta})\right\Vert\\
&\leq &\ds K_4(Z_\ell) \Vert \tilde\theta-\hat\theta \Vert=\{{\rm{E}}[K_4(Z)]+o_{as}(1)\} \times o_{as}(1),
\end{eqnarray*}

\noindent where we have used A6-(ii) and the fact that
$\tilde{\theta}$ and $\hat\theta$ are consistent estimators for
$\theta_0$.

\section{Step 2\label{step2}} We prove $\tilde{A}-\hat A=o_{as}(1)$. The term
$\tilde{A}-\hat A$ is
\begin{eqnarray}
\tilde{A}-\hat
A&=&\left(\htr(\tilde{\theta})\Omega\htt(\overline{\theta})\right)-
\left(\Htr(\hat\theta)\Omega \Htt(\overline{\theta}^*) \right)\nonumber\\
&=&\left[\htr(\tilde{\theta})\Omega \left(\htt(\tilde{\theta})+o_{as}(1) \right)\right]
-\left[\Htr(\hat\theta)\Omega \left(\Htt(\hat\theta)+o_{as}(1)\right)\right]\nonumber\\
&=&\left(\htr(\tilde\theta)\Omega \htt(\tilde\theta)\right)-\left(\Htr(\hat\theta)\Omega\Htt(\hat\theta)\right)+o_{as}(1)\nonumber\\
&=&\left[\left(\htr(\tilde\theta)-\Htr(\hat\theta)\right)\Omega\right]\left(\htt(\tilde\theta)+\Htt(\hat\theta)\right).\label{eq:a7}
\end{eqnarray}

\noindent where the second equality comes from the following
\begin{eqnarray*}
\left\Vert \htt(\overline{\theta})-\htt(\tilde\theta)\right\Vert&\leq&\ds
\Vert m_3(V_{p\ell},Z_\ell;\overline{\theta})-m_3(V_{p\ell},Z_\ell;\tilde{\theta})\Vert\\
&\leq& \ds K_4(Z_\ell) \Vert \overline{\theta}-\tilde\theta \Vert
=\{{\rm{E}}[K_4(Z)]+o_{as}(1)\} o_{as}(1)=o_{as}(1),
\end{eqnarray*}
where we use A6-(ii), the fact that $\tilde\theta\leq\overline{\theta}\leq \theta_0$ and that
$\tilde\theta \stackrel{a.s}{\longrightarrow}\theta_0$. Similarly we can show that
\begin{eqnarray*}
\Htt(\overline{\theta}^*)=\Htt(\hat\theta)+o_{as}(1),
\end{eqnarray*}
since $\hat\theta\leq\overline{\theta}^*\leq \theta_0$ and $\hat\theta \stackrel{a.s}{\longrightarrow}\theta_0$.
Now, for the last line in (\ref{eq:a7}) we observe that by Step 1, the first factor in (\ref{eq:a7}) is $o_{as}(1)$  and
the second factor can be expressed as follows
\begin{eqnarray*}
\lefteqn{\left\Vert\left(\htt(\tilde{\theta})+\Htt(\hat\theta)\right)\right\Vert=\left\Vert\ds
[m_3(V_{p\ell},Z_\ell;\tilde\theta)+ m_3(\hat V_{p\ell},Z_\ell;\hat\theta)]\right \Vert}\\
&\leq& \ds \Vert [m_3(V_{p\ell},Z_\ell;\tilde\theta)+ m_3(\hat V_{p\ell},Z_\ell;\hat\theta)] \Vert\leq \ds \Vert m_3(V_{p\ell},Z_\ell;\tilde\theta) \Vert + \ds \Vert m_3(\hat V_{p\ell},Z_\ell;\hat\theta)\Vert\\
&\leq& \ds \sup_{\theta\in\Theta} \Vert m_3(V_{p\ell},Z_\ell;\theta) \Vert + \ds \Vert m_3(\hat V_{p\ell},Z_\ell;\hat\theta)-
m_3(\hat V_{p\ell},Z_\ell;\theta_0)+ m_3(\hat V_{p\ell},Z_\ell;\theta_0)\Vert \\
&\leq& \ds K_5(V_{p\ell},Z_\ell)+ \ds \Vert m_3(\hat V_{p\ell},Z_\ell;\hat\theta)- m_3(\hat V_{p\ell},Z_\ell;\theta_0)\Vert\\
&&+\ds \Vert m_3(\hat V_{p\ell},Z_\ell;\theta_0)-m_3(V_{p\ell},Z_\ell;\theta_0) +m_3(V_{p\ell},Z_\ell;\theta_0)\Vert\\
&\leq& \{{\rm{E}}[K_5(V,Z)]+o_{as}(1)\}+\ds K_4(Z_\ell) \Vert\hat\theta-\theta_0 \Vert\\
&&+\ds \Vert m_3(\hat V_{p\ell},Z_\ell;\theta_0)-m_3(V_{p\ell},Z_\ell;\theta_0) \Vert+\ds \Vert m_3(V_{p\ell},Z_\ell;\theta_0) \Vert\\
&\leq&\{{\rm{E}}[K_5(V,Z)]+o_{as}(1)\}+\{{\rm{E}}[K_4(Z)]+o_{as}(1)\}o_{as}(1)+ \ds K_3(Z_\ell) \vert\hat V_{p\ell}-V_{p\ell} \vert\\
&&+\ds \sup_{\theta\in\Theta} \Vert m_3(V_{p\ell},Z_\ell;\theta) \Vert\\
&\leq& \{{\rm{E}}[K_5(V,Z)]+o_{as}(1)\}+\{{\rm{E}}[K_4(Z)]+o_{as}(1)\}o_{as}(1)
+ \{{\rm{E}}[K_3(Z)]+o_{as}(1)\}\sup_{p\ell} \vert \hat V_{p\ell}-V_{p\ell}\vert\\
&&+\{{\rm{E}}[K_5(V,Z)]+o_{as}(1)\} =2\{{\rm{E}}[K_5(V,Z)]+o_{as}(1)\} <\infty
\end{eqnarray*}
where we use assumption A6-(ii),(iv),(v) and the two results stated at the beginning of this Appendix.
Therefore the second factor in the last line of (\ref{eq:a7}) converges to a finite limit, and
since the first factor is $o_{as}(1)$ the desired result follows.

\section{Step 3\label{step3}} The final step is to prove $\sqrt{L}(S_L(\theta_0)-\hat
S_L(\theta_0))=O_p(1)$.
Since this step is the longest and the most tedious, to facilitate reading we divide this step further into two sub-steps: Step 3.1 and Step 3.2, and before we provide the formal proof we give a detailed description of all the steps involved. 

Let $B=\sqrt{L}[S_L(\theta_0)-\hat S_L(\theta_0)]= B_1+B_2$.
In Step 3.1 we show that $B_1=O_p(1)+o_{as}(1)$ and in Step 3.2 we show $B_2=o_{as}(1)$, see Equation (\ref{B_2}).
Of these two, Step 3.1 is more involved, but we can break down the proof into following steps:

\setcounter{equation}{0}
\renewcommand{\theequation}{B-\arabic{equation}}

\begin{eqnarray*}
B_1&\leq& B_{11}+B_{12} \quad(\textrm{See Equation (\ref{B11+B12})}) \\
&=&B_{111}+B_{112}+B_{12}\quad(\textrm{See Equation (\ref{B11})})  \\
&\leq& C D +B_{112}+B_{12} \quad (\because C < \infty, D=o_p(1))  \\
&\leq& o(1)+B_{1121}+B_{1122}+B_{12} \quad (\textrm{See Equation (\ref{B112}) and $B_{1121}=o_p(1)$}) \\
&\leq& o(1)+o_p(1)+A+B+B_{12}  \quad (\textrm{See Equation (\ref{B1122}) and $B_{1122}=A+B$})\\
&\leq& o(1)+o_p(1)+A_1-A_2+B+B_{12}\quad\!\!(\textrm{See Equation (\ref{A1_A2}), $A_{1} \leq A_{11}+ A_{12}$ and $ A_{11}$ and $A_{12}$ are $o_{as}(1/\sqrt{L})$})   \\
&\leq& o(1)+o_p(1)+o_{as}(1/\sqrt{L})-A_2+B+B_{12}\quad\!\!(\because\textrm{$A_{2} \leq A_{21} + A_{22}$ and $ A_{21}$ and $A_{22}$ are $o_{as}(1/\sqrt{L})$})\\
&\leq& o(1)+o_p(1)+o_{as}(1/\sqrt{L})-o_{as}(1/\sqrt{L})+B+B_{12} \quad (\because B_2=B_{121}\times B_{122})  \\
&\leq& o(1)+o_p(1)+o_{as}(1/\sqrt{L})-o_{as}(1/\sqrt{L})+B+B_{121}+B_{122}\quad\!\!(\because B_{121} < \infty,  B_{122}=o(1) \Rightarrow B_{12}=o(1)) \\
&\leq& o(1)+o_p(1)+o_{as}(1/\sqrt{L})-o_{as}(1/\sqrt{L})+B+o(1)\quad(\because B=O_p(1)) \\
&=& o(1)+o_p(1)+o_{as}(1/\sqrt{L})-o_{as}(1/\sqrt{L})+O_p(1)+o(1)\quad (\textrm{See equation (\ref{B1122})}).
\end{eqnarray*}

%where,
%
%\begin{description}
%\item[(B-1)]  See equation (\ref{B11+B12})
%
%\item[(B-2)]  See equation (\ref{B11})
%
%\item[(B-3)] $C < \infty$ and $D=o_p(1)$
%
%
%\item[(B-4)]  See (\ref{B112}). We prove $B_{1121}=o_p(1)$
%
%\item[(B-5)]  See (\ref{B1122}). We write $B_{1122}=A+B$
%
%\item[(B-6)] See (\ref{A1_A2}). We write $A_{1} \leq A_{11}+ A_{12}\vert$. We show $ A_{11}  =o_{as}(1/\sqrt{L})$ and similar argument is applied to $A_{12}$.
%
%\item[(B-7)]   We write $A_{2} \leq A_{21} + A_{22}$. We show $ A_{21} =o_{as}(1/\sqrt{L})$ and similar argument is applied to $A_{22}$.
%
%\item[(B-8)] We write $B_2=B_{121} B_{122}$
%
%
%\item[(B-9)] We show $B_{121} < \infty$ and $B_{122}=o(1)$. Hence $B_{12}=o(1)$.
%
%\item[(B-10)] We show $B=O_p(1)$.
%
%\item[(B-11)]  See equation (\ref{B1122}).
% 
% 
%\end{description}
%

We formalize the proof below.
First we prove that the term $\sqrt{L}(S_L(\theta_0)-\hat S_L(\theta_0))$ is
\begin{eqnarray*}
B=\sqrt{L}(S_L(\theta_0)-\hat S_L(\theta_0))&=&\sqrt{L}\Bigg(\ds \mo-\ds\mhato \Bigg)\\
&=&\sqrt{L}\ds \Big[\mo - \mhato \Big]=O_p(1)+o_{as}(1).
\end{eqnarray*}
The above expression can be rewritten as
\begin{eqnarray}
B&=&-\sqrt{L}\ds\left[\moneo (\hat{V}_{p\ell}- V_{p\ell}) \right]+\sqrt{L}\ds \left[ \moneo-\moneob\right](\hat{V}_{p\ell}- V_{p\ell})\nonumber\\
&=&B_1+B_2,
\label{B_step3}
\end{eqnarray}

\noindent where the second equality comes from a Taylor expansion of
order one and the following
\begin{eqnarray*}
\mo-\mhato&=&\moneob ({V}_{p\ell}-\hat V_{p\ell})\\
&=&\moneo (\hat{V}_{p\ell}-V_{p\ell})+\moneob({V}_{p\ell}-\hat V_{p\ell})-\moneo(\hat{V}_{p\ell}-V_{p\ell})\\
&=&-\moneo(\hat{V}_{p\ell}-V_{p\ell})+[\moneo-\moneob](\hat{V}_{p\ell}-V_{p\ell}).
\end{eqnarray*}
\subsection*{Step 3.1} We consider $B_1$ in (\ref{B_step3}) and moreover we observe that for each $I$ we can write
\begin{eqnarray}
\Vert B_1\Vert&=&\left\Vert\sqrt{L}\dsnew m_1(V_{p\ell},X_\ell,I;\theta_0) (\hat V_{p\ell}-V_{p\ell})\right\Vert\nonumber\\
&=&\Bigg \Vert\sqrt{L}\dsnew m_1(V_{p\ell},X_\ell,I;\theta_0) \frac{1}{I-1}\left[\frac{\Gtildei}{\gtildei}-\frac{\Gi}{\gi}\right]\Bigg\Vert\nonumber\\
&=&\Bigg \Vert\sqrt{L}\dsinew m_1(V_{p\ell},X_\ell,I;\theta_0)\Bigg\{\frac{\Gtildei}{\gi} -\frac{\Gi}{g^2_0(B_{p\ell|X_\ell,I})}\nonumber\\
&&\gtildei+\frac{\Gi}{\gi}\frac{1}{\gtildei \gi} \Big[\gtildei-\gi\Big]^2\nonumber\\
&&-\frac{1}{\gtildei \gi}\left[\Gtildei-\Gi\right]\left[\gtildei-\gi\right]\Bigg\}\Bigg\Vert\nonumber\\
&\leq& \Bigg\Vert \sqrt{L}\dsinew m_1(V_{p\ell},X_\ell,I;\theta_0) \Bigg[\frac{\Gtildei}{\gtildei}-\frac{\Gi} {g^2_0(B_{p\ell} | X_\ell,I)}\nonumber\\
&&\gtildei\Bigg]\Bigg\Vert+\Bigg\Vert \sqrt{L}\dsinew m_1(V_{p\ell},X_\ell,I;\theta_0)\nonumber \\
&&\Bigg(\frac{\Gi}{\gi}\frac{1}{\gtildei\gi}\left[\gtildei-\gi\right]^2\nonumber\\
&&-\frac{1}{\gtildei \gi}\left[\Gtildei-\Gi\right]\nonumber\\
&&\Big[\gtildei-\gi\Big]\Bigg)\Bigg\Vert= \Vert B_{11}\Vert+\Vert B_{12}\Vert
\label{B11+B12}
\end{eqnarray}

\noindent where the third line uses the following identity:
$$\displaystyle{\frac{\tilde{a}}{\tilde{b}}-\frac{a}{b}=
\frac{\tilde{a}-\frac{a}{b}\tilde{b}}{b}
+\frac{a}{b}\frac{1}{\tilde{b}b}[\tilde{b}-b]^2-\frac{1}{\tilde{b}b}[\tilde{a}-a][\tilde{b}-b]}.$$

\medskip

The term $B_{11}$ can be written as
\begin{eqnarray}
B_{11}&=&\sqrt{L}\dsinew m_1(V_{p\ell},X_\ell,I;\theta_0)\left[\frac{\Gtildei-\frac{\Gi}{\gi}\gtildei}{\gi}\right]\nonumber\\
&=&\sqrt{L}(R_L+\frac{L(L-1)}{L^2}U_L) =B_{111}+B_{112},
\label{B11}
\end{eqnarray}
where
\begin{eqnarray*}
R_L&=&\dslsnew \frac{m_1(V_{p\ell},X_\ell,I;\theta_0)}{\gi}\Bigg[\omega_{I,R+1,j}^G K_{G,h_{G}}(0) \Unit(B_{p\ell}\leq B_{p\ell})\\
&&-\frac{\Gi}{\gi}\omega_{I,R,j}^g K_{1g,h_{g}}(0) K_{2g,h_{g}}(0)\Bigg],\\
U_L&=&\tslminus  \frac{m_1(V_{p\ell},X_\ell,I;\theta_0)}{\gi}\\
&&\Bigg[\omega_{I,R+1,j}^G K_{G,h_{G}}(X_j-X_\ell)\Unit(B_{qj}\leq B_{p\ell})-\frac{\Gi}{\gi}\omega_{I,R,j}^g K_{1g,h_{g}}(X_{j}-X_{\ell}) K_{2g,h_{g}}(B_{qj}-B_{p\ell})\Bigg].
\end{eqnarray*}
To see how to obtain the last line in (\ref{B11}), we observe that the
term within brackets in the first line of (\ref{B11}) can be expressed
as
\begin{eqnarray}
&&\frac{\Gtildei-\frac{\Gi}{\gi}\gtildei}{\gi}=\frac{1}{\gi}\left[\Gtildei-\frac{\Gi}{\gi}\gtildei \right]\nonumber\\
&=&\frac{1}{\gi}\nonumber\\
&&\Bigg[\dskGj\nonumber\\
&&-\frac{\Gi}{\gi}\dskgj\Bigg]\nonumber\\
&=&\frac{1}{\gi}\Bigg[\dsj \ikGqj-\frac{\Gi}{\gi}\nonumber\\
&&\dsj \kgqj\Bigg]
\label{A.11}
\end{eqnarray}

\noindent where we have used the following notations:
\begin{eqnarray}
K_{G,h_{G}}(X_j-X_\ell)&=&\frac{1}{h_{G}}K_G\left(\frac{X_j-X_\ell}{h_G}\right), \label{A.12}\\
K_{1g,h_g}(X_j-X_\ell)&=&\frac{1}{h_g}K_{1g}\left(\frac{X_j-X_\ell}{h_g}\right),\\
K_{2g,h_g}(B_{qj}-B_{p\ell})&=&\frac{1}{h_g}K_{2g}\left(\frac{B_{qj}-B_{p\ell}}{h_g}\right),\\
\omega^G_{I,R+1,j}&=&e_1^T \left(\frac{X_{I,R+1}^T W_x^G X_{I,R+1}}{n_I}\right)^{-1} X_{R+1,j},\\
\omega^g_{I,R,j}&=&e_1^T \left(\frac{X_{I,R}^T W_x^g X_{I,R}}{n_I}\right)^{-1} X_{R,j}.
\label{A.16}
\end{eqnarray}
Now using (\ref{A.11}) in the first line of (\ref{B11}), we get
\begin{eqnarray}
B_{11}&=&\sqrt{L}\Bigg(\ts\frac{m_1(V_{p\ell},X_\ell,I;\theta_0)}{\gi}\Bigg[\omega_{I,R+1,j}^G K_{G,h_G}(X_j-X_\ell)\Unit(B_{qj}\leq B_{p\ell})\nonumber\\
&& -\frac{\Gi}{\gi}\kgqj\Bigg]\Bigg).
\label{B11}
\end{eqnarray}
The term between parenthesis in (\ref{B11}) can be decomposed as follows:\\
1) Diagonal terms $(\ell=j, p=q)$
\begin{eqnarray}
R_L&=&\dsls \frac{m_1(V_{p\ell},X_\ell,I;\theta_0)}{\gi}\Bigg[\omega_{I,R+1,j}^G K_{G,h_{G}}(0) \Unit(B_{p\ell}\leq B_{p\ell})\nonumber\\
&&-\frac{\Gi}{\gi}\omega_{I,R,j}^g K_{1g,h_{g}}(0) K_{2g,h_{g}}(0)\Bigg],
\label{RL}
\end{eqnarray}
2) Off-diagonal terms $(\ell\neq j)$
\begin{eqnarray}
\frac{L(L-1)}{L^2}U_L&\!\!\!\!=\!\!\!\!&\tsd \frac{m_1(V_{p\ell},X_\ell,I;\theta_0)}{\gi}\nonumber\\
&&\Bigg[\omega_{I,R+1,j}^G K_{G,h_{G}}(X_j-X_\ell)\Unit(B_{qj}\leq B_{p\ell})\nonumber\\
&&-\frac{\Gi}{\gi}\omega_{I,R,j}^g  K_{1g,h_{g}}(X_j-X_\ell)  K_{2g,h_{g}}(B_{qj}-B_{p\ell})\Bigg].
\label{UL}
\end{eqnarray}
From (\ref{RL}) and (\ref{UL}) we have the expression in the last line of
(\ref{B11}). It remains to show that $B_{11}=B_{111}+B_{112}=o_{as}(1)$.
We consider first $B_{111}=\sqrt{L}R_L$ in (\ref{B11}). Specifically,
\begin{eqnarray*}
B_{111}&=&\sqrt{L}\Vert R_L\Vert=\sqrt{L}\Bigg\Vert\Bigg(\dsls \frac{m_1(V_{p\ell},X_\ell,I;\theta_0)}{\gi}\Bigg[\omega_{I,R+1,j}^G K_{G,h_{G}}(0)\\
&& -\frac{\Gi}{\gi}\omega_{I,R,j}^g K_{1g,h_{g}}(0) K_{2g,h_{g}}(0)\Bigg]\Bigg)\Bigg\Vert\\
&=&\sqrt{L}\Bigg\Vert\dsils  \frac{m_1(V_{p\ell},X_\ell,I;\theta_0)}{\gi}\Bigg[\omega_{I,R+1,j}^G \frac{ K_{G,h_{G}}(0)}{L}-\frac{\Gi}{\gi}\\
&& \omega_{I,R,j}^g \frac{ K_{1g,h_{g}}(0) K_{2g,h_{g}}(0) }{L}\Bigg]\Bigg\Vert\\
&\leq&\left(\dsils \Vert m_1(V_{p\ell},X_\ell,I;\theta_0) \Vert^2 \right)^{\frac{1}{2}}\\
&&\sqrt{L}\Bigg(\dsils \frac{1}{\gi^2}\Bigg[\omega_{I,R+1,j}^G \frac{ K_{G,h_{G}}(0)}{L}- \frac{\Gi}{\gi}\\
&&\omega_{I,R,j}^g \frac{ K_{1g,h_{g}}(0) K_{2g,h_{g}}(0)}{L}\Bigg]^2\Bigg)^{\frac{1}{2}}\\
&=&C D,
\end{eqnarray*}

\noindent where the inequality comes from Cauchy-Schwartz. First we
show that $C^2<\infty$. Using A6-(vi), $0<(1/(I-1))<1$ for each $I \in {\cal I} $ and $L/n_I=L/(I L_I)<\infty$ we get
\begin{eqnarray*}
C^2&=&\dsils \Vert m_1(V_{p\ell},X_\ell,I;\theta_0) \Vert^2
\leq\frac{L}{n_I}\dsnew \frac{1}{I(I-1)} \sup_{\theta \in \Theta}\Vert m_1(V_{p\ell},X_\ell,I;\theta)\Vert^2\\
&<&\dsnew K_7(V_{p\ell},X_\ell,I)^2={\rm E}[K_7(V,X,I)^2]+o_{as}(1)<\infty.
\end{eqnarray*}

It remains to consider the D term above. Namely,
\begin{eqnarray*}
D&\leq&\sqrt{L} \Bigg(\dsils \frac{1}{\gi^2}\Bigg[\omega_{I,R+1,j}^G \frac{K_G(0)}{Lh_{G}}-\frac{\Gi}{\gi}\\
&& \omega_{I,R,j}^g
\frac{K_{1g,h_{g}}(0) K_{2g,h_{g}}(0)}{Lh_{g}^2}\Bigg]^2\Bigg)^{\frac{1}{2}}\\
&=&\sqrt{L}\Bigg(\dsils\frac{1}{\gi^2}\Bigg[O_p(1) O_p\left(\frac{1}{Lh_G}\right)-\frac{\Gi}{\gi} \\
&&O_p(1) O_p\left(\frac{1}{Lh_{g}^2}\right)\Bigg]^2
\Bigg)^{\frac{1}{2}}
<\sqrt{L}\kappa_1\left[O_p\left(\frac{1}{Lh_{G}}\right)- \kappa_2 O_p\left(\frac{1}{Lh_{g}^2}\right)\right]\\
&=&\kappa_1 \left[O_p\left(\frac{1}{\sqrt{L}h_{G}}\right)- \kappa_2 O_p\left(\frac{1}{\sqrt{L}h_{g}^2}\right) \right]=\kappa_1[o_p(1)-\kappa_2 o_p(1)]
=o_p(1),
\end{eqnarray*}

\noindent where after the first equality we use (\ref{A.12})- (\ref{A.16}). The second line follows from observing that
\begin{eqnarray*}
\omega_{I,R+1,j}^G&=&e_1^T \left(\frac{X_{I,R+1}^T W_x^G X_{I,R+1}}{n_I}\right)^{-1} X_{R+1,j}=e_1^T\left[\frac{1}{n_I h_G}\sum_{\iota=1}^{n_I} {\mathbf{x}}_\iota^T {\mathbf{x}}_\iota K_G\left(\frac{x_\iota-x_j}{h_g}\right) \right]^{-1} e_1=O_p(1),
\end{eqnarray*}
\noindent and similarly for $\omega_{I,R,j}^g$.
The third line uses the fact that densities are bounded away from
zero and $0<(1/I(I-1))<1$ for all $I$. The last line follows from Assumption A4.AN. Thus, $B_{111}=C D=o(1)$ as
desired.
Let $Y_{p\ell}=(B_{p\ell},X_\ell)$ and for each $I$ define
$r_L(Y_{p\ell},I)=E[\pl|(Y_{p\ell},I)]$, where
$p_L(\cdot,\cdot)$ is a symmetric function, and 
$$\theta_L=E[r_L(Y_{p\ell},I)]=E[\pl]; \quad\!\!\hat U_L=\theta_L+\dstwo [r_L(Y_{p\ell},I)-\theta_L].$$
Next, we consider $B_{112}$ in (\ref{B11})
\begin{eqnarray}
B_{112}&=&\frac{L(L-1)}{L^2}\sqrt{L}U_L =
\frac{L(L-1)}{L^2}\sqrt{L}(U_L-\hat U_L)+\frac{L(L-1)}{L^2}\sqrt{L}\hat U_L =B_{1121}+B_{1122},
\label{B112}
\end{eqnarray}
where $U_L$ can be written as a U-statistic. 

Namely,
\begin{eqnarray*}
U_L&=&\tslminus \frac{m_1(V_{p\ell},X_\ell,I;\theta_0)}{\gi} \\
&&\Bigg[\omega_{I,R+1,j}^G K_{G,h_G}(X_j-X_\ell)\Unit(B_{qj}\leq B_{p\ell})\\
&&-\frac{\Gi}{\gi}\omega_{I,R,j}^g K_{1g,h_g}(X_j-X_\ell)K_{2g,h_g}(B_{qj}-B_{p\ell})\Bigg]\\
&=&\tslminusone\Bigg\{\imfl \frac{m_1(V_{p\ell},X_\ell,I;\theta_0)}{\gi}\\
&&\Bigg[\omega_{I,R+1,j}^G K_{G,h_G}(X_j-X_\ell)\Unit(B_{qj}\leq B_{p\ell})\\
&&-\frac{\Gi}{\gi}\omega_{I,R,j}^g K_{1g,h_g}(X_j-X_\ell)K_{2g,h_g}(B_{qj}-B_{p\ell})\Bigg]\Bigg\}\\
&=&\tslminusonetwo\Bigg[\frac{m_1(V_{p\ell},X_\ell,I;\theta_0) K^{**}(B_{p\ell},B_{qj},X_\ell,X_j,I)}{2}\\
&&+\frac{m_1(V_{qj},X_j,I;\theta_0) K^{**}(B_{qj},B_{p\ell},X_j,X_\ell,I)}{2}\Bigg]\\
&=&\left(\begin{array}{c}L\\2\end{array}\right)^{-1}\tsnd
p_L\left((B_{p\ell},X_\ell,I),(B_{qj},X_j,I)\right).
\end{eqnarray*}
Now we prove $B_{1121}=\sqrt{L}(U_L-\hat U_L)=o_p(1)$. By Lemma 3.1 in
\cite{PowellStockStoker1989} it is enough to show that ${\rm E}[\Vert
p_L((Y_{p\ell},I),(Y_{qj},I))\Vert^2]=o(L)$.
We will show that ${\rm E}[\Vert
p_L((Y_{p\ell},I),(Y_{qj},I))\Vert^2|I]=o(L)$, which implies the above condition.

{\footnotesize\begin{eqnarray}
\lefteqn{{\rm E}[\Vert \pl \Vert^2|I]=\int \Vert \pl \Vert^2 \gz \gzj \dypl \dypj} \nonumber\\
&=&\frac{1}{4}\int \Bigg\Vert \frac{L}{n_I(I-1)}\izmone \Bigg[\hGd \kG \nonumber\\
&&- \ratioGg \hgdone \kgz \Bigg]\nonumber\\
&&+\frac{L}{n_I(II-1)}\izjmone\Bigg [\hGd \kGj\nonumber\\
&&- \ratioGgj \hgdone \kgzj\Bigg] \Bigg \Vert^2 \nonumber\\
&&\gz \gzj \dypl \dypj\nonumber\\
&=&\frac{1}{4}\int \Bigg \Vert \Bigg[\frac{L}{n_II(I-1)} \izmone \hGd \kG\nonumber\\
&&+\frac{L}{n_I(I-1)}\izjmone \hGd \kGj \Bigg]\nonumber\\
&&+ \Bigg[\frac{L}{n_I(I-1)}\izmone \ratioGg \hgdone \omega_{I,R,j}^g K_{1g}\left(\frac{X_j-X_\ell}{h_{g}}\right) \nonumber\\ &&K_{2g}\left(\frac{B_{qj}-B_{p\ell}}{h_{g}}\right)+\frac{L}{n_I(I-1)}\izjmone \ratioGgj \hgdone \omega_{I,R,j}^g K_{1g}\left(\frac{X_\ell-X_j}{h_{g}}\right) \nonumber\\
&&K_{2g}\left(\frac{B_{p\ell}-B_{qj}}{h_{g}}\right) \Bigg]\Bigg \Vert^2\gz \gzj \dypl \dypj\nonumber\\
&=&\frac{1}{4}\int \Bigg \Vert \hGd \frac{L}{n_I}\Bigg[\frac{1}{(I-1)}\izmone  \kG \nonumber\\
&&+ \frac{1}{(I-1)}\izjmone  \kGj\Bigg]\nonumber\\
&&+ \hgdone \frac{L}{n_I}\Bigg [\frac{1}{(I-1)}\izmone \ratioGg \kgz\nonumber\\
&&+\frac{1}{I-1}\izjmone \ratioGgj \kgzj \Bigg]\Bigg \Vert^2\nonumber\\
&&\gz \gzj \dypl \dypj\nonumber\\
&\leq&\frac{1}{2}\int \Bigg\{\Bigg\Vert \frac{L}{n_I}\hGd \Bigg [\frac{1}{(I-1)}\izmone \kG\nonumber\\
&&+\frac{1}{(I-1)}\izjmone \kGj\Bigg]\Bigg\Vert^2\nonumber\\
&&\!\!+\!\Bigg\Vert\frac{L}{n_II}\hgdone \Bigg[\frac{1}{(I-1)}\izmone \ratioGg \kgz\nonumber\\
&&+\frac{1}{(I-1)}\izjmone \ratioGgj\kgzj \Bigg] \Bigg\Vert^2\Bigg\}\nonumber\\
&&\gz \gzj \dypl \dypj\nonumber\\
&=&C+D,
\label{A.21}
\end{eqnarray}}
where the inequality comes from using $(a+b)^2\leq
2(a^2+b^2)$. Therefore, now we need to show that both C and D are $o(L)$.
We consider first the C term in (\ref{A.21}), and note that we can write
$V_{p\ell}=\xi(B_{p\ell},X_\ell,I)$. It gives

\begin{eqnarray*}
C\!\!\!\!&=&\!\!\!\hGtwod \int \left(\frac{L}{n_I}\frac{1}{(I-1)}\right)^2\Bigg\Vert \izmonexi \kG \\
&&+\izjmone \kGj \Bigg\Vert^2 \gz \gzj \dypl \dypj\\
&=&\frac{1}{2} \int \left(\frac{L}{n_I}\frac{1}{(I-1)}\right)^2\Bigg\Vert \imoneratioKGu \omega_{I,R+1,j}^G K_G(-u_2)\\
&&\Unit(B_{qj}\leq u_1 h_G+B_{qj})+\ratioigj \omega_{I,R+1,j}^G K_G(u_2)\Unit(u_1 h_G+B_{qj}\leq B_{qj} )\Bigg \Vert^2 \\
&&\guhG \gzj du \dypj\\
&\leq& \int \Bigg[\Bigg\Vert \imoneratioKGu \omega_{I,R+1,j}^G K_G(-u_2)\Unit(B_{qj}\leq u_1 h_G+B_{qj})\Bigg\Vert^2 \\
&&+\left\Vert  \ratioigj \omega_{I,R+1,j}^G K_G(u_2)\Unit(u_1 h_G+B_{qj}\leq B_{qj} ) \right\Vert^2\Bigg]\\
&&\guhG \gzj du\dypj\\
&=&C_1+C_2,
\end{eqnarray*}

\noindent where we have used the change of variable
$u=\displaystyle{\frac{Y_{p\ell}-Y_{qj}}{h_{G}}=\left(\frac{B_{p\ell}-B_{qj}}{h_G},\frac{X_\ell-X_j}{h_G}\right)=(u_1,u_2)}$,
and the inequality comes from using $(a+b)^2\leq 2(a^2+b^2)$ and $\left(\frac{L}{n_I}\frac{1}{(I-1)}\right)^2<\infty$.

Next we consider $C_1$
\begin{eqnarray*}
C_1&=& \int \Bigg\Vert \imoneratioKGu \omega_{I,R+1,j}^G K_G(-u_2)\Unit(B_{qj}\leq u_1 h_G+B_{qj})\Bigg\Vert^2 \\
&&\guhG \gzj du \dypj\\
&\!\!\!=&\!\!\! \!\!\int \left\Vert \imoneratioKGu \right\Vert^2 \left\Vert \omega_{I,R+1,j}^G K_G(-u_2)\Unit(B_{qj}\leq u_1 h_G+B_{qj})\right\Vert^2\\
&&\guhG \gzj du \dypj\\
&\!\!\!=& \!\!\! \!\!\int \left\Vert \moneu\right\Vert^2 \left\Vert \omega_{I,R+1,j}^G K_G(-u_2)\Unit(B_{qj}\leq u_1 h_G+B_{qj})\right\Vert^2\\
&& \frac{\guhG g(Y_{qj}|I)}{g_0(u_1h_{G}+B_{qj}|u_2h_{G}+X_j,I)^2} du  \dypj\\
&\leq& \int \Vert\moneu\Vert^2 \left\Vert\omega_{I,R+1,j}^G K_G(-u_2)\Unit(B_{qj}\leq u_1 h_G+B_{qj})\right\Vert^2\\
&&\gzj du\dypj,
\end{eqnarray*}

\noindent where the last inequality comes from the assumption that densities are
bounded.
By the Lebesgue Dominated Convergence (LDC) Theorem and A6-(vi), the
above integral converges to
\begin{eqnarray*}
\int \left\Vert O_p(1)K_G(-u_2)\right\Vert^2 du \int \Vert
m_1(V_{qj},X_j,I;\theta_0)\Vert^2 \gzj\dypj <\infty.
\end{eqnarray*}
Hence, $C_1=o(L)$ as $L \rightarrow\infty$.
A similar argument can be used to show that $C_2=o(L)$ as $L \rightarrow\infty$. Therefore, $C=C_1+C_2=o(L)$.

%Next we consider $C_2$
%
%\begin{eqnarray*}
%C_2&=&\hGd \int \left\Vert \ratioigj W_G^n(u_1,0)K^*_G(u_2) \right\Vert^2\guhG \gzj du_2 du_1 \dypj\\
%&\leq& \hGd \int \Vert m_1(V_{qj},Z_j;\theta_0)\Vert^2
%\left\Vert W_G^n(u_1,0)K^*_G(u_2)\right\Vert^2 \guhG du_2 du_1 \dypj
%\end{eqnarray*}
%
%\noindent where the last inequality comes from the assumption that densities are
%bounded.
%
%
%By the LDC Theorem and A6-(vi), the above integral converges to
%
%
%\begin{eqnarray*}
%\int \left\Vert W_G^n(u_1,0)K^*_G(u_2)\right\Vert^2 du_2 du_1 \int \Vert
%m_1(V_{qj},Z_j;\theta_0)\Vert^2  \gzj \dypj <\infty.
%\end{eqnarray*}
%
%
%Hence, $C_2=o(L)$ if and only if $Lh_{1G}^d h_{2G}=L h_G^{d+1}\rightarrow\infty$ as
%implied by A4.AN and shown above.

\medskip

Next we consider the D term in (\ref{A.21}). Namely
\begin{eqnarray*}
D&=&\frac{1}{2}\int \Bigg\Vert\frac{L}{n_I}\hgdone \Bigg[\frac{1}{(I-1)}\izmone \ratioGg \omega_{I,R,j}^g K_{1g}\left(\frac{X_j-X_\ell}{h_{g}}\right)\\
&&K_{2g}\left(\frac{B_{qj}-B_{p\ell}}{h_{g}}\right)+\frac{1}{(I-1)}\izjmone \ratioGgj\omega_{I,R,j}^g \\
&&K_{1g}\left(\frac{X_\ell-X_j}{h_{g}}\right) K_{2g}\left(\frac{B_{p\ell}-B_{qj}}{h_{g}}\right) \Bigg] \Bigg\Vert^2\gz \gzj \dypl \dypj\\
&=&\hgtwod \left(\frac{L}{n_I}\frac{1}{(I-1)}\right)^2 \int \Bigg\Vert \izmonexi \ratioGg \omega_{I,R,j}^g \\
&&K_{1g}\left(\frac{X_j-X_\ell}{h_{g}}\right)K_{2g}\left(\frac{B_{qj}-B_{p\ell}}{h_{g}}\right)\\
&&+\izjmone \ratioGgj\kgzj  \Bigg\Vert^2\\
&&\gz \gzj \dypl \dypj\\
&\leq&\hgdonehalf \int \Bigg\Vert \imoneratioKGu  \ratioGgju\\
&&\omega_{I,R,j}^g K_{1g}(-u_2) K_{2g}(-u_1)\\
&&\!\!\!\!\! +\izjmone \ratioGgj\omega_{I,R,j}^g K_{1g}(u_2) K_{2g}(u_1) \Bigg\Vert^2\!\! \gju \gzj du \dypj\\
&\leq& \hgdone \int \Bigg[\Bigg\Vert \imoneratioKGu \ratioGgju\omega_{I,R,j}^g \\
&&K_{1g}(-u_2) K_{2g}(-u_1)\Bigg\Vert^2+\left\Vert \izjmone \ratioGgj \omega_{I,R,j}^g K_{1g}(u_2) K_{2g}(u_1)\right\Vert^2\Bigg]\\
&&\gju \gzj du \dypj\\
&=&D_1+D_2,
\end{eqnarray*}

\noindent where we have used the change of variable
$u=\displaystyle{\frac{Y_{p\ell}-Y_{qj}}{h_{g}}=\left(\frac{B_{p\ell}-B_{qj}}{h_g},\frac{X_\ell-X_j}{h_g}\right)=(u_1,u_2)}$,
and the first inequality follows from the fact that $\left(\frac{L}{n_I}\frac{1}{(I-1)}\right)^2<\infty$ and the second inequality uses $(a+b)^2\leq 2 (a^2+b^2)$.

We consider first $D_1$. Specifically,
\begin{eqnarray*}
D_1&=&\hgdone \int \Big\Vert \izjmoneu \Gzju \Big\Vert^2  \\
&&\Big\Vert\omega_{I,R,j}^g K_{1g}(-u_2) K_{2g}(-u_1) \Big\Vert^2\frac{\gju g_0(Y_{qj}|I)}{g_0(u_1h_{g}+B_{qj}|u_2h_{g}+X_j,I)^4} du \dypj\\
&\leq &\hgdone \int \Vert \izjmoneu \Vert^2 \left\Vert \omega_{I,R,j}^g K_{1g}(-u_2) K_{2g}(-u_1)\right\Vert^2 \\
&&\gzj du \dypj,
\end{eqnarray*}
\noindent where the inequality uses the fact that
$G(\cdot|\cdot,I)$ is bounded and that densities are
bounded from above.
By the LDC Theorem and A6-(vi) the above integral converges to
\begin{eqnarray*}
\int \left\Vert O_p(1) K_{1g}(-u_2) K_{2g}(-u_1)\right\Vert^2 du \int \left\Vert
m_1(V_{qj},X_j,I;\theta_0) \right\Vert^2 \gzj \dypj<\infty.
\end{eqnarray*}
Hence, $D_1=o(L)$ if and only if $Lh_{g}^2 \rightarrow\infty$ as
implied by A4.AN-(ii) since,
$
Lh_{g}^2=\sqrt{L}\sqrt{L} h_{g}^{2}\rightarrow\infty.
$
%Next we consider the $D_2$ term. Namely,
%
%\begin{eqnarray*}
%D_2&=&\hgdone \int \left\Vert m_1(V_{qj},Z_j;\theta_0)\Gzj W_g^n(u_1,0)K^*_g(u_2) \right\Vert^2 \\
%&&\frac{\gju g_0(Y_{qj}|I_j)}{g_0(Y_{qj}|I_j)^4} du_2 du_1 \dypj\\
%&\leq&\hgdone \int \Vert m_1(V_{qj},Z_j;\theta_0)\Gzj \Vert^2\left\Vert W_g^n(u_1,0)K^*_g(u_2) \right\Vert^2\\
%&&\frac{\gju}{g(Y_{qj}I_j)^3} du_2 du_1  \dypj\\
%&\leq&\hgdone \int \Vert
%m_1(V_{qj},Z_j;\theta_0)\Vert^2\left\Vert W_g^n(u_1,0)K^*_g(u_2) \right\Vert^2
%\gju du_2 du_1\dypj.
%\end{eqnarray*}
%
%
%The above integral converges to
%
%\begin{eqnarray*}
%\int \left\Vert W_g^n(u_1,0)K^*_g(u_2)\right\Vert^2 du_2 du_1 \int \Vert
%m_1(V_{qj},Z_j;\theta_0) \Vert^2\gzj \dypj <\infty.
%\end{eqnarray*}
A similar argument can be used to show that $D_2=o(L)$. That is, $D_2=o(L)$ if and only if $Lh_{g}^2  \rightarrow\infty$, as
implied by A4.AN-(ii).
%Hence, $D_2=o(L)$ if and only if $Lh_{1g}^d h_{2g}=L h_g^{d+1} \rightarrow\infty$, as
%implied by A4.AN and shown above.
Therefore,
$C+D=C_1+C_2+D_1+D_2=o(L)$ and the desired result follows, i.e by
Lemma 3.1 in Powell, Stock and Stoker (1989) $\sqrt{L}(U_L-\hat
U_L)=o_p(1)$.

\bigskip

Next we consider the second term in (\ref{B112})
\begin{eqnarray}
B_{1122}&=&\frac{L(L-1)}{L^2} \sqrt{L}\hat U_L =\frac{L(L-1)}{L^2} \sqrt{L}\left\{\theta_L+\dstwo [r_L(Y_{p\ell},I)-\theta_L]\right\} \notag\\
&=&\frac{L(L-1)}{L^2}\sqrt{L}E[\pl] +\frac{L(L-1)}{L^2}\sqrt{L}\!\!\dstwo [r_L(Y_{p\ell},I)-\theta_L].
\label{B1122}
\end{eqnarray}
Next, we show that the first term in (\ref{B1122}) is $o_{as}(1)$.
%By the Central Limit Theorem (CLT), the second term above is $O_p(1)$ (See Lemma \ref{Lemma A1} below). 
%Next step is to show that the first term in (\ref{B1122}) is $o_{as}(1)$. We consider the expectation in the first term above. Namely
%Therefore, it remains to show that the first term is $o_{as}(1)$. 
Consider only the expectation part in the first term in (\ref{B1122}). Namely
\begin{eqnarray}
\lefteqn{{\rm E}[\pl]}\nonumber\\
&=&\frac{1}{2}\frac{L}{n_I}\sumj\int\Bigg\{\frac{m_1(V_{p\ell},X_\ell,I;\theta_0)}{g_0(B_{p\ell}|X_\ell,I)}
\Bigg[\omega_{I,R+1,j}^G K_{G,h_G}(X_j-X_\ell)\Unit(B_{qj}\leq B_{p\ell})\nonumber\\
&&-\frac{\Gi}{\gi}\omega_{I,R,j}^g  K_{1g,h_{g}}(X_j-X_\ell) K_{2g,h_{g}}(B_{qj}-B_{p\ell})\Bigg]\nonumber\\
&&+\frac{m_1(V_{qj},X_j,I;\theta_0)}{g_0(B_{qj}|X_j,I)}\Bigg[\omega_{I,R+1,j}^G K_{G,h_G}(X_\ell-X_j)\Unit(B_{p\ell}\leq B_{qj})\nonumber\\
&&-\frac{\Gj}{\gj} \omega_{I,R,j}^g  K_{1g,h_{g}}(X_\ell-X_j) K_{2g,h_{g}}(B_{p\ell}-B_{qj})\Bigg]\Bigg\} \nonumber\\
&&\gjoint \gjointj \dypl\dypj \nonumber\\
&=&\frac{1}{2}\frac{L}{n_I}\sumj\int \hGd \Bigg[\frac{m_1(V_{p\ell},X_\ell,I;\theta_0)}{g_0(B_{p\ell}|X_\ell,I)}\kG\nonumber\\
&&+\frac{m_1(V_{qj},X_j,I;\theta_0)}{g_0(B_{qj}|X_j,I)}\kGj\Bigg]\nonumber \\
&& \gjoint \gjointj \dypl\dypj \nonumber \\
&&-\frac{1}{2}\frac{L}{n_I}\sumj\int\hgdone \Bigg[\frac{m_1(V_{p\ell},X_\ell,I;\theta_0)}{g_0(B_{p\ell}|X_\ell,I)} \ratioGg \omega_{I,R,j}^g K_{1g}\left(\frac{X_j-X_\ell}{h_{g}}\right) \nonumber\\
&&K_{2g}\left(\frac{B_{qj}-B_{p\ell}}{h_{g}}\right)+\frac{m_1(V_{qj},X_j,I;\theta_0)}{g_0(B_{qj}|X_j,I)}\ratioGgj\omega_{I,R,j}^g K_{1g}\left(\frac{X_\ell-X_j}{h_{g}}\right) \nonumber\\
&& K_{2g}\left(\frac{B_{p\ell}-B_{qj}}{h_{g}}\right)\Bigg]\gjoint \gjointj \dypl\dypj \nonumber\\
&=&A_1-A_2.
\label{A1_A2}
\end{eqnarray}

We consider first $A_1$.
\begin{eqnarray*}
\Vert A_1\Vert&=&\Bigg\Vert\frac{1}{2}\frac{L}{n_I}\sumj\int \hGd \Bigg
[\frac{m_1(\xi(B_{p\ell},X_\ell,I),X_\ell,I;\theta_0)}{g_0(B_{p\ell}|X_\ell,I)}\omega_{I,R+1,j}^G K_G\left(\frac{X_j-X_\ell}{h_{G}}\right) \\
&& \Unit(B_{qj}\leq B_{p\ell})+\frac{m_1(V_{qj},X_j,I;\theta_0)}{g_0(B_{qj}|X_j,I)}\kGj\Bigg]\\
&&\gjoint \gjointj \dypl\dypj\Bigg\Vert\leq \Vert A_{11}\Vert+\Vert A_{12}\Vert
\end{eqnarray*}

It is enough to show that $\Vert A_{11}\Vert=o_{as}(1/\sqrt{L})$ since the
same argument can be used to show that $\Vert
A_{12}\Vert=o_{as}(1/\sqrt{L})$. We observe the following
\begin{eqnarray}
\Vert A_{11}\Vert&=&\Bigg\Vert\frac{1}{2}\frac{L}{n_I}\sumj\int \frac{1}{h_G}
\Bigg[\frac{m_1(\xi(B_{p\ell},X_\ell,I),X_\ell,I;\theta_0)}{g_0(B_{p\ell}|X_\ell,I)}\omega_{I,R+1,j}^G K_G\left(\frac{X_j-X_\ell}{h_{G}}\right)\nonumber\\
&&\Unit(B_{qj}\leq B_{p\ell})\Bigg] \gjoint \gjointj \dypl\dypj\Bigg\Vert\nonumber\\
&\leq&\Bigg\Vert\sum_{I}\int \frac{1}{h_G}
\frac{m_1(\xi(B_{p\ell},X_\ell,I),X_\ell,I;\theta_0)}{g_0(B_{p\ell}|X_\ell,I)}\kG\nonumber\\
&&\gjoint \gjointj \dypl\dypj\Bigg\Vert\nonumber\\
&\leq&\Bigg\Vert\sum_{I}\int \frac{1}{h_G}m_1(\xi(B_{p\ell},X_\ell,I),X_\ell,I;\theta_0)\kG\nonumber\\
&&\gjoint \dypl\dypj\Bigg\Vert\nonumber\\
&=&h_G \Bigg\Vert\sum_{I}\int  m_1(\xi(uh_G+Y_{qj},I),u_2h_G+X_j,I;\theta_0)\omega_{I,R+1,j}^G K_G(-u_2)\nonumber \\
&&\Unit(B_{qj}\leq u_1h_G+B_{qj}) g_0(uh_G+Y_{qj},I) du \dypj\Bigg\Vert\nonumber\\
&\leq&h_G \Bigg\Vert\sum_{I}\int  m_1(\xi(uh_G+Y_{qj},I),u_2h_G+X_j,I;\theta_0)K_G(-u_2)\nonumber \\
&&\Unit(B_{qj}\leq u_1h_G+B_{qj}) g_0(uh_G+Y_{qj},I) du \dypj\Bigg\Vert,
\label{A24}
\end{eqnarray}
\noindent where we have used the change of variable $u=\displaystyle{\frac{Y_{p\ell}-Y_{qj}}{h_{G}}=\left(\frac{B_{p\ell}-B_{qj}}{h_G},\frac{X_\ell-X_j}{h_G}\right)=(u_1,u_2)}$ and the fact that densities are bounded. The last
inequality comes from observing that $\omega_{I,R+1,j}^G=O_p(1)$.
Now consider the expectation inside the norm of the term above
evaluated at $h_G=0$, namely

\begin{eqnarray*}
\sum_{I}\iint m_1(\xi(Y_{qj},I),X_j,I;\theta_0)g_0(Y_{qj},I) \dypj
&=& \sum_{I}\int \left[\int m_1(\xi(Y_{qj},I),X_j,I;\theta_0)
g_0(B_{qj}|X_j,I) dB_{qj}\right] \\
&&f_m(X_j,I) dX_j\\
&=& \sum_{I}{\rm{E}}[m_1(V,X,I;\theta_0)],
\end{eqnarray*}

\noindent where we use A3-(ii) and the Law of Iterated Expectations. The last line in the expression above follows from observing that the integral inside can be solved by using 
integration by parts twice, as follows:
\begin{eqnarray*}
\lefteqn{\int_{\underline{B}(X_j)}^{\overline{B}(X_j,I)} m_1(\xi(B_{qj},X_j,I),X_j,I;\theta_0) g_0(B_{qj}|X_j,I) dB_{qj}}\\
&=& m_1(\xi(\overline{B}(X_j,I),X_j,I),X_j,I;\theta_0)G_0(\overline{B}(X_j,I)|X_j,I)-
m_1(\xi(\underline{B}(X_j),X_j,I),X_j,I;\theta_0) \\
&&G_0(\underline{B}(X_j)|X_j,I)-\int_{\underline{B}(X_j)}^{\overline{B}(X_j,I)} m_{11}(\xi(B_{qj},X_j,I),X_j,I;\theta_0) G_0(B_{qj}|X_j,I) dB_{qj}\\
&=& m_1(\overline{V},X_j,I;\theta_0)-\int_{\underline{B}(X_j)}^{\overline{B}(X_j,I)} m_{11}(\xi(B_{qj},X_j,I),X_j,I;\theta_0) G_0(B_{qj}|X_j,I) dB_{qj}\\
&=& m_1(\overline{V},X_j,I;\theta_0)-m_1(\xi(\overline{B}(X_j,I),X_j,I),X_j,I;\theta_0) G_0(\overline{B}(X_j,I)|X_j,I)\\
&&+m_1(\xi(\underline{B}(X_j),X_j,I),X_j;\theta_0) G_0(\underline{B}(X_j)|X_j,I)+\int_{\underline{B}(X_j)}^{\overline{B}(X_j,I)} m_{1}(\xi(B_{qj},X_j,I),X_j,I;\theta_0) g_0(B_{qj}|X_j,I) dB_{qj}\\
&=& m_1(\overline{V},X_j,I;\theta_0)-m_1(\overline{V},X_j,I;\theta_0)+\int_{\underline{B}(X_j)}^{\overline{B}(X_j,I)}
m_{1}(\xi(B_{qj},X_j,I),X_j,I;\theta_0)
g_0(B_{qj}|X_j,I) dB_{qj}\\
&=&\int_{\underline{V}(X_j,I)}^{\overline{V}(X_j,I)} m_1(V_{qj},X_j,I;\theta_0) f(V_{qj}|X_{j},I) dV_{qj}
= {\rm{E}}[m_1(V,X,I;\theta_0)|X,I],
\end{eqnarray*}

\noindent where the fifth equality uses
$G_0(B_{qj}|X_j,I)=F(\xi(B_{qj},X_j,I)|X_j,I)$, so that
$g_0(B_{qj}|X_j,I)=f(V_{qj}|X_j,I) \xi_1(B_{qj},X_j,I)$.
Therefore at $h_G=0$ the integral inside the norm in (\ref{A24}) exists by A6-(vii). Thus, we can apply a Taylor expansion of order $R+1$ in the RHS of (\ref{A24})
around $h_G$ to obtain
\begin{eqnarray*}
\Vert A_{11}\Vert&\leq& h_G \sum_{I} \left\Vert d_1 h_G+d_2\frac{h_G^2}{2}+\ldots+d_R \frac{h_G^R}{R!}+O(h_G^{R+1})\right\Vert\\
&=&\sum_{I} \left\Vert d_1 h_G^2 +d_2\frac{h_G^3}{2}+\ldots+d_R \frac{h_G^{R+1}}{R!}+O(h_G^{R+2})\right\Vert.
\end{eqnarray*}
We note that the remainder term vanishes, i.e.
$\sqrt{L}h_G^{R+2}=o(1)$, and also that
$\sqrt{L}h_G^{R+1}=o(1)$, by A4.AN-(I). The
remaining $R-1$ terms also vanish by A3-(iii), i.e, since the kernels are of order $R-1$. To see this observe
that the $k$th coordinate of $d_\rho$, $\rho=1,\ldots,R-1$ is
\begin{eqnarray*}
d_{k_\rho}&=&\frac{\partial^\rho}{\partial h_G^\rho}\int [H_k(uh_G+\overline{Y})-H_k(uh_G+\underline{Y})]K_G(-u_2)du\Big{|}_{h_G=0}\\
&=&\sum_{k_1,\ldots,k_\rho=1}^{2} \int (u_{k_1}\ldots u_{k_\rho})K_G(-u_2)
\frac{\partial^\rho}{\partial Y_{k_1}\ldots \partial Y_{k_\rho}}H_k(\overline{Y}) du\\
&&-\sum_{k_1,\ldots,k_\rho=1}^{2} \int (u_{k_1}\ldots u_{k_\rho})K_G(-u_2)
\frac{\partial^\rho}{\partial Y_{k_1}\ldots \partial Y_{k_\rho}}H_k(\underline{Y})du=0,
\end{eqnarray*}
where $dH_k/dY(y)= m_{1,k}(\xi(y,I),x,I;\theta_0)
g_0(y,I)$. The third equality uses A3-(iii), that is since
$K_G(\cdot)$ is a higher order kernel, all moments of order strictly
smaller than $R-1$ vanish.
It remains to consider now $A_2$ in (\ref{A1_A2}). Namely
\begin{eqnarray*}
\Vert A_2\Vert&=&\Bigg\Vert\frac{1}{2}\frac{L}{n_I}\sumj\int\hgdone \Bigg[\frac{m_1(V_{p\ell},X_\ell,I;\theta_0)}{g_0(B_{p\ell}|X_\ell,I)} \ratioGg \kgz\\
&&+\frac{m_1(V_{qj},X_j,I;\theta_0)}{g_0(B_{qj}|X_j,I)}\ratioGgj\kgzj\Bigg] \\
&&\gjoint \gjointj \dypl\dypj\Bigg\Vert \leq \Vert A_{21}\Vert +\Vert A_{22}\Vert.
\end{eqnarray*}
We show only that $A_{21}=o_{as}(1/\sqrt{L})$ since
a similar argument can be used to show that $A_{22}=o_{as}(1/\sqrt{L})$. 
We observe the following
\begin{eqnarray*}
\Vert A_{21}\Vert&=&\Bigg\Vert\frac{1}{2}\frac{L}{n_I}\sumj\int \frac{1}{h_g^{2}}\frac{m_1(V_{p\ell},X_\ell,I;\theta_0)}
{g_0(B_{p\ell}|X_\ell,I)}\ratioGg\\
&&\kgz\gjoint \gjointj \dypl\dypj\Bigg\Vert\\
&\leq&\Bigg\Vert\sum_{I}\int \frac{1}{h_g^{2}}\frac{m_1(V_{p\ell},X_\ell,I;\theta_0)}{g_0(B_{p\ell}|X_\ell,I)}\ratioGg \omega_{I,R,j}^g K_{1g}\left(\frac{X_j-X_\ell}{h_{g}}\right) \\
&&K_{2g}\left(\frac{B_{qj}-B_{p\ell}}{h_{g}}\right)\gjoint \gjointj \dypl\dypj\Bigg\Vert\\
&\leq&\Bigg\Vert\sum_{I}\int \frac{1}{h_g^{2}}m_1(\xi(B_{p\ell},X_\ell,I),X_\ell,I;\theta_0)\kgz\\
&&\gjoint \gjointj \dypl\dypj\Bigg\Vert\\
&=&\Bigg\Vert\sum_{I}\int m_1(\xi(uh_g+Y_{qj},I),u_2+X_j,I;\theta_0)\omega_{I,R,j}^g K_{1g,h_g}(-u_2)
K_{2g,h_g}(-u_1)\\
&&g_0(uh_g+Y_{qj},I)du dY_{qj}\Bigg\Vert\\
&\leq&\Bigg\Vert\sum_{I}\int m_1(\xi(uh_g+Y_{qj},I),u_2+X_j,I;\theta_0) K_{1g,h_g}(-u_2)
K_{2g,h_g}(-u_1)
g_0(uh_g+Y_{qj},I)du dY_{qj}\Bigg\Vert,
\end{eqnarray*}

\noindent where we have used that $(1/2)(L/n_I)1/(I-1)\leq \infty$ and also that densities are bounded.
The last equality uses the change of variable $u=(Y_{p\ell}-Y_{qj})/h_g$ and the last inequality comes from observing that $\omega_{I,R,j}^g=O_p(1)$.
We observe that $A_{21}$ can be expanded as a Taylor series of order
$R$ in the bandwidth $h_g$. Moreover, $A_{21}|_{h_g=0}<\infty$ by A6-(vii) as
we have already shown above for $A_{11}|_{h_G=0}<\infty$. %Using the Law of Iterated Expectations and
%integration by parts yields
%
%\begin{eqnarray*}
%A_{21}|_{h=0}&=&\frac{1}{2}\sumj\int m_1(V_{qj},X_j,I_j;\theta_0) G_0(Y_{qj},I_j) \dypj\\
%&=&\frac{1}{2}\sumj\int \left[\int m_1(V_{qj},X_j,I_j;\theta_0) G_0(B_{qj}|X_j,I_j)dB_{qj} \right] f_m(X_j,I_j)dX_j\\
%&=&\frac{1}{2}\sumj\int\Big[m(\xi(B_{qj},X_j,I_j),X_j,I_j;\theta_0) G_0(B_{qj}|X_j,I_j)\Big|_{\underline{B}(X_j)}^{\overline{B}(X_j,I_j)} \\
%&&-\int_{\underline{B}(X_j)}^{\overline{B}(X_j,I_j)} m(\xi(B_{qj},X_j,I_j),X_j,I_j;\theta_0) g_0(B_{qj}|X_j,I_j)dB_{qj}\Big]f_m(X_j,I_j)dX_j\\
%&=&\frac{1}{2}\sumj\int \Big[m(\overline{V},X_j,I_j;\theta_0)-\int_{\underline{V}}^{\overline{V}} m(V_{qj},X_j,I_j;\theta_0)\frac{g_0(B_{qj}|X_j,I_j)}{\xi_1{B_{qj},X_j,I_j}} dV_{qj}\Big]f_m(X_j,I_j)dX_j\\
%&=&\frac{1}{2}\sumj\int m(\overline{V},X_j,I_j;\theta_0) f_m(X_j,I_j)dX_j\\
%&=&0,
%\end{eqnarray*}
%
%\noindent where the fourth equality uses
%$G_0(B_{qj}|X_j,I_j)=F(\xi(B_{qj},X_j,I_j)|X_j,I_j)$, so that
%$g_0(B_{qj}|X_j,I_j)=f(V_{qj}|X_j,I_j) \xi_1(B_{qj},X_j,I_j)$.
Then, we can apply a Taylor expansion around $h_g$ to
obtain
\begin{eqnarray*}
\Vert A_{21}\Vert \leq \sum_{I_j}\left\Vert c_1 h_g+c_2 \frac{h_g^2}{2}+\ldots+c_{R-1} \frac{h_g^{R-1}}{(R-1)!}+O(h_g^{R})\right\Vert.
\end{eqnarray*}
We note that the remainder term vanishes, i.e.
$\sqrt{L}h_g^{R}=o(1)$ by A4.AN-(ii). The remaining $R-1$ terms also vanish by A3-(iii). To
see this observe that the $k$th coordinate of $c_\rho$,
$\rho=1,\ldots,R-1$ is
\begin{eqnarray*}
c_{k_\rho}&=&\frac{\partial^\rho}{\partial h_g^\rho}\int [H_k(uh_g+\overline{Y})-H_k(uh_g+\underline{Y})]K_{1g}(-u_2) K_{2g}(-u_1) du|_{h_g=0}\\
&=&\sum_{k_1,\ldots,k_\rho=1}^{2} \int (u_{k_1}\ldots u_{k_\rho})K_{1g}(-u_2) K_{2g}(-u_1)
\frac{\partial^\rho}{\partial Y_{k_1}\ldots \partial Y_{k_\rho}}H_k(\overline{Y}) du\\
&&-\sum_{k_1,\ldots,k_\rho=1}^{2} \int (u_{k_1}\ldots u_{k_\rho})K_{1g}(-u_2) K_{2g}(-u_1)
\frac{\partial^\rho}{\partial Y_{k_1}\ldots \partial Y_{k_\rho}}H_k(\underline{Y})du=0,
\end{eqnarray*}
where $dH_k/dY(y)= m_{1,k}(\xi(y,I),x,I;\theta_0)
g_0(y,I)$. The third equality uses A3-(iii), that is since
$K_{1g}(\cdot)$ and $K_{2g}(\cdot)$ are higher order kernels, all moments of order strictly
smaller than $R-1$ vanish.
This shows that the first term in (\ref{B1122}) indeed is $o_{as}(1)$. 
We still have to show that the second term in (\ref{B1122}) is $O_p(1)$. 
In fact we will not only show that we will also provide the asymptotic linear representation, which gives us the asymptotic variance. 
%But for the sake of clarity we defer the result until Lemma \ref{Lemma A1}, where we also derive the  asymptotic linear representation of our estimator. 
From Equation (\ref{B1122}) we have
\begin{eqnarray*}
\sqrt{L}(\hat\theta-\theta_0)-\sqrt{L}(\tilde{\theta}-\theta_0)=\frac{L(L-1)}{L^2}\frac{2}{\sqrt{L}}
\sum_{\{\ell:I_\ell=I\}}^L \frac{1}{I}\sum_{p=1}^{I}
[r_L(Y_{p\ell},I)-\theta_L]
\end{eqnarray*}
where, $Y_{p\ell}=(B_{p\ell},X_\ell)$ and $r_L(Y_{p\ell},I)={\rm{E}}[\pl|(Y_{p\ell},I)]$ and $\theta_L={\rm{E}}[r_L(Y_{p\ell},I)]={\rm{E}}[\pl]$. 
First, we show that $$r_L(Y_{p\ell},I)=\displaystyle{-\sum_{I}\frac{1}{I(I-1)}
N(Y_{p\ell},I)f_m^{-1}(X_\ell,I)g_0(Y_{p\ell},I)+t_L(Y_{p\ell},I)}.$$
We observe that
\begin{eqnarray*}
r_L(Y_{p\ell},I)&=&{\rm{E}}[\pl|(Y_{p\ell},I)]\\
\!\!\!&=&\!\! \!\Bigg\{\begin{array}{ll}
\int p_L((B_{p\ell},X_\ell,I),(B_{qj},X_j,I)) g_0(B_{qj},X_j,I) dY_{qj} & \mbox{if $\ell\neq j$} \\
\int p_L((B_{pj},X_j,I),(B_{qj},X_j,I))
g_0((B_{pj},X_j,I),(B_{qj},X_j,I)|(B_{pj},X_j,I)) dY_{qj}  &
\mbox{if $\ell= j$.}
\end{array}\Bigg.
\end{eqnarray*}

We consider first the case $\ell\neq j$.
\begin{eqnarray*}
&&r_L(Y_{p\ell},I)=\frac{1}{2}\frac{L}{n_I}\sumj \Bigg\{\int\Bigg[\frac{m_1(V_{p\ell},X_\ell,I;\theta_0)}{g_0(B_{p\ell}|X_\ell,I)}\hGd \kG\\
&&+\frac{m_1(V_{qj},X_j,I;\theta_0)}{g_0(B_{qj}|X_j,I)} \hGd\kGj\Bigg] g_0(Y_{qj},I) \dypj \Bigg\}\\
&&-\frac{1}{2} \frac{L}{n_I}\sumj \Bigg\{\int\Bigg[\frac{m_1(V_{p\ell},X_\ell,I;\theta_0)}{g_0(B_{p\ell}|X_\ell,I)}\frac{G_0(B_{p\ell}|X_\ell,I)}{g_0(B_{p\ell}|X_\ell,I)}\hgdone \omega_{I,R,j}^g K_{1g}\left(\frac{X_j-X_\ell}{h_{g}}\right) \\
&&K_{2g}\left(\frac{B_{qj}-B_{p\ell}}{h_{g}}\right)+\frac{m_1(V_{qj},X_j,I;\theta_0)}{g_0(B_{qj}|X_j,I)}
\frac{G_0(B_{qj}|X_j,I)}{g_0(B_{qj}|X_j,I)}\hgdone \omega_{I,R,j}^g K_{1g}\left(\frac{X_\ell-X_j}{h_{g}}\right)\\ &&K_{2g}\left(\frac{B_{p\ell}-B_{qj}}{h_{g}}\right)\Bigg]g_0(Y_{qj},I)\dypj \Bigg\}\\
&=&\frac{1}{2}\frac{L}{n_I}\sumj \Bigg\{\int\Bigg[ M(Y_{p\ell},I)\hGd \kG\\
&&+M(Y_{qj},I)\hGd\kGj\Bigg] g_0(Y_{qj},I) \dypj \Bigg\}\\
&&-\frac{1}{2}\frac{L}{n_I} \sumj \Bigg\{\int \Bigg[N(Y_{p\ell},I)\hgdone \kgz\\
&&+ N(Y_{qj},I)\hgdone \kgzj\Bigg]g_0(Y_{qj},I) \dypj\Bigg\}\\
&\!\!=&\!\!\!\frac{1}{2}\frac{L}{n_I}\sumj \int h_G\Big[ M(Y_{p\ell},I)\omega_{I,R+1,j}^G K_G(u_2) \Unit(B_{p\ell}\leq u_1h_G+B_{p\ell})+M(uh_G+Y_{p\ell},I)\\
&&\omega_{I,R+1,j}^G K_G(-u_2)\Unit(u_1h_G+B_{p\ell}\leq B_{p\ell})\Big]g_0(uh_G+Y_{p\ell},I)du\\
&&-\frac{1}{2}\frac{L}{n_I}\sumj \int\Big[N(Y_{p\ell},I)\omega_{I,R,j}^g K_{1g}(u_2)K_{2g}(u_1)+N( uh_g+Y_{p\ell},I)\\
&&\omega_{I,R,j}^g K_{1g}(-u_2)K_{2g}(-u_1)\Big]g_0(uh_g+Y_{p\ell},I)du.
\end{eqnarray*}

We note that as $h=(h_G,h_g) \rightarrow 0$ we have
\begin{eqnarray*}
r_L(Y_{p\ell},I) &\longrightarrow& -\frac{1}{2} \frac{1}{I}\sumj \int \Big[N(Y_{p\ell},I) f^{-1}_m(X_\ell,I) K_{1g}(u)] K_{1g}(u_2)K_{2g}(u_1)\\
&&+ N(Y_{p\ell},I_j)f^{-1}_m(X_\ell,I)  K_{1g}(-u_2)K_{2g}(-u_1)\Big]g_0(Y_{p\ell},I)du\\
&&=-\frac{1}{I}\sumj N(Y_{p\ell},I) f^{-1}_m(X_\ell,I)g_0(Y_{p\ell},I)
\end{eqnarray*}

\noindent where we have used the following
\begin{eqnarray*}
\omega_{I,R,j}^g&=&e_1^T \left[\frac{1}{n_I h_g}\sum_{\iota=1}^{n_I} {\mathbf{{x_\iota x_\iota}}}^T K_{1g}
\left(\frac{X_\iota-X_\ell}{h_g}\right) \right]^{-1} [1\quad (X_j-X_\ell)\ldots (X_j-X_\ell)^{R-1}]^T\\
&=&e_1^T \left[\frac{1}{n_I h_g}\sum_{\iota=1}^{n_I} {\mathbf{{x_\iota x_\iota}}}^T K_{1g}
\left(\frac{X_\iota-X_\ell}{h_g}\right) \right]^{-1} [1\quad (-u_2h_g)\ldots (-u_2h_g)^{R-1}]^T\\
&\stackrel{p}{\longrightarrow}& e_1^T \left[{\rm{E}\left({\mathbf{{x_\iota x_\iota}}}^T K_{1g}
\left(\frac{X_\iota-X_\ell}{h_g}\right)\right) } \right]^{-1} e_1=f^{-1}_m(X_\ell,I)
\end{eqnarray*}

\noindent therefore we define
\begin{eqnarray*}
r_L(Y_{p\ell},I_j)=-\frac{1}{I}\sumj N(Y_{p\ell},I) f^{-1}_m(X_\ell,I) g_0(Y_{p\ell},I)+t_L(Y_{p\ell},I).
\end{eqnarray*}

We consider now the reminder term $t_L(Y_{p\ell},I)$
\begin{eqnarray*}
t_L(Y_{p\ell},I)&=&r_L(Y_{p\ell},I)+\frac{1}{I}\sumj N(Y_{p\ell},I) f^{-1}_m(X_\ell,I)g_0(Y_{p\ell},I)\\
&=&\frac{1}{2}\frac{L}{n_I}\sumj \int h_G\Big[ M(Y_{p\ell},I)\omega_{I,R+1,j}^G K_G(u_2) \Unit(B_{p\ell}\leq u_1h_G+B_{p\ell})\\
&&+M(uh_G+Y_{p\ell},I)\omega_{I,R+1,j}^G K_G(-u_2)\Unit(u_1h_G+B_{p\ell}\leq B_{p\ell})\Big]g_0(uh_G+Y_{p\ell},I)du\\
&&-\frac{1}{2}\frac{L}{n_I}\sumj \int\Big[N(Y_{p\ell},I)\omega_{I,R,j}^g K_{1g}(u_2)K_{2g}(u_1)+N( uh_g+Y_{p\ell},I)\\
&&\omega_{I,R,j}^g K_{1g}(-u_2)K_{2g}(-u_1)\Big]g_0(uh_g+Y_{p\ell},I)du\\
&&+\frac{1}{I}\sumj N(Y_{p\ell},I) f^{-1}_m(X_\ell,I) g_0(Y_{p\ell},I)
\end{eqnarray*}

Now, using $\int K_{1g}(u_2)K_{2g}(u_1)du =1$, we can write
\begin{eqnarray*}
\frac{1}{I}\sumj N(Y_{p\ell},I) f^{-1}_m(X_\ell,I)g_0(Y_{p\ell},I)
&=&\frac{1}{2}\frac{1}{I} \sumj\int K_{1g}(u_2)K_{2g}(u_1) N(Y_{p\ell},I) f^{-1}_m(X_\ell,I)g_0(Y_{p\ell},I) du\\
&&+\frac{1}{2}\frac{1}{I}\sumj \int K_{1g}(-u_2)K_{2g}(-u_1) N(Y_{p\ell},I) f^{-1}_m(X_\ell,I)g_0(Y_{p\ell},I) du
\end{eqnarray*}

\noindent therefore we can write the reminder term as follows
\begin{eqnarray*}
t_L(Y_{p\ell},I_j)&=&\frac{1}{2}\frac{L}{n_I}\sumj \int h_G\Big[ M(Y_{p\ell},I)\omega_{I,R+1,j}^G K_G(u_2) \Unit(B_{p\ell}\leq u_1h_G+B_{p\ell})\\
&&\!\!\!+M(uh_G+Y_{p\ell},I)\omega_{I,R+1,j}^G K_G(-u_2)\Unit(u_1h_G+B_{p\ell}\leq B_{p\ell})\Big]g_0(uh_G+Y_{p\ell},I)du\\
&&-\frac{1}{2}\sumj \int \frac{L}{iL_i} N(Y_{p\ell},I) \omega_{I,R,j}^g K_{1g}(u_2)K_{2g}(u_1) g_0(uh_g+Y_{p\ell},I)du\\
&&\!\!\!\!-\!\!\!\!\frac{1}{2}\sumj \int \frac{L}{iL_i} N(uh_g+Y_{p\ell},I)\omega_{I,R,j}^g K_{1g}(-u_2)K_{2g}(-u_1)g_0(uh_g+Y_{p\ell},I)du\\
&&+\frac{1}{2}\sumj \int \frac{1}{I} N(Y_{p\ell},I) f^{-1}_m(X_\ell,I)K_{1g}(u_2)K_{2g}(u_1) g_0(Y_{p\ell},I) du\\
&&+\frac{1}{2}\sumj \int \frac{1}{I} N(Y_{p\ell},I) f^{-1}_m(X_\ell,I)K_{1g}(-u_2)K_{2g}(-u_1) g_0(Y_{p\ell},I) du\\
&=&\frac{1}{2}\frac{L}{n_I}\sumj \int h_G\Big[ M(Y_{p\ell},I)\omega_{I,R+1,j}^G K_G(u_2) \Unit(B_{p\ell}\leq u_1h_G+B_{p\ell})\\
&&+M(uh_G+Y_{p\ell},I)\omega_{I,R+1,j}^G K_G(-u_2)\Unit(u_1h_G+B_{p\ell}\leq B_{p\ell})\Big]g_0(uh_G+Y_{p\ell},I)du\\
&&-\frac{1}{2}\frac{1}{I}\sumj \int N(Y_{p\ell},I) K_{1g}(u_2)K_{2g}(u_1)\Bigg[\frac{L}{L_I}\omega_{I,R,j}^gg_0(uh_g+Y_{p\ell},I)-f^{-1}_m(X_\ell,I)g_0(Y_{p\ell},I)\Bigg] du\\
&&-\frac{1}{2}\frac{1}{I}\sumj \int K_{1g}(-u_2)K_{2g}(-u_1)\\
&&\Bigg[\frac{L}{L_I}\omega_{I,R,j}^g N(uh_g+Y_{p\ell},I)g_0(uh_g+Y_{p\ell},I)-f^{-1}_m(X_\ell,I)N(Y_{p\ell},I)g_0(Y_{p\ell},I)\Bigg] du\\
&=&\frac{1}{2}\frac{L}{n_I}\sumj \int h_G\Big[ M(Y_{p\ell},I)\omega_{I,R+1,j}^G K_G(u_2) \Unit(B_{p\ell}\leq u_1h_G+B_{p\ell})\\
&&+M(uh_G+Y_{p\ell},I)\omega_{I,R+1,j}^G K_G(-u_2)\Unit(u_1h_G+B_{p\ell}\leq B_{p\ell})\Big]g_0(uh_G+Y_{p\ell},I)du\\
&&-\frac{1}{2}\frac{1}{I}\sumj \int N(Y_{p\ell},I) K_{1g}(u_2)K_{2g}(u_1) \left[f^{-1}_m(X_\ell,I)+o_{as}(1)\right][g_0(uh_g+Y_{p\ell},I)-g_0(Y_{p\ell},I)]du\\
&&-\frac{1}{2}\frac{1}{I}\sumj \int K_{1g}(-u_2)K_{2g}(-u_1) \left[f^{-1}_m(X_\ell,I)+o_{as}(1)\right] \\
&&[N(uh_g+Y_{p\ell},I)g_0(uh_g+Y_{p\ell},I)-N(Y_{p\ell},I)g_0(Y_{p\ell},I)]du
\end{eqnarray*}

\noindent thus using the above expression we have
\begin{eqnarray*}
\frac{2}{\sqrt{L}} \sum_{\{\ell:I_\ell=I\}}^L \frac{1}{I}\sum_{p=1}^{I}[r_L(Y_{p\ell},I)-\theta_L]&=&\frac{2}{\sqrt{L}} \sum_{\{\ell:I_\ell=I\}}^L \frac{1}{I}\sum_{p=1}^{I} \Bigg\{-\frac{1}{I}\sumj N(Y_{p\ell},I) f^{-1}_m(X_\ell,I) g_0(Y_{p\ell},I)\\
&&+{\rm{E}}\Big[\frac{1}{I}\sumj N(Y_{p\ell},I) f^{-1}_m(X_\ell,I) g_0(Y_{p\ell},I)\Big]
+t_L(Y_{p\ell},I)-{\rm{E}}[t_L(Y_{p\ell},I)]\Bigg\}\\
&=&-\frac{2}{\sqrt{L}} \sum_{\{\ell:I_\ell=I\}}^L \frac{1}{I}\sum_{p=1}^{I}
\Bigg\{\frac{1}{I}\sumj N(Y_{p\ell},I) f^{-1}_m(X_\ell,I) g_0(Y_{p\ell},I)\\
&&-{\rm{E}}\Big[\frac{1}{I}\sumj N(Y_{p\ell},I) f^{-1}_m(X_\ell,I) g_0(Y_{p\ell},I)\Big] \Bigg\}\\
&&+\frac{2}{\sqrt{L}} \sum_{\{\ell:I_\ell=I\}}^L \frac{1}{I}\sum_{p=1}^{I}\Big[t_L(Y_{p\ell},I)-{\rm{E}}[t_L(Y_{p\ell},I)]\Big].
\end{eqnarray*}
We denote the second term above by $T_L$ and we observe that
${\rm{E}}[T_L]=0$. We now show that ${\rm{var}}[T_L]=o_{as}(1)$.
\begin{eqnarray}
{\rm{var}}[T_L]&=&4\frac{L_I}{L} {\rm{var}}\left[\frac{1}{I}\sum_{p=1}^{I}
t_1(Y_{p1},I)\right]= 4\frac{L_I}{L}{\rm{E}}\left\{{\rm{var}}\left[\frac{1}{I}\sum_{p=1}^{I}
t_1(Y_{p1},I)\Bigg|I \right]\right\} + 4\frac{L_I}{L}
{\rm{var}}\left\{{\rm{E}}\left[\frac{1}{I}\sum_{p=1}^{I}
t_1(Y_{p1},I)\Bigg|I \right]\right\}\nonumber\\
&=&4\frac{L_I}{L}{\rm{E}}\left\{\frac{1}{I}{\rm{var}}\left[
t_1(Y_{p1},I)\Bigg|I \right]\right\}+4\frac{L_I}{L} {\rm{var}}\Big\{
{\rm{E}}\left[ t_1(Y_{p1},I)|I
\right]\Big\}= A+B.
\label{varTL}
\end{eqnarray}
We consider first the $k$th coordinate of the conditional variance
inside the A term above, namely
\begin{eqnarray*}
{\rm{var}}\left[ t_{1_k}(Y_{p1},I)\Big|I\right]&\leq&{\rm{E}}\left[ t_{1_k}(Y_{p1},I)^2\Big|I\right]
\leq O\left(h_G^{2}\right)+O\left(h_g^{2(R-1)}\right),
\end{eqnarray*}

\noindent where the last inequality comes from observing that
\begin{eqnarray*}
t_{1_k}(Y_{p1},I)&=&\frac{1}{2}\frac{L}{iL_i}\sumj \int h_G\Big[ M(Y_{p1},I)\omega_{I,R+1,j}^G K_G(u_2) \Unit(B_{p1}\leq u_1h_G+B_{p1})\\
&&+M(uh_G+Y_{p1},I)\omega_{I,R+1,j}^G K_G(-u_2)\Unit(u_1h_G+B_{p1}\leq B_{p1})\Big]g_0(uh_G+Y_{p1},I)du\\
&&-\frac{1}{2}\frac{1}{I}\sumj \int N(Y_{p1},I) K_{1g}(u_2)K_{2g}(u_1) f^{-1}_m(X_1,I)\\
&&[g_0(uh_g+Y_{p1},I)-g_0(Y_{p1},I)]du\\
&&-\frac{1}{2}\frac{1}{I}\sumj \int K_{1g}(-u_2)K_{2g}(-u_1) f^{-1}_m(X_1,I) [N(uh_g+Y_{p1},I)\\
&&g_0(uh_g+Y_{p1},I)-N(Y_{p1},I)g_0(Y_{p1},I)]du+o_{as}(1)
= a+b+c+o_{as}(1).
\end{eqnarray*}
Therefore, applying $(a+b)^2\leq 2 (a^2+b^2)$ twice yields $(a+b+c)^2\leq \kappa(a^2+b^2+c^2)$, and thus
\begin{eqnarray*}
{\rm{E}}\left[ t_{1_k}(Y_{p1},I)^2\Big|I\right]&=&{\rm{E}}[(a+b+c)^2|I]+o_{as}(1)\leq 4 {\rm{E}}[a^2+b^2+c^2|I]+o_{as}(1)\\
&=& O\left(h_G^{2}\right)+O\left(h_g^{2(R-1)}\right)+O\left(h_g^{2(R-1)}\right),
\end{eqnarray*}
where the order of the last two terms after the last equality follows from $(R-1)^{th}$ Taylor Expansion around $Y_{p1}$ and the kernels are of order $R-1$ by A.3-(iii).
Next, we consider $B$ in (\ref{varTL}):
\begin{eqnarray*}
\frac{B}{4}&=&\frac{L_I}{L}{\rm{var}}\Big\{
{\rm{E}}\left[ t_1(Y_{p1},I)|I
\right]\Big\}\leq{\rm{E}}\Big\{
{\rm{E}}\left[ t_1(Y_{p1},I)|I
\right]^2\Big\}\leq{\rm{E}}\Big\{
{\rm{E}}\left[ t_1(Y_{p1},I)^2|I
\right]\Big\}\\
&\leq& O\left(h_G^{2}\right)+O\left(h_g^{2(R-1)}\right)+O\left(h_g^{2(R-1)}\right),
\end{eqnarray*}
where the last inequality follows from the same argument used above. Hence, by Chebyshev Inequality $T_L=o_p(1)$.
We consider now the case $\ell=j$ and observe the following
\begin{eqnarray*}
r_L(Y_{pj},I)&=&{\rm{E}}[\plj|(B_{pj},X_j,I)]\\
&=&\frac{1}{2}\frac{L}{n_I}\sumj \int
\frac{1}{h_G}\Bigg[ M(Y_{pj},I)
\omega_{I,R+1,j}^G K_G(0)\Unit(B_{qj}\leq B_{pj})\\
&&+ M(Y_{qj},I) \omega_{I,R+1,j}^G K_G(0)\Unit(B_{pj}\leq B_{qj})\Bigg]g_0((Y_{pj},I),(Y_{qj},I)|(Y_{pj},I)) \dypj\\
&&-\frac{1}{2} \frac{L}{n_I}\sumj \int \frac{1}{h_g^{2}}\Bigg[
N(Y_{pj},I) \omega_{I,R,j}^g K_{1g}(0)K_{2g}\left(\frac{B_{qj}-B_{pj}}{h_g}\right) \\
&&+ N(Y_{qj},I)  \omega_{I,R,j}^g K_{1g}(0) K_{2g}\left(\frac{B_{pj}-B_{qj}}{h_g}\right)\Bigg]g_0((Y_{pj},I),(Y_{qj},I)|(Y_{pj},I))\dypj.
\end{eqnarray*}
Making the change of variables $u=(Y_{qj}-Y_{pj})/h_G$ and $\tilde u=(Y_{qj}-Y_{pj})/h_g$ gives 
\begin{eqnarray*}
r_L(Y_{pj},I)&=&\frac{1}{2}\frac{L}{n_I}\sumj \int h_G \Big[ M(Y_{pj},I)\omega_{I,R+1,j}^G K_G(0) \Unit(u_1h_G+B_{pj}\leq B_{pj})+M(uh_G+Y_{pj},I)\\
&&\omega_{I,R+1,j}^G K_G(0) \Unit(B_{pj} \leq u_1h_G+B_{pj})\Big]g_0((Y_{pj},I),(uh_G+Y_{pj},I)|(Y_{pj},I))du\\
&&-\frac{1}{2}\frac{L}{iL_i}\sumj \int\Big[N(Y_{pj},I)\omega_{I,R,j}^g K_{1g}(0)K_{2g}(\tilde u_1)+N(\tilde uh_g+Y_{pj},I)\\
&&\omega_{I,R,j}^g K_{1g}(0)K_{2g}(-\tilde u_1)\Big]g_0((Y_{pj},I),(uh_g+Y_{pj},I)|(Y_{pj},I))d\tilde u.
\end{eqnarray*}
Next, we observe that as $h=(h_G,h_g) \rightarrow 0$ we have
\begin{eqnarray*}
r_L(Y_{pj},I) &\rightarrow& -\frac{1}{2}\frac{1}{I}\sumj \int \Big[N(Y_{pj},I) f_m^{-1}(X_j,I)K_{1g}(0)K_{2g}(\tilde u_1)\\
&&+N(Y_{pj},I) f_m^{-1}(X_j,I)K_{1g}(0)K_{2g}(-\tilde u_1)\Big]g_0((Y_{pj},I),(Y_{pj},I)|(Y_{pj},I))d\tilde u\\
&=&-\frac{1}{I}\sumj N(Y_{pj},I)f_m^{-1}(X_j,I) g_0(Y_{pj},I),
\end{eqnarray*}
and, as before, we define
\begin{eqnarray*}
r_L(Y_{pj},I)=-\frac{1}{I}\sumj N(Y_{pj},I)f_m^{-1}(X_j,I) g_0(Y_{pj},I)+t_L(Y_{pj},I).
\end{eqnarray*}
The rest of the proof is analogous to the one for the case $\ell\neq j$.
Next, we consider $B_{12}$ in (\ref{B11+B12}):
\begin{eqnarray*}
\Vert B_{12}\Vert&=&\Bigg\Vert\dsi m_1(V_{p\ell,Z_\ell;\theta_0}) \sqrt{L}\Bigg[\frac{1}{\gtilde\g}\Bigg(\frac{\G}{\g}\\
&&\!\!\left[\gtilde-\g\right]^2\!\!\!-\!\!\left[\Gtilde-\G\right]\!\!\left[\gtilde-\g\right]\!\Bigg)\!\Bigg]\!\Bigg\Vert\\
&\leq&\left(\dsi \Vert \moneo\Vert^2 \right)^{\frac{1}{2}}\\
&& \!\!\!\!\!\!\Bigg\{\dsi L \Bigg[\frac{1}{\gtilde \g}\!\!\Bigg(\frac{\G}{\g}\left[\gtilde-\g\right]\\
&& -\left[\Gtilde-\G\right]\Big[\gtilde-\g\Big]\Bigg)\Bigg]^2\Bigg\}^{\frac{1}{2}}=B_{121} B_{122},
\end{eqnarray*}
\noindent where the inequality comes from Cauchy-Schwartz.
First we show that $ B_{121}^2 <\infty$ then we show $B_{122}=o(1)$.
\begin{eqnarray*}
B_{121}^2&=&\dsi \Vert\moneo\Vert^2 \leq\dsi \sup_{\theta \in \Theta} \Vert \monet\Vert^2\\
&\leq &\dsi K_7(V_{p\ell},Z_\ell)^2\leq\ds K_7(V_{p\ell},Z_\ell)^2={\rm E}[K_7(V,Z)^2]+o_{as}(1)<\infty
\end{eqnarray*}
\noindent where the second inequality follows from A6-(vi) and $0<1/(I_\ell-1)\leq1$.
Next we show that $B_{122}=o(1)$.
\begin{eqnarray*}
B_{122}\!\!\!\!\!&=&\!\!\!\!\Bigg\{\dsi  \!\! L \Bigg[\frac{1}{\gtilde \g}\Bigg(\frac{\G}{\g}\Big[\gtilde-\g\Big]^2\\
&&-\left[\Gtilde-\G\right]\Big[\gtilde-\g\Big]\Bigg)\Bigg]^2\Bigg\}^{\frac{1}{2}}\\
&< &\sqrt{L} \left\{ \dsi \left[\kappa_1 \left(\kappa_2 O\left(\frac{1}{{r_g}^2}\right)-O\left(\frac{1}{r_G}\right)O\left(\frac{1}{r_g}\right)\right)\right]^2\right\}^{\frac{1}{2}}\\
&\leq&\sqrt{L}\kappa_1 \left[\kappa_2 O\left(\frac{1}{{r_g}^2}\right)-O\left(\frac{1}{r_G}\right)O\left(\frac{1}{r_g}\right)\right]=\kappa_1\left[\kappa_2
O\left(\frac{\sqrt{L}}{{r_g}^2}\right)-O\left(\frac{\sqrt{L}}{r_G
r_g}\right)\right]=o(1),
\end{eqnarray*}

\noindent where we have used: 
$\Bigg\vert\displaystyle{ \frac{1}{\gtilde \g}}\Bigg\vert
<\kappa_1<\infty$, and $\Bigg\vert \displaystyle{\frac{\G}{\g}}\Bigg\vert<\kappa_2<\infty$ since the densities are bounded away from zero and $\gtilde \stackrel{a.s}{\longrightarrow} \g $ from Proposition B2 in GPV (2000);
%(ii) $\Bigg\vert \displaystyle{\frac{\G}{\g}}\Bigg\vert<\kappa_2<\infty$ since $g_0(\cdot,\cdot)$ is bounded away from zero; 
%since densities are bounded away from zero and $\gtilde \stackrel{a.s}{\longrightarrow} \g $; \\
 \begin{eqnarray*}
O\left(\frac{1}{r_g^2}\right)&=&\Big\vert\gtilde-\g\Big
\vert^2=O\left(h_{1g}^{2R}+h_{2g}^{2R}+\frac{\log L}{L h_{1g} h_{2g}}\right),\\
O\left(\frac{1}{r_G}\right)O\left(\frac{1}{r_g}\right)&=&\Big\vert\Gtilde-\G\Big \vert
\Big\vert\displaystyle{\gtilde-\g}\Big
\vert\\
&=&O\left(h_{G}^{R+1}+\sqrt{\frac{\log L}{L h_{G}}}\quad\right)
O\left(h_{1g}^R+h_{2g}^R+\sqrt{\frac{\log L}{L h_{1g}h_{2g}}}\quad\right);
\end{eqnarray*}
and $\forall\ell, 0<1/(I_\ell-1)\leq1.$ Therefore, $B_{12}=o(1)=o(1)$ in (\ref{B11+B12}).
Next, we consider $B_2$ in (\ref{B_step3})
\subsection*{Step 3.2}
\begin{eqnarray}
\Vert B_2\Vert&\leq& \sqrt{L} \ds
\Vert \moneo-\moneob \Vert\vert\hat{V}_{p\ell}- V_{p\ell}\vert \notag\\
&\leq& \sqrt{L}\ds  K_6(Z_\ell) \vert V_{p\ell}-V^*_{p\ell}\vert\vert\hat{V}_{p\ell}-V_{p\ell}\vert\leq\sqrt{L} \ds  K_6(Z_\ell) (\hat{V}_{p\ell}-V_{p\ell})^2 \notag\\
&\leq&\sqrt{L} \sup_{p,\ell}(\hat{V}_{p\ell}-V_{p\ell})^2 \ls K_6(Z_\ell)=\sqrt{L} O_{as}\left(\frac{1}{r^2}\right) O_{as}(1)\leq O_{as}\left(\frac{\sqrt{L}}{r^2}\right) O_{as}(1)\notag\\
&=&O_{as}\left(\frac{L^{1/4}}{r}\right) O_{as}(1)=o_{as}(1),
\label{B_2}
\end{eqnarray}
\noindent where the second inequality follows from A6-(v), the third
from $\hat{V}_{p\ell}\leq V^*_{p\ell}\leq V_{p\ell}$ and the last from A4.AN.\end{proof}

\newpage
\bibliographystyle{econometrica}
%\bibliography{/Users/gaurabaryal/Dropbox/collusion/tex/references}
\bibliography{../../collusion/tex/references}

\end{document}